\documentclass[twocolumn]{aastex63}
\pdfoutput=1 
\usepackage{amsmath,amstext}
\usepackage[T1]{fontenc}
\usepackage{apjfonts} 
\usepackage[figure,figure*]{hypcap}
\usepackage{upgreek}
\usepackage{float}
\usepackage{rotating}
\usepackage{boldline}
\usepackage{listings}
\usepackage{hyperref}
\usepackage{xcolor}

\usepackage{xspace}
\newcommand{\Teff}{$T_\mathrm{eff}$\xspace}
\newcommand{\Lbol}{$L_\mathrm{bol}$\xspace}
\usepackage{booktabs}

\graphicspath{{./}{figures/}}

\received{July 14, 2020}
\revised{September 21, 2020}
\accepted{\today}
\submitjournal{ApJ}

\shorttitle{Retrieval of SDSS J1416$+$1348AB}
\shortauthors{Gonzales et al.}

\begin{document}

\title{Retrieval of the d/sdL7+T7.5p binary SDSS J1416$+$1348AB}

\correspondingauthor{Eileen Gonzales}
\email{egonzales@amnh.org}

\author[0000-0003-4636-6676]{Eileen C. Gonzales}
\altaffiliation{51 Pegasi b Fellow}
\altaffiliation{LSSTC Data Science Fellow}
\affiliation{Department of Astronomy and Carl Sagan Institute, Cornell University, 122 Sciences Drive, Ithaca, NY 14853, USA}
\affiliation{Department of Astrophysics, American Museum of Natural History, New York, NY 10024, USA}
\affiliation{The Graduate Center, City University of New York, New York, NY 10016, USA}
\affiliation{Centre for Astrophysics Research, School of Physics, Astronomy and Mathematics, University of Hertfordshire, Hatfield AL10 9AB}
\affiliation{Department of Physics and Astronomy, Hunter College, City University of New York, New York, NY 10065, USA}

\author[0000-0003-4600-5627]{Ben Burningham}
\affiliation{Centre for Astrophysics Research, School of Physics, Astronomy and Mathematics, University of Hertfordshire, Hatfield AL10 9AB}

\author[0000-0001-6251-0573]{Jacqueline K. Faherty}
\affiliation{Department of Astrophysics, American Museum of Natural History, New York, NY 10024, USA}

\author{Colleen Cleary}
\affiliation{Department of Astrophysics, American Museum of Natural History, New York, NY 10024, USA}

\author[0000-0001-6627-6067]{Channon Visscher}
\affiliation{Chemistry \& Planetary Sciences, Dordt University, Sioux Center, IA}
\affiliation{Center for Extrasolar Planetary Systems, Space Science Institute, Boulder, CO}

\author[0000-0002-5251-2943]{Mark Marley}
\affiliation{NASA Ames Research Center, Moffett Field, CA 94035, USA}

\author[0000-0003-3444-5908]{Roxana Lupu}
\affiliation{BAER Institute/ NASA Ames Research Center, Moffett Field, CA 94035, USA}

\author{Richard Freedman}
\affiliation{Seti Institute, Mountain View, CA }
\affiliation{NASA Ames Research Center, Moffett Field, CA 94035, USA}

\begin{abstract}
We present the distance-calibrated spectral energy distribution (SED) of the d/sdL7 SDSS J14162408$+$1348263A (J1416A) and an updated SED for SDSS J14162408$+$1348263B (J1416B). We also present the first retrieval analysis of J1416A using the \textit{Brewster} retrieval code base and the second retrieval of J1416B. We find that the primary is best fit by a non-grey cloud opacity with a power-law wavelength dependence, but is indistinguishable between the type of cloud parameterization. J1416B is best fit by a cloud-free model, consistent with the results from \cite{Line17}. Most fundamental parameters derived via SEDs and retrievals are consistent within 1$\sigma$ for both J1416A and J1416B. The exceptions include the radius of J1416A, where the retrieved radius is smaller than the evolutionary model-based radius from the SED for the deck cloud model, and the bolometric luminosity which is consistent within 2.5$\sigma$ for both cloud models. The pair's metallicity and Carbon-to-Oxygen (C/O) ratio point towards formation and evolution as a system. By comparing the retrieved alkali abundances while using two opacity models, we are able to evaluate how the opacities behave for the L and T dwarf. Lastly, we find that relatively small changes in composition can drive major observable differences for lower temperature objects.

\end{abstract}


\section{Introduction} \label{sec:intro}
Brown dwarfs are a class of astronomical objects that straddle the mass boundary between stars and planets with masses below $\leq75$ $M_\mathrm{Jup}$ \citep{Saum96, Chab97} and effective temperatures of $250-3000$ K, corresponding to late-type M, L, T, or Y spectral types \citep{Burg02a,Kirk05,Cush11}. Due to electron degeneracy, they never reach a core temperature high enough for stable Hydrogen burning, but instead, contract and cool through their lifetimes progressing through spectral classifications as they age. 

Field aged-brown dwarfs anchor the spectral type scheme, however, low-gravity, low-metallicity, and color outliers expand the standard scheme. Low-metallicity sources, known as subdwarfs, have unusually blue near-infrared (NIR) $J-K$ colors \citep{Burg03c,Burg09a} compared to equivalent field sources. Spectral features distinguishing them from field dwarfs include enhanced metal-hydride absorption bands (e.g. FeH), weak or absent metal oxides (TiO, CO, VO), and enhanced collisionally-induced H$\,_{2}$ absorption (\citealt{Burg03c} and references therein). Subdwarfs also exhibit substantial radial velocities, high proper motions, and inclined, eccentric, and sometimes retrograde Galactic orbits indicating membership in the Galactic Halo \citep{Dahn08, Burg08a, Cush09}. To date, as classified by \cite{Zhang2017a, Zhang2018a, Zhang2018b, Zhang2019a} there are approximately 66 L subdwarfs and 41 T subdwarfs. Although most T subdwarfs are not classified as such in previous literature \citep[see Table 3 of ][]{Zhang2019a}. To be identified as a T subdwarf in \cite{Zhang2019a}, T dwarfs need to have a suppressed $K$ band spectrum.  

Presently there is only one subdwarf L+T system, SDSS J14162408$+$1348263AB (hereafter J1416AB), and it is ideally suited for low-metallicity bd-bd binary atmospheric characterization via retrievals. In this paper, we determine and examine fundamental parameters and atmospheric features of J1416AB via two methods: (1) by coupling the empirical bolometric luminosity, from the distance-calibrated spectral energy distribution (SED), with evolutionary models and (2) atmospheric retrievals where we explore similarities and differences between the pair to determine their formation and evolution and to understand their individual atmospheric structure. 

In Section \ref{sec:literature} we present literature data on J1416AB. Section \ref{sec:1416data} presents data used for creating distance-calibrated SEDs and the retrievals, as well as the resultant fundamental parameters derived from creating the SED. Section \ref{sec:RetrievalModel} describes our retrieval framework and  setup for J1416AB. Retrieval results for J1416A and J1416B are discussed in Sections~\ref{sec:Retrieval_Models_A} and \ref{sec:Retrieval_Models_B} respectively. Fundamental parameters derived from SED and retrieval methods are compared in Section~\ref{sec:FundParmComp1416} to the literature and evolutionary models. Lastly, Section~\ref{sec:discussion1416} brings together the individual retrievals of J1416AB to discuss the alkali abundance, metallicity, and Carbon-to-Oxygen (C/O) ratios derived and what we can interpret for the system as a whole.

\section{Literature Data on SDSS J1416AB}\label{sec:literature}  
At the time of discovery, J1416AB was one of the few known widely separated L+T systems, thus allowing for the properties of both to be examined in tandem. This system is a benchmark as features of the primary indicate an old age for the system. Here we present the literature data for the independent discoveries of the L and T dwarfs.

\subsection{Literature Data on SDSS J1416A}
SDSS J141624.08+134826.7 (here after J1416A) was discovered independently via a variety of methods by \cite{Burn10,Schm10a} and \cite{Bowl10}. It was initially overlooked in color based searches due to its unusually blue NIR color ($J-K=1.03\pm 0.03$) \citep{Schm10a}, suggesting a low metallicity and/or high surface gravity \citep{Burn10}. The spectral type of J1416A is agreed to be bluer than normal in the literature; however, the spectral type varies with classifications of: d/sdL7 by \cite{Burn10}, sdL7 by \cite{Kirk10,Kirk16,Zhang2017a}, and as a blue L dwarf by both \cite{Schm10a} (L5 optical, L4 NIR) and \cite{Bowl10} (L6 optical, L6p NIR). There are currently three optical spectra (\citealt{Schm10a} SDSS and MagE and \citealt{Kirk16} Palomar), 3 NIR spectra (Spex Prism: \citealt{Schm10a, Bowl10}, Spex SXD: \citealt{Schm10a}), and one L band spectrum \citep{Cush10} available of J1416A. 

The most precise proper motions and parallax for J1416A is provided by $Gaia$ DR2 \citep{GaiaDR1,GaiaDR2,Lind18}, with previous measurements by \cite{Schm10a} (proper motions) and \cite{Dupu12a,Fahe12} (parallax). Radial velocity measurements have been reported by SDSS DR7 (\citealt{Abaz09}), \cite{Schm10a}, and \cite{Bowl10}. $UVW$ kinematic measurements place J1416A in the thin disk \citep{Schm10a, Bowl10} and in Table~\ref{tab:1416data} we present updated $UVW$ kinematics using $Gaia$ DR2 proper motions and parallax paired with the radial velocity from \cite{Schm10a}. 

Many studies aimed to determine the fundamental properties of J1416A by fitting its spectrum to self-consistent grid models \citep{Burn10,Schm10a, Bowl10, Cush10}. Its atmosphere was determined to be relatively dust-free \citep{Burn10} and like other blue L dwarfs it possibly had a thin or patchy cloud deck with large grains which could cause the observed blue NIR colors. Additionally, J1416A might have an older age and higher surface gravity \citep{Schm10a, Bowl10}. \cite{Cush10} found evidence for vertical mixing in the atmosphere due to the lack of CH$_4$ absorption at 3.3~$\mu$m. Temperature estimates of J1416A vary from $1500-2200$~K \citep{Burn10, Bowl10, Schm10a, Cush10}, while the literature agrees on a surface gravity of 5.5 dex \citep{Burn10, Bowl10, Cush10}, and a weakly or unconstrained age \citep{Burn10, Bowl10, Schm10a}.

J1416A was examined for variability in \cite{Khan13}, \cite{Metc15}, and \cite{Miles17}. \cite{Khan13} found marginal evidence of variability detected in one night of their observations using Gemini camera $J$ and $K'$ bands on the Shane Telescope. \cite{Metc15} monitored J1416A using Spitzer ch1 (14 hours) and ch2 (7 hours) as part of their Weather on other Worlds survey to look for variability attributed to patchy clouds, finding no evidence for variability. In \cite{Miles17}, variability correlated to activity was tested using the Gemini Multi-object spectrograph (GMOS-N) with the R831-G5302 grating, but no evidence for variability was found.

Values for J1416A from the literature and those determined in this work are listed in Table~\ref{tab:1416data}. All literature values are also listed in Table~\ref{tab:LitFunParamsA} for comparison in Section~\ref{sec:FundParmComp1416}.

\subsection{Literature Data on SDSS J1416B}
ULAS J141623.94+1348836.30 (hereafter J1416B) was discovered by \cite{Burn10} through a cross match of SDSS and UKIRT  finding a separation of 9" between the A and B component. J1416B was also independently discovered by \cite{Scho10} with a projected separation of 75AU, which we have updated (now 83.7 AU) using the \textit{Gaia} DR2 parallax and the angular separation from \cite{Burn10}. Like J1416A, J1416B has unusual features of a late-T dwarf. Particularly: the CH$_4-J$-early peculiarity (where the CH$_4$-$J$ index on the red side of the $J$ band peak suggests an earlier spectral type than the H$_2$O-$J$ index on the blue side of the $J$ band peak), the very blue $H-K$ color, and the extremely red $H-[4.5]$ color, leading to its classification as a T7.5p \citep{Burn10}. At the time of its discovery, J1416B was both the bluest $H-K$ and reddest $H-[4.5]$ T dwarf. The CH$_4-J$-early peculiarity of 1416B pointed towards either low metallicity or high surface gravity \citep{Burn10}. It was noted that J1416B forms a sequence with other low-metallicity and high-gravity T dwarfs and because of the extremely red $H-[4.5]$ color it could not be ruled out as a binary itself \citep{Burn10}. \cite{Burg10b} classified J1416B as a T7.5 but noted strong water and methane bands, a possible detection of ammonia between $1-1.3\,\mu$m, and a broadened $Y$-band peak and suppressed $K$ band indicative of high gravity or low metallicity in its spectrum. \cite{Kirk16} regarded J1416B as an sdT7.5 in relation to J1416A and \cite{Zhang2017a} also classified it as sdT7.5 via their subdwarf metallicity classification scheme. Presently, J1416B has three NIR prism spectra (IRCS, SpeX, and FIRE) from \citealt{Burn10} and \citealt{Burg10b,Burg10c} respectively.

Fundamental parameters of J1416B were determined through comparison to grid models \citep{Burg10b, Burg10c}, spectral energy distribution (SED) fitting \citep{Fili15}, and atmospheric retrieval \citep{Line17}. \cite{Burg10b} found that J1416B was well matched to the archetype blue T dwarf 2MASS J09393548$-$2448279. They determined a \Teff$=650\pm50$~K, log\,$g=5.2\pm0.4$, [Fe/H]\,$\leq-0.3$, and K$_{zz}=10^{4}$ using the \cite{Saum08} models and used Baraffe evolutionary models to find an age range of $2-10$~Gyr, mass between $22-47$~M$_\mathrm{Jup}$, and radius of 0.83~$R_\mathrm{Jup}$. Both cloudless and cloudy models were fit to the spectrum of J1416B in \cite{Burg10c}, with cloudy models producing a marginally better fit to the data bringing the temperature closer to that inferred by its mid-infrared colors. \cite{Fili15} improved upon the fundamental parameters from \cite{Burg10b, Burg10c} by determining semi-empirical parameters based on its distance-calibrated SED. Most recently, \cite{Line17} retrieved its thermal profile and derived fundamental parameters, metallicity ([M/H]), and a C/O ratio. Values for J1416B from the literature and those determined in this work are listed in Table~\ref{tab:1416data}. J1416B was studied in \cite{Metc15} for variability with no evidence found in Spitzer ch1 and ch2. All literature values are also listed in Table~\ref{tab:LitFunParamsB} for comparison in Section~\ref{sec:FundParmComp1416}.

\startlongtable 
\begin{deluxetable*}{l c c c c c c}
\tablecaption{Properties of the J1416$+$1348AB System\label{tab:1416data}}
\tablehead{\colhead{Property} & \colhead{J1416A} &\colhead{} &\colhead{}& \colhead{J1416B} &\colhead{} &\colhead{}\\
            \colhead{} & \colhead{Value} &\colhead{Reference} &\colhead{} &\colhead{Value} &\colhead{Reference} & \colhead{}} 
  \startdata
  \hline
  Spectral type & d/sdL7 & 1 && T7.5p& 1\\ \hline
  &&&\multicolumn{1}{c}{\textbf{Astrometry}} && \\ \hline
  R.A. & $14^h 16^m 24.08^s$ & 2 && $14^h 16^m 23.94^s$ & 1\\ 
  Decl. & $+13 ^\circ 48' 26''.3$ & 2 && $+13 ^\circ 48' 36''.3$ & 1\\
  R.A. (epoch 2015.0)& $214.1 \pm 0.30$ & 3 && $\cdots$ & $\cdots$\\ 
  Decl. (epoch 2015.0) & $+13.81 \pm 0.22$ & 3 && $\cdots$ & $\cdots$\\
  $\pi$ (mas) & $107.56 \pm 0.30$  & 3 && $107.56 \pm 0.30$  & 3\\ 
  $\mu_\alpha$ (mas yr$^{-1}$) & $85.69\pm0.69$ & 3 && $221 \pm 33$ & 1\\  
  $\mu_\delta$ (mas yr$^{-1}$) & $129.07\pm0.47$ & 3 && $115\pm45$ & 1\\ 
  $V_{r}$ (km s$^{-1}$) & $-42.2 \pm 1.24$ & 4 && $\cdots$ & $\cdots$\\ 
  $V_\mathrm{tan}$ (km s$^{-1}$) & $-42.2 \pm 5.1$ & 4 && $\cdots$ & $\cdots$\\ 
  $U$ (km s$^{-1}$)\tablenotemark{a} & $-17.84\pm 0.50$ & 5 && $\cdots$ & $\cdots$\\ 
  $V$ (km s$^{-1}$)\tablenotemark{a} & $5.81\pm 0.04$ & 5 && $\cdots$ & $\cdots$\\ 
  $W$ (km s$^{-1}$)\tablenotemark{a} & $-38.4 \pm 1.1$& 5 && $\cdots$ & $\cdots$\\ \hline
  &&&\multicolumn{1}{c}{\textbf{Photometry}} && \\ \hline
  SDSS $r$ (mag) & $20.69 \pm 0.04$ & 6 && $\cdots$ & $\cdots$\\
  SDSS $i$ (mag) & $18.38 \pm 0.01$ & 6 && $25.21 \pm 0.26$ & 7\\
  SDSS $z$ (mag) & $15.917 \pm 0.007$ & 6 && $20.87 \pm 0.09$& 7\\
  PS $r$ (mag) & $20.42 \pm 0.01$ & 8 && $\cdots$ & $\cdots$\\
  PS $i$ (mag) & $18.35 \pm 0.01$ & 8 && $\cdots$ & $\cdots$\\
  PS $z$ (mag) & $16.3 \pm 0.01$ & 8 && $\cdots$ & $\cdots$\\
  PS $y$ (mag) & $15.21 \pm 0.01$ & 8 && $19.8 \pm 0.06$ & 8\\
  2MASS $J$ (mag) & $13.148 \pm 0.021$ & 2 && $\cdots$ & $\cdots$\\
  2MASS $H$ (mag) & $12.456 \pm 0.027$ & 2 && $\cdots$ & $\cdots$\\
  2MASS $K_{s}$ (mag) & $12.114 \pm 0.021$ & 2 && $\cdots$ & $\cdots$\\
  $Y_\mathrm{MKO}$ (mag) & $14.25 \pm 0.01$ & 1 && $18.13 \pm 0.02$ & 1\\
  $J_\mathrm{MKO}$ (mag) & $12.99 \pm 0.01$ & 1 &&  $17.35 \pm 0.02$ & 1\\
  $H_\mathrm{MKO}$ (mag) & $12.47 \pm 0.01$ & 1 &&  $17.62 \pm 0.02$ & 1\\
  $K_\mathrm{MKO}$ (mag) & $12.05 \pm 0.01$ & 1 &&  $18.93 \pm  0.17$ & 1\\
  WISE $W1$ (mag) & $11.364 \pm 0.022$ & 9 && $16.12 \pm 0.20$ &10\\
  WISE $W2$ (mag) & $11.026 \pm 0.02$ & 9 && $12.791 \pm 0.038$ &10\\
  WISE $W3$ (mag) & $10.26 \pm 0.055$ & 9 && $12.19 \pm 0.23$ &10\\
  IRAC [3.6] (mag) & $10.99 \pm 0.07$ & 1 && $14.69 \pm 0.05$ & 1\\ 
  IRAC [4.5] (mag) & $10.98 \pm 0.05$ & 1 && $12.76 \pm 0.03$ & 1\\ \hline
  &&&\multicolumn{1}{c}{\textbf{System}} && \\ \hline
  &&&\multicolumn{1}{c}{Value \phm{stringz} Reference} && \\ \hline
  Separation (") & $\cdots$ & $\cdots$ &\multicolumn{1}{c}{9  \phm{stringzzzzzzzzz} 1}& $\cdots$ & $\cdots$\\ 
  Separation (AU) & $\cdots$ & $\cdots$ &\multicolumn{1}{c}{83.7  \phm{stringzzzzzz} 5}& $\cdots$ & $\cdots$\\ \hline 
  &&&\multicolumn{1}{c}{\textbf{Parameters from SED}\tablenotemark{b}} && \\ \hline 
  \Lbol & $-4.18 \pm 0.011$ & 5 && $-5.80 \pm 0.07$ & 5\\ 
  \Teff (K) & $1694\pm 74$ & 5 && $660 \pm 62$ & 5\\
  Radius ($R_\mathrm{Jup}$) & $0.92\pm0.08$ & 5 && $0.94 \pm 0.16$ & 5\\ 
  Mass ($M_\mathrm{Jup}$ )& $60\pm18$ & 5 && $33 \pm 22$ & 5\\
  log $g$ (dex) & $5.22\pm0.22$ & 5 && $4.83 \pm 0.51$ & 5\\                                       
  Age (Gyr) & $0.5-10$ & 5 && $0.5-10$ & 5\\
  Distance (pc) & $9.3\pm0.3$ & 5 && $9.3 \pm 0.03$ & 5\\ \hline
  &&&\multicolumn{1}{c}{\textbf{Retrieved Parameters\tablenotemark{c}\tablenotemark{d}}} && \\ \hline
  & Value & Reference & Model & Value & Reference & Model \\ \hline
  &&&\multicolumn{1}{c}{\textit{Allard Alkalies}} && \\ \hline
  log $g$ (dex) & $5.26\substack{+0.32 \\ -0.33}$ & 5 & power-law deck cloud& $5.00\substack{+0.28 \\ -0.41}$ & 5 & cloud-free\\ 
  \Lbol & $-4.23 \pm 0.01$ & 5 & power-law deck cloud& $-5.93\substack{+0.05 \\ -0.04}$ & 5 & cloud-free\\ 
  \Teff (K) & $1891.47\substack{+42.56 \\ -41.38}$ & 5 & power-law deck cloud& $659.05\substack{+15.33 \\ -13.21}$ & 5 & cloud-free\\ 
  Radius ($R_\mathrm{Jup}$) & $0.7\pm0.04$ & 5 & power-law deck cloud& $0.81\substack{+0.07 \\ -0.06}$ & 5 & cloud-free\\ 
  Mass ($M_\mathrm{Jup}$) & $36.82\substack{+31.92 \\ -18.71}$ & 5 & power-law deck cloud& $26.01\substack{+22.68 \\ -16.07}$ & 5 & cloud-free\\
  {C/O}\tablenotemark{e} & $0.59\substack{+0.11 \\ -0.21}$ & 5 & power-law deck cloud& $0.52\substack{+0.09 \\ -0.07}$ & 5 & cloud-free\\
  {C/O}$_\mathrm{AB}$\tablenotemark{f} & $0.59\substack{+0.11 \\ -0.21}$ & 5 & power-law deck cloud& $0.53\substack{+0.10 \\ -0.08}$ & 5 & cloud-free\\
  {[M/H]}\tablenotemark{g} & $-0.19\substack{+0.21 \\ -0.23}$ & 5 & power-law deck cloud & $-0.38\substack{+0.15 \\ -0.17}$ & 5 & cloud-free\\
  {[M/H]}$_\mathrm{AB}$\tablenotemark{h} & $-0.17\substack{+0.21 \\ -0.23}$ & 5 & power-law deck cloud & $-0.35\substack{+0.15 \\ -0.17}$ & 5 & cloud-free\\ 
  {[M/H]}$_\mathrm{Line17}$\tablenotemark{i} & $\cdots$ & $\cdots$ & $\cdots$ & $-0.36\substack{+0.14 \\ -0.18}$ & 5 & cloud-free\\\hline 
  log $g$ (dex) & $5.18\substack{+0.28 \\ -0.36}$ & 5 & power-law slab cloud & $\cdots$ & $\cdots$ & $\cdots$\\ 
  \Lbol & $-4.21 \pm 0.01$ & 5 &  power-law slab cloud & $\cdots$ & $\cdots$ & $\cdots$\\ 
  \Teff (K) & $1821.53\substack{+64.58 \\ -102.79}$ & 5&  power-law slab cloud & $\cdots$ & $\cdots$ & $\cdots$\\
  Radius ($R_\mathrm{Jup}$) & $0.77\substack{+0.10 \\ -0.06}$ & 5&  power-law slab cloud & $\cdots$ & $\cdots$ & $\cdots$\\ 
  Mass ($M_\mathrm{Jup}$) & $36.96\substack{+30.48 \\ -18.71}$ & 5&  power-law slab cloud & $\cdots$ & $\cdots$ & $\cdots$\\
  {C/O}\tablenotemark{e} & $0.58\substack{+0.11 \\ -0.21}$ & 5 & power-law slab cloud & $\cdots$ & $\cdots$ & $\cdots$\\
  {C/O}$_\mathrm{AB}$\tablenotemark{f} & $0.58\substack{+0.11 \\ -0.21}$ & 5 & power-law slab cloud & $\cdots$ & $\cdots$ & $\cdots$\\
  {[M/H]}\tablenotemark{g} & $-0.35\substack{+0.20 \\ -0.26}$ & 5 & power-law slab cloud & $\cdots$ & $\cdots$ & $\cdots$ \\ 
  {[M/H]}$_\mathrm{AB}$\tablenotemark{h} & $-0.33\substack{+0.20 \\ -0.26}$ & 5 & power-law slab cloud & $\cdots$ & $\cdots$ & $\cdots$ \\\hline
  &&&\multicolumn{1}{c}{\textit{Burrows Alkalies}} && \\ \hline
  log $g$ (dex) & $5.42\substack{+0.23 \\ -0.29}$ & 5 & power-law deck cloud& $4.77\substack{+0.32 \\ -0.34}$ & 5 & cloud-free\\ 
  \Lbol & $-4.22 \pm 0.01$ & 5 & power-law deck cloud& $-5.90\pm 0.04$ & 5 & cloud-free\\ 
  \Teff (K) & $1904.69\substack{+39.99\\-42.49}$ & 5 & power-law deck cloud& $653.05\substack{+16.01 \\ -13.23}$ & 5 & cloud-free\\ 
  Radius ($R_\mathrm{Jup}$) & $0.69\pm0.04$ & 5 & power-law deck cloud& $0.86\pm 0.06$ & 5 & cloud-free\\ 
  Mass ($M_\mathrm{Jup}$) & $51.76\substack{+28.21 \\ -24.33}$ & 5 & power-law deck cloud& $17.22\substack{+15.67 \\ -9.07}$ & 5 & cloud-free\\
  {C/O}\tablenotemark{e} & $0.60\substack{+0.10 \\ -0.16}$ & 5 & power-law deck cloud& $0.50\substack{+0.10 \\ -0.07}$ & 5 & cloud-free\\
  {C/O}$_\mathrm{AB}$\tablenotemark{f} & $0.60\substack{+0.10 \\ -0.16}$ & 5 & power-law deck cloud& $0.50\substack{+0.11 \\ -0.07}$ & 5 & cloud-free\\
  {[M/H]}\tablenotemark{g} & $-0.11\substack{+0.18 \\ -0.21}$ & 5 & power-law deck cloud & $-0.50\substack{+0.16 \\ -0.14}$ & 5 & cloud-free\\ 
  {[M/H]}$_\mathrm{AB}$\tablenotemark{h} & $-0.09\substack{+0.18 \\ -0.21}$ & 5 & power-law deck cloud & $-0.47\substack{+0.16 \\ -0.14}$ & 5 & cloud-free\\
  {[M/H]}$_\mathrm{Line17}$\tablenotemark{i} & $\cdots$ & $\cdots$ & $\cdots$ & $-0.47\substack{+0.15 \\ -0.14}$ & 5 & cloud-free\\\hline
  log $g$ (dex) & $5.31\substack{+0.24 \\ -0.34}$ & 5 & power-law slab cloud & $\cdots$ & $\cdots$ & $\cdots$\\ 
  \Lbol & $-4.22\substack{+0.02 \\ -0.01}$ & 5 &  power-law slab cloud & $\cdots$ & $\cdots$ & $\cdots$\\ 
  \Teff (K) & $1859.07\substack{+61.09 \\ -110.17}$ & 5&  power-law slab cloud & $\cdots$ & $\cdots$ & $\cdots$\\
  Radius ($R_\mathrm{Jup}$) & $0.73\substack{+0.11 \\ -0.06}$ & 5&  power-law slab cloud & $\cdots$ & $\cdots$ & $\cdots$\\ 
  Mass ($M_\mathrm{Jup}$) & $45.73\substack{+27.91 \\ -22.22}$ & 5&  power-law slab cloud & $\cdots$ & $\cdots$ & $\cdots$\\
  {C/O}\tablenotemark{e} & $0.57\substack{+0.11 \\ -0.26}$ & 5  &  power-law slab cloud & $\cdots$ & $\cdots$ & $\cdots$\\
  {C/O}$_\mathrm{AB}$\tablenotemark{f} & $0.57\substack{+0.11 \\ -0.26}$ & 5  &  power-law slab cloud & $\cdots$ & $\cdots$ & $\cdots$\\
  {[M/H]}\tablenotemark{g} & $-0.30\substack{+0.21 \\ -0.27}$ & 5 &  power-law slab cloud & $\cdots$ & $\cdots$ & $\cdots$ \\
  {[M/H]}$_\mathrm{AB}$\tablenotemark{h} & $-0.29\substack{+0.21 \\ -0.27}$ & 5 &  power-law slab cloud & $\cdots$ & $\cdots$ & $\cdots$ \\
  \enddata
  \tablenotetext{a}{We do not correct for LSR.}
  \tablenotetext{b}{Using \cite{Saum08} low-metallicity (M/H$=-0.3$) evolutionary models, assuming an age of $0.5-10$~Gyrs.}
  \tablenotetext{c}{\Lbol, \Teff, radius, mass, C/O ratio, {[Fe/H]}, and {[M/H]} are not directly retrieved parameters, but are calculated using the retrieved $R^2/D^2$ and log $g$ values along with the predicted spectrum. C/O ratio not relative to the Sun, it is absolute. Solar C/O is 0.55.}
  \tablenotetext{d}{J1416A is best fit using Allard alkalies, while J1416B is best fit with Burrows. We conclude the Allard alkali opacities provide the best fit across both sources.}
  \tablenotetext{e}{Atmospheric C/O using constrained gases. J1416A (both models): H$_2$O, CO, CH$_4$, and VO. J1416B: H$_2$O and CH$_4$ (Same gases as used in \cite{Line17} here without the rainout correction).}
  \tablenotetext{f}{Atmospheric C/O using only the gases in common between J1416AB: H$_2$O, CH$_4$, and CO.}
  \tablenotetext{g}{Metallicity determined using all constrained gases, J1416A: H$_2$O, CO, CH$_4$, VO, CrH, FeH, and Na+K. J1416B: H$_2$O, CH$_4$, NH$_3$, Na+K.}
  \tablenotetext{h}{Metallicity determined using only the gases in common between J1416AB: H$_2$O, CH$_4$, CO, and Na+K. }
  \tablenotetext{i}{Metallicity using the same gases as \cite{Line17}: H$_2$O and CH$_4$ and without the rainout correction.}
\tablerefs{(1) \cite{Burn10}, (2) \cite{Cutr03}, (3) \cite{GaiaDR1,GaiaDR2,Lind18}, (4) \cite{Schm10a}, (5) This Paper, (6) \cite{Abaz09}, (7) \cite{Legg12}, (8) \cite{Cham16}, (9) \cite{Cutr13}, (10) \cite{Cutr12}}
\end{deluxetable*}

\section{Data used and results from generating the SED} \label{sec:1416data}  

\begin{deluxetable*}{l c c c c}
\tablecaption{Spectra used to construct SEDs and for Retrievals \label{tab:SpectraReferences1416}}
\tablehead{\colhead{Name} & \colhead{Spectrum} &\colhead{Obs. Date} & \colhead{Ref.} & \colhead{Use}}  
  \startdata
  J1416A & SpeX SXD, LXD1.9 & 2009--06--29, 2010--01--29 & 1 & SED\\
  J1416A & Spex prism & 2009-06-28 & 3 & Retrieval\\
  J1416B & Spex prism & 2001--10--23 & 2 & Both\\
  \enddata
  \tablerefs{(1) \cite{Cush10}, (2) \cite{Burg10b}, (3) \cite{Schm10a}}
\end{deluxetable*} 

\begin{figure*}
\gridline{\fig{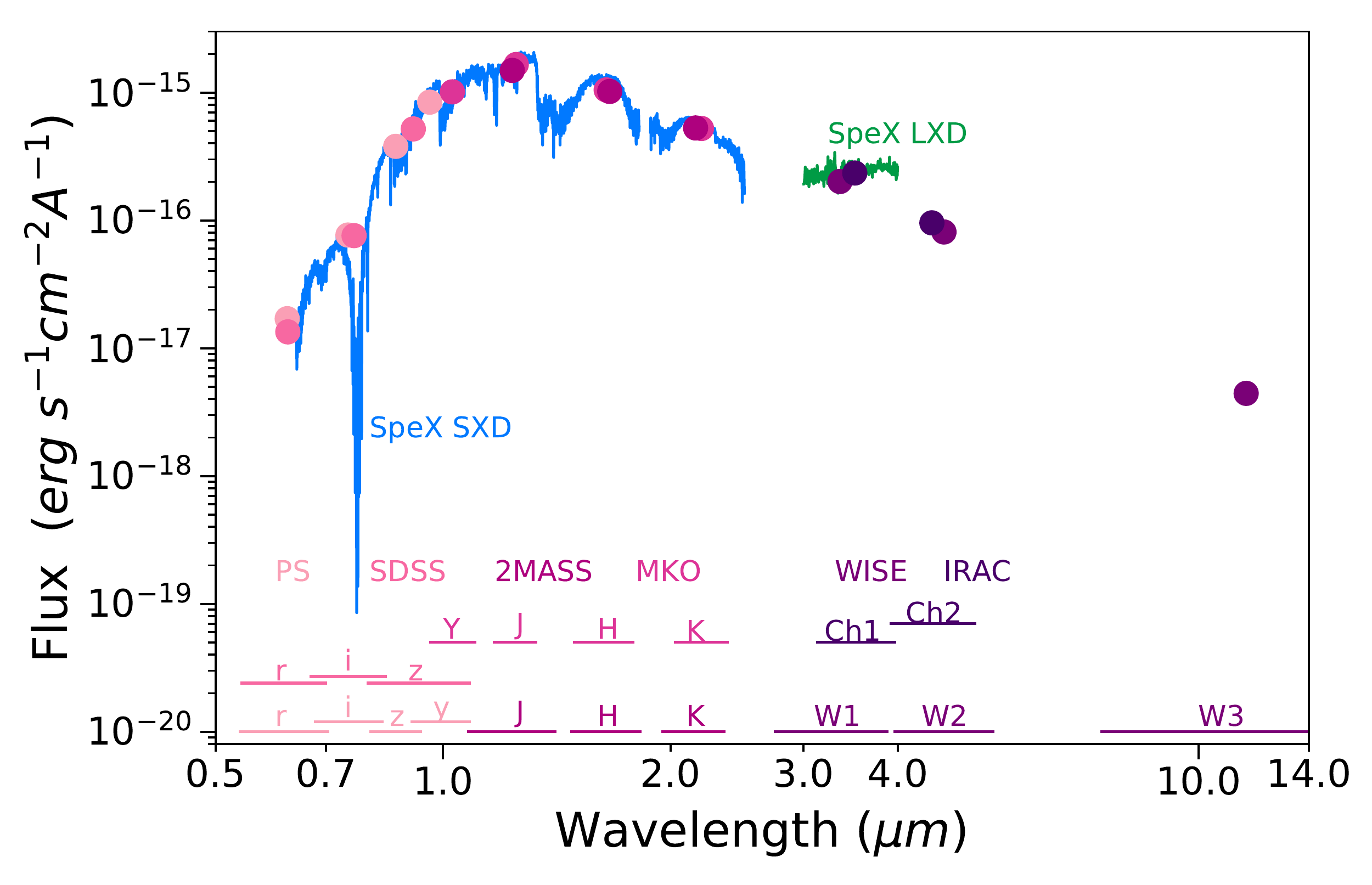}{0.5\textwidth}{\large(a)}
          \fig{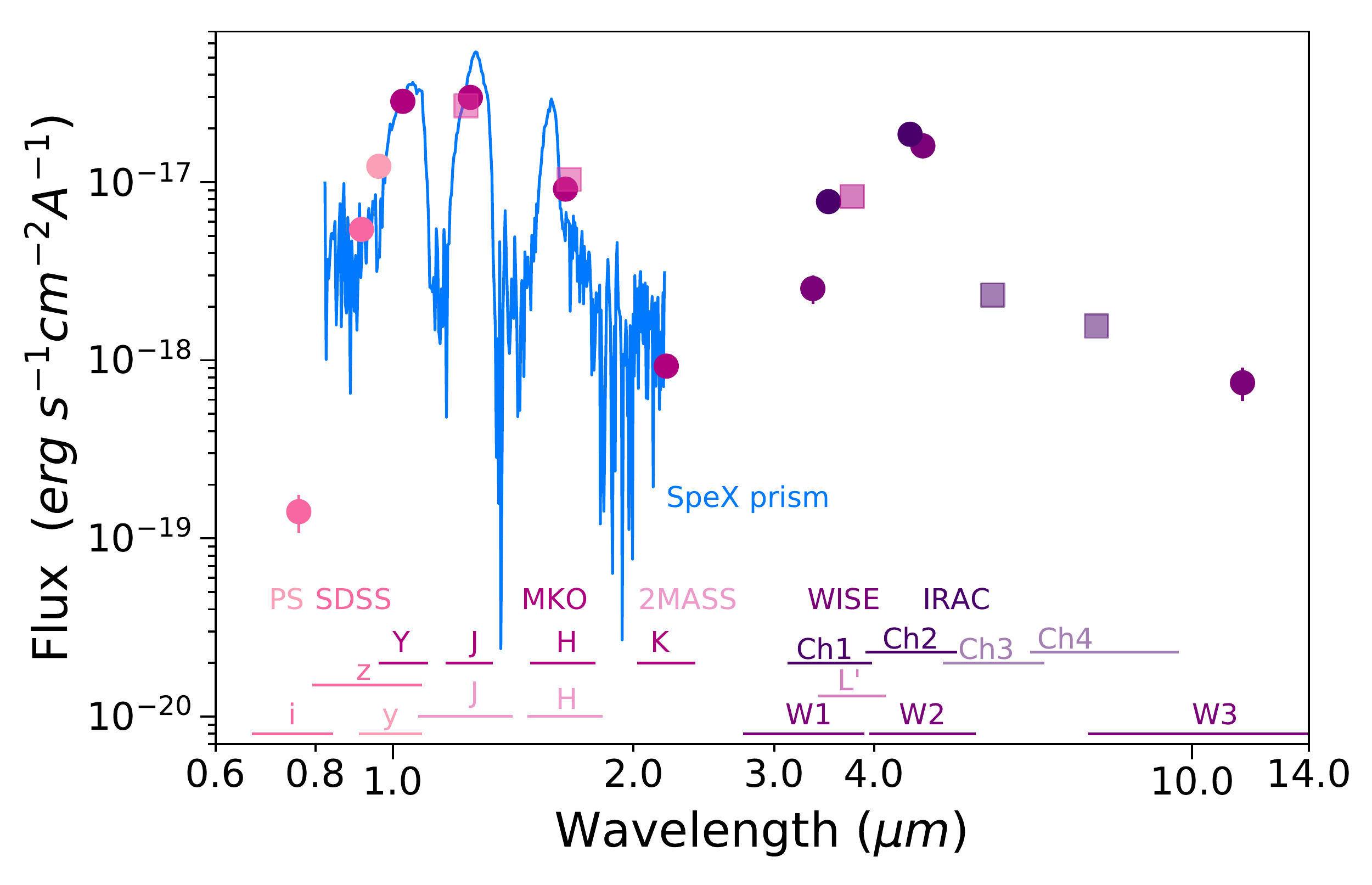}{0.5\textwidth}{\large(b)}} 
\caption{SEDs of J1416$+$1348AB. Photometry (shades of pink and purple) are labeled by instrument or filter system. The horizontal lines at the bottom show the wavelength coverage for the corresponding photometric measurement. Error bars on the photometric points are smaller than the point size. Observation references can be found in Tables \ref{tab:1416data} and \ref{tab:SpectraReferences1416}. (a) Distance-calibrated SED of J1416A. Spex SXD in blue, SpeX LXD in green. No estimated photometry (b) Distance-calibrated SED of J1416B. Spex prism in blue. Estimated synthetic photometry shown as transparent squares.}
\label{fig:1416ABSEDs}
\vspace{0.5cm} 
\end{figure*}

The fundamental parameters for J1416AB were determined using the technique of \cite{Fili15}, where we create a distance-calibrated SED using the spectra, photometry, and parallax\footnote{SEDkit is available on GitHub at \url{https://github.com/hover2pi/SEDkit}. The Eileen branch was used for this work.}. The SED of J1416A uses the SpeX short-cross-dispersed (SXD) and long-cross-dispersed (LXD) spectrum from \cite{Cush10}, while J1416B uses the SpeX prism spectrum from \cite{Burg10b}. The photometry and \textit{Gaia} parallax used for both sources are listed in Table \ref{tab:1416data}. Table~\ref{tab:SpectraReferences1416} lists the spectra used in the SEDs and the retrieval models, which differ for J1416A due to the current time constraints on data resolution for our retrieval model.

To generate the SED of J1416A, the SpeX SXD and LXD spectra were stitched, linearly interpolating to fill gaps in the data, into a composite spectrum and then scaled to the absolute magnitudes of the observed photometry. For J1416B we scale the SpeX prism spectrum to the absolute magnitudes of observed and synthetic (those calculated based on empirical relations) photometry. Synthetic photometry for J1416B is included because if we linearly interpolated between W2 and W3, without including the synthetic MIR IRAC Ch3 and Ch4 photometry calibrated based on field dwarfs, we would likely overestimate the MIR flux compared to most T dwarfs, causing a noticeable change in the \Teff. As there are no known low-metallicity T dwarfs with IRAC Ch3 or Ch4 MIR photometry, we cannot place a level of error on their difference from field T dwarfs. The SEDs of J1416A and J1416B are shown in Figure~\ref{fig:1416ABSEDs}, with the synthetic magnitudes used for J1416B plotted as transparent squares in Figure~\ref{fig:1416ABSEDs}(b).  

The bolometric luminosity (\Lbol) was determined by integrating under the distance-calibrated SED from 0 to 1000 $\mu$m, using a distance of $9.3\pm0.3$~pc based on the \cite{GaiaDR2} parallax measurement. The effective temperature (\Teff) was calculated using the Stefan-Boltzmann law with the resultant inferred radius from the cloudless \cite{Saum08} low metallicity (-0.3 dex) evolutionary model. The low metallicity models were chosen for the assumed radius due to the literature spectral type classification of sd for both components. Additionally as done in \cite{Fili15}, the \cite{Chab00}, \cite{Bara03}, and cloud-free \cite{Saum08} evolutionary models were also used to determine the radius. The final radius range was set as the maximum and minimum from all model predictions as done in \citealt{Fili15}. An age range of $0.5-10$ Gyr for the system was chosen to conservatively encompass possible field and subdwarf ages. Additional details on the SED generation can be found in \cite{Fili15}. Fundamental parameters derived for J1416A and J1416B using this approach are listed in Table \ref{tab:1416data} and are compared to the literature in Section~\ref{sec:FundParmComp1416} (Also see Tables~\ref{tab:LitFunParamsA} and \ref{tab:LitFunParamsB}).

\section{The \textit{Brewster} Retrieval Framework \label{sec:RetrievalModel}} 
Our retrievals use the \textit{Brewster} framework \citep{Burn17} with a modified setup from the one in \cite{Burn17} in order to optimize for low metallicity atmospheres. A summary of the \textit{Brewster} framework with our modifications is discussed below. We differ from \cite{Burn17} with a higher resolution for opacity sampling, using a second method (thermochemical equilibrium with rainout) for determining gas abundances, and expanded temperature and mass priors. A more detailed description of \textit{Brewster} can be found in \cite{Burn17}.

\subsection{The forward Model}
The forward model in \textit{Brewster} uses the two-stream radiative transfer technique of \cite{Toon89}, including scattering, as first introduced by \cite{Mckay89} and subsequently used by e.g. \citet{Marl96, Saum08, Morl12}. We use a 64 pressure layer (65 levels) atmosphere with geometric mean pressures between $\log P = -4$ and $2.3$ bars in 0.1~dex spaced intervals. The temperature in each layer is characterized by the three exponential functions as done following the \cite{Madh09} parameterization, splitting the atmosphere in three zones where the pressure and temperature are related by:   

\begin{equation}
\begin{aligned}
    P_0 < P < P_1&: P_0e^{\alpha_1(T-T_0)^{1/2}}\; (\mathrm{Zone\; 1}) \\
    P_1 < P < P_3&: P_2e^{\alpha_2(T-T_2)^{1/2}}\; (\mathrm{Zone\; 2}) \\
    P > P_3&: T = T_3\;\;\;\;\;\;\;\;\;\;\;\;\,  (\mathrm{Zone\; 3}) 
\end{aligned}
\label{eqn:madhu}
\end{equation}
where $P_0$ and $T_0$ are the pressure and temperature at the top of the atmosphere and the atmosphere becomes isothermal at pressure $P_3$ with temperature $T_3$. Since $P_{0}$ is fixed in our model and continuity at the zonal boundaries requires fixing two parameters, we consider six free parameters: $\alpha_{1}$, $\alpha_{2}$, $P_1$, $P_2$, $P_3$, and $T_3$. A thermal inversion can occur when $P_{2} > P_{1}$, however, this is ruled out by setting $P_{2} = P_{1}$ thus further simplifying this to five free parameters.

\subsection{Gas Opacities} 
Layer optical depths due to absorbing gases are calculated using opacities sampled at a resolving power R~$= 10000$ taken from \cite{Free08,Free14}.  Line wing profiles based on the unified line shape theory \citep{nallard2007a, nallard2007b} are used to account for the broadening of the D resonance doublets of \ion{Na}{1} ($\sim 0.59~\mu$m) and \ion{K}{1} ($\sim 0.77~\mu$m) in brown dwarf spectra. Tabulated line profiles (Allard N., private communication) are calculated for the \ion{Na}{1} and \ion{K}{1} D1 and D2 lines broadened by collisions with H$_{2}$ and He, for temperatures in the $500-3000$~K range and perturber (H$_{2}$ or He) densities up to $10^{20}$ cm$^{-3}$ with two collisional geometries considered for broadening by H$_{2}$. Within 20~cm$^{-1}$ of the line center a Lorentzian profile with a width calculated from the same theory. While there are updated versions of these opacities \citep{nallard16,nallard19,Phil20}, we did not have access to them for this work. We also use the Na and K alkali opacities from \cite{Burrows_Alkalies} to be consistent with \cite{Line17} in the J1416B retrievals.

Across our temperature-pressure regime, the line opacities are tabulated in 0.5~dex steps for pressure and in steps ranging from 20~K to 500~K as we move from 75~K to 4000~K in temperature where we then linearly interpolate this to our working pressure grid. We include free-free continuum opacities for H$^{-}$ and H$_2^{-}$ and bound-free continuum opacity for H$^{-}$, which are influenced by the H$^-$ metallicity and determined from the thermochemical equilibrium grid (see Section~\ref{sec:gas_abundances}). Continuum opacities for H$_{2}$-H$_{2}$ and H$_{2}$-He collisionally induced absorption, using cross-sections from \citet{Rich12} and \citet{Saum12} are included, as well as Rayleigh scattering due to H$_{2}$, He and CH$_{4}$ but we neglect the remaining gases. Neutral H gas fraction abundance determined from the thermochemical equilibrium grid. The atmosphere is assumed to be dominated by H$_2$ and He, with proportions of (0.84H$_2+$ 0.16He) based on Solar abundances. After including the retrieved gases, neutral H, H$^{-}$, and electrons, H$_2$ and He are assumed to make up the remainder of the gas in a layer. The former is drawn from the thermochemical equilibrium grids discussed later in this section.

\subsection{Determining Gas Abundances \label{sec:gas_abundances}} 
As done in \cite{Burn17}, we use the uniform-with-altitude mixing ratios method for absorbing gases and retrieve these directly, also known as ``free'' retrievals, for all of our retrieval models. While simple, the uniform-with-altitude mixing method cannot capture important variations in gas abundance with altitude for some species (i.e. see metal-oxides and metal-hydrides of J1416A and the alkalies for J1416B) which can vary by several orders of magnitude in the photosphere and are expected to have a large contribution to the flux we observe. Freely retrieving abundances that vary with altitude would be preferred; however, the resultant large number of parameters to solve for in this approach is computationally difficult. To address this issue we use a second method, the chemical equilibrium method, which instead retrieves [Fe/H] and C/O. Gas fractions in each layer of this method are pulled from tables of thermochemical equilibrium abundances as a function T, P, [Fe/H], C/O ratio along with the thermal profile of a given state-vector. The thermochemical equilibrium grids we use were calculated using the NASA Gibbs minimization CEA code \citep{McBr94}, based on previous thermochemical models \citep{Fegl94,Fegl96,Lodd99,Lodd02,Lodd02b,Lodd10,Lodd06,Viss06,Viss10a,Viss12,Moses12,Moses13} and recently utilized to explore gas and condensate chemistry over a range of conditions in substellar atmospheres \citep{Morl12,Morl13,Skem16,Kata16,Wake17}. The chemical grids in this work determine equilibrium abundances of atmospheric species over pressures ranging from 1 microbar to 300 bar, temperatures between $300-4000$~K, metallicities ranging from $-1.0 < [\mathrm{Fe/H}] < +2.0$, and C/O abundance ratios of $0.25$ to $2.5$x the solar abundance.

\subsection{Cloud Model \label{sec:cloudmodel}}
The cloud model follows that of \cite{Burn17}, with options for a ``deck'' or ``slab'' cloud parameterization.  Both clouds are defined similarly where the cloud's opacity is distributed among layers in pressure space, with the optical depth either grey or as a power-law ($\tau = \tau_0\lambda^\alpha$, where $\tau_0$ is the optical depth at 1~$\mu$m).

The deck cloud is parameterized by: (1) a cloud top pressure, $P_{top}$, the point at which the cloud passes $\tau=1$ (looking down), (2) the decay height, $\Delta \log P$, over which the optical depth falls to lower pressures as $d\tau/dP \propto \exp((P-P_{deck}) / \Phi)$ where $\Phi = (P_{top}(10^{\Delta \log P} - 1))/(10^{\Delta \log P})$, and (3) the cloud particle single-scattering albedo. The deck cloud becomes optically thick at $P_{top}$. At $P >P_{top}$, the optical depth increases following the decay function until it reaches $\Delta \tau_{layer} = 100$. With this decay function, the deck cloud can quickly become opaque with increasing pressure and therefore we obtain essentially no atmospheric information from deep below the cloud top. Because of this, it is important to note that the pressure-temperature (PT) profile (and spread) below the deck is an extension of the gradient (and spread) at the cloud top pressure.

Unlike the deck cloud, it is possible to see the bottom of the slab cloud and thus include an additional parameter for determining the total optical depth at 1$\mu$m ($\tau_{cloud}$), bringing the total number of parameters for the slab cloud to 4. The optical depth is distributed through the slab cloud extent as $d\tau / dP \propto P$ (looking down), reaching its total value at the bottom (highest pressure) of the slab. In principle the slab can have any optical depth, however, we restrict our prior as $0.0 \leq \tau_{cloud} \leq 100.0$. Because it is possible to see to the bottom of the slab cloud a physical extent in log-pressure ($\Delta \log P$) is determined, instead of the decay scale as done for the deck cloud.

If the deck or slab cloud is non-grey, an additional parameter for the power ($\alpha$) in the optical depth is included.

\subsection{Retrieval Model}\label{sec:Retmodelsetup}
As described in \cite{Burn17}, we use EMCEE \citep{emcee} to sample posterior probabilities. Table \ref{tab:Priors} shows our priors for both J1416A and J1416B. We differ from the \cite{Burn17} setup by extending the thermal profile temperature up to 6000~K for both J1416A and J1416B and extending the mass prior up to 100~$M_\mathrm{Jup}$ for J1416A (up to only 80~$M_\mathrm{Jup}$ for J1416B) to expand the surface gravity in an effort to encompass likely ranges for subdwarfs. In our retrievals of J1416A and J1416B we use their distance-calibrated SpeX prism spectra (output from generating our SED) trimmed to the $1.0-2.5\,\mu$m region and set the distance to 10~pc. This spectrum calibration differs from \cite{Burn17}, where they calibrated the spectrum to the 2MASS $J$-band photometry and used the true distance in their initialization. 

\begin{deluxetable*}{l c c}
\tablecaption{Priors for J1416$+$1348AB retrieval models\label{tab:Priors}} 
\tablehead{\colhead{Parameter} &\phm{stringzzzzzzzz} &\colhead{Prior}}
  \startdata
  gas volume mixing ratio && uniform, log $f_{gas} \geq -12.0$, $\sum_{gas} f_{gas} \leq 1.0$ \\
  thermal profile ($\alpha_{1}, \alpha{2}, P1, P3, T3$) && uniform, $0.0\, \mathrm{K} < T < 6000.0\, \mathrm{K}$\\ 
  scale factor ($R^2/D^2$) && uniform, $0.5\,R_\mathrm{Jup} \leq\, R\, \leq 2.0\,R_\mathrm{Jup}$ \\
  gravity (log\,$g$)\tablenotemark{a} && uniform, $1\,M_\mathrm{Jup} \leq\; gR^2/G\; \leq 100\,M_\mathrm{Jup}$ \\
  cloud top\tablenotemark{b} && uniform, $-4 \leq \mathrm{log}\, P_{CT} \leq 2.3$\\ 
  cloud decay scale\tablenotemark{c} && uniform,$0< \mathrm{log}\,\Delta\, P_{decay}<7$\\
  cloud thickness\tablenotemark{d} && uniform, log\,$P_{CT} \leq\,$log $(P_{CT}+\Delta P)\, \leq2.3$\\
  cloud total optical depth at $1\mu$m && uniform,  $0.0 \geq \tau_{cloud} \geq 100.0$ \\
  single scattering albedo ($\omega_0$) && uniform, $0.0 \leq \omega_0 \leq 1.0$\\
  wavelength shift && uniform, $-0.01 < \Delta \lambda <0.01 \mu$m \\
  tolerance factor && uniform, log($0.01 \times min(\sigma_{i}^2)) \leq b \leq$ log$(100 \times max(\sigma_{i}^2)) $\\
  \enddata
  \tablenotetext{a}{Gravity prior upper limit only to 80~$M_\mathrm{Jup}$ for J1416B.}
  \tablenotetext{b}{For the deck cloud this is the pressure where $\tau_{cloud} = 1$, for a slab cloud this is the top of the slab.}
  \tablenotetext{c}{Decay height for deck cloud above the $\tau_{cloud} = 1.0$ level.}
  \tablenotetext{d}{Thickness and $\tau_{cloud}$ only retrieved for slab cloud.}
\end{deluxetable*}

For each retrieval of J1416A and J1416B we initialize 16 walkers per parameter in a tight gaussian for the gases, surface gravity, wavelength shift between the model and data ($\Delta\lambda$), and scale factor where $R \approx 1.0\, R_\mathrm{Jup}$. Gases are centered around the approximate solar composition equilibrium chemistry values for gas volume mixing ratios, while the surface gravity is initiated centered around the SED-derived value. The tolerance parameter has a flat distribution across the entire prior range. For cloud parameters, the cloud top pressure and power-law are initialized as tight Gaussians, while the optical depth, albedo, and cloud thickness are flat across the entire prior range. As in \cite{Burn17}, we use the five parameter thermal profile, as we do not expect a temperature inversion for either of these objects, and use the \cite{Saum08} \Teff$=1700$~K log\,$g=5.0$ model to initialize $\alpha_{1}$, $\alpha_{2}$, $P_1$, $P_2$, $P_3$, and $T_3$ for both J1416A and J1416B. Differences in the individual setups between J1416A and J1416B are discussed in the following subsection.

\subsubsection{J1416A}
To explore the atmosphere of J1416A, we retrieved for the following gases: H$_2$O, CO, CO$_2$, CH$_4$, TiO, VO, CrH, FeH, K, and Na. As done in \cite{Burn17} and \cite{Line15} we tie K and Na together as a single element in the state-vector assuming a Solar ratio taken from \cite{Aspl09}.  Additionally, we include the H$^-$ bound-free and free-free continuum opacities to account for the possibility of the profile going above 3000K in the photosphere. As stated above the log\,$g$ mass prior ranges from $0-100$~M$_\mathrm{Jup}$. The multiple cloud parameterizations are tested building up from the cloudless to the 4 parameter power-law slab cloud model. We also test both the uniform-with-altitude mixing ratios and chemical equilibrium methods for determining the gas abundances.

\subsubsection{J1416B}
The retrieval setup and initialization for 1416B is similar to J1416A with the following exceptions: (1) as J1416B is much cooler we retrieve only H$_2$O, CH$_4$, CO$_2$, NH$_3$, K and Na (where Na and K are tied together), and (2) we do not include the H$^-$ bound-free and free-free continuum opacities as the profile is cooler than the L dwarf and does not warrant them. As the T dwarf should be less massive the log\,$g$ mass prior ranges from $0-80$~M$_\mathrm{Jup}$. We also differ from the retrieval setup of that in \cite{Line17} by (1) excluding CO$_2$ and H$_2$S in our gas list as \cite{Line17} could only derive upper limits and (2) testing both the Allard and Burrows alkali opacities.

\subsection{Model Selection}
A variety of parameters were tested in our retrievals of J1416A and J1416B with some aspects remaining constant throughout (the gases included in each model) while others differed. The aspects that were allowed to differ in our retrievals include cloud parameterization, gas abundance method, and alkali opacities. To compare all of our retrievals, model selection was assessed using the Bayesian Information Criterion (BIC) where the lowest BIC is preferred. We use the following intervals from \cite{Kass95} for selecting between two models, with evidence against the higher BIC as:

\begin{itemize}
    \item 0< $\Delta$BIC <2: no preference worth mentioning
    \item 2< $\Delta$BIC <6: positive 
    \item 6< $\Delta$BIC <10: strong 
    \item $\Delta$BIC >10: very strong 
\end{itemize}

A variety of cloud assumptions are explored in our retrievals by building up from the least complex cloud-free model to the most complex slab cloud model. Prior to moving from the cloud-free to cloudy models, we tested the impact of assuming different metallicities when determining the neutral H, H$^{-}$, and electron abundances used for the continuum opacity calculations as both targets are expected to be low metallicity. We found using low-metallicity ([M/H]$=0.3$) ion fractions to be indistinguishable from the solar metallicity ion fractions and thus proceeded using the solar ion abundances for the cloudy models. 

Once the ``winning'' model was determined, we tested two additional methods for calculating gas abundances: (1) the thermochemical equilibrium assumption and (2) alternate opacities based on the Burrows and Allard line broadening treatments (Burrows for J1416A, and Allard for J1416B) in the uniform-with-altitude assumption. We examined both Allard and Burrows alkali opacities as there is no agreement in the literature as to which is the preferred choice in retrievals or grid models (\citealt{Saum08, Todo16, Burn17, Line17, Nowak20}, Marley et al. in prep). As done in \cite{Burn17} we started with the Allard opacities for the L dwarf and as done in \cite{Line17} we started with the Burrows opacities for the T dwarf. By testing the alternative line profile treatments, we aim to establish the impact of this choice on the derived alkali abundances for J1416AB.

\section{Retrieval Model of J1416A}\label{sec:Retrieval_Models_A}

The $\Delta$BIC for all tested models for J1416A are shown in Table \ref{tab:1416AModels}. The best-fitting model is parameterized as a power-law deck cloud. However, this model is indistinguishable from the power-law slab cloud ($\Delta$BIC$=1.40$), meaning both models provide similarly good fits to the spectroscopic features observed in J1416A. The retrieval results of the power-law deck and power-law slab cloud models are discussed in \ref{sec:1416ADeck89} and \ref{sec:1416ASlab89} respectively. The ``winning'' deck and slab cloud models were also indistinguishable when instead, using the Burrows alkali opacities. Section~\ref{sec:Alkalies} provides further discussion of the preferred choice of alkali opacities for comparing J1416A to J1416B.

\begin{deluxetable*}{l c c }
\tablecaption{$\Delta$BIC for J1416A retrieval models\label{tab:1416AModels}} 
\tablehead{\colhead{Model} & \colhead{Number of Parameters} & \colhead{$\Delta$BIC}}
  \startdata
  Cloud Free & 18 & 8.8 \\
  Cloud Free, [M/H]= -0.3 for ions & 18 & 9.0 \\
  Grey Deck cloud & 21 & 11.3 \\
  Grey Slab cloud & 22 & 18.5 \\
  Power-law Deck cloud & 22 & 0 \\
  Power-law Deck cloud Chemical Equilibrium & 15 & 9.2 \\
  Power-law Deck cloud, Burrows Alkali & 22 & 0.7 \\
  Power-law Slab cloud & 23 & 1.4 \\
  Power-law Slab cloud, Chemical Equilibrium & 16 & 20.0 \\
  Power-law Slab cloud, Burrows Alkali & 23 & 2.1 \\    
  \enddata
  \tablecomments{Unless otherwise listed default alkali opacities used are from Allard.}
\end{deluxetable*}

\subsection{Best-fit Model: Power-law deck cloud \label{sec:1416ADeck89}}
\subsubsection{Pressure-Temperature Profile and Contribution Function\label{sec:PT_cont_deck}}
\begin{figure*}
\gridline{\fig{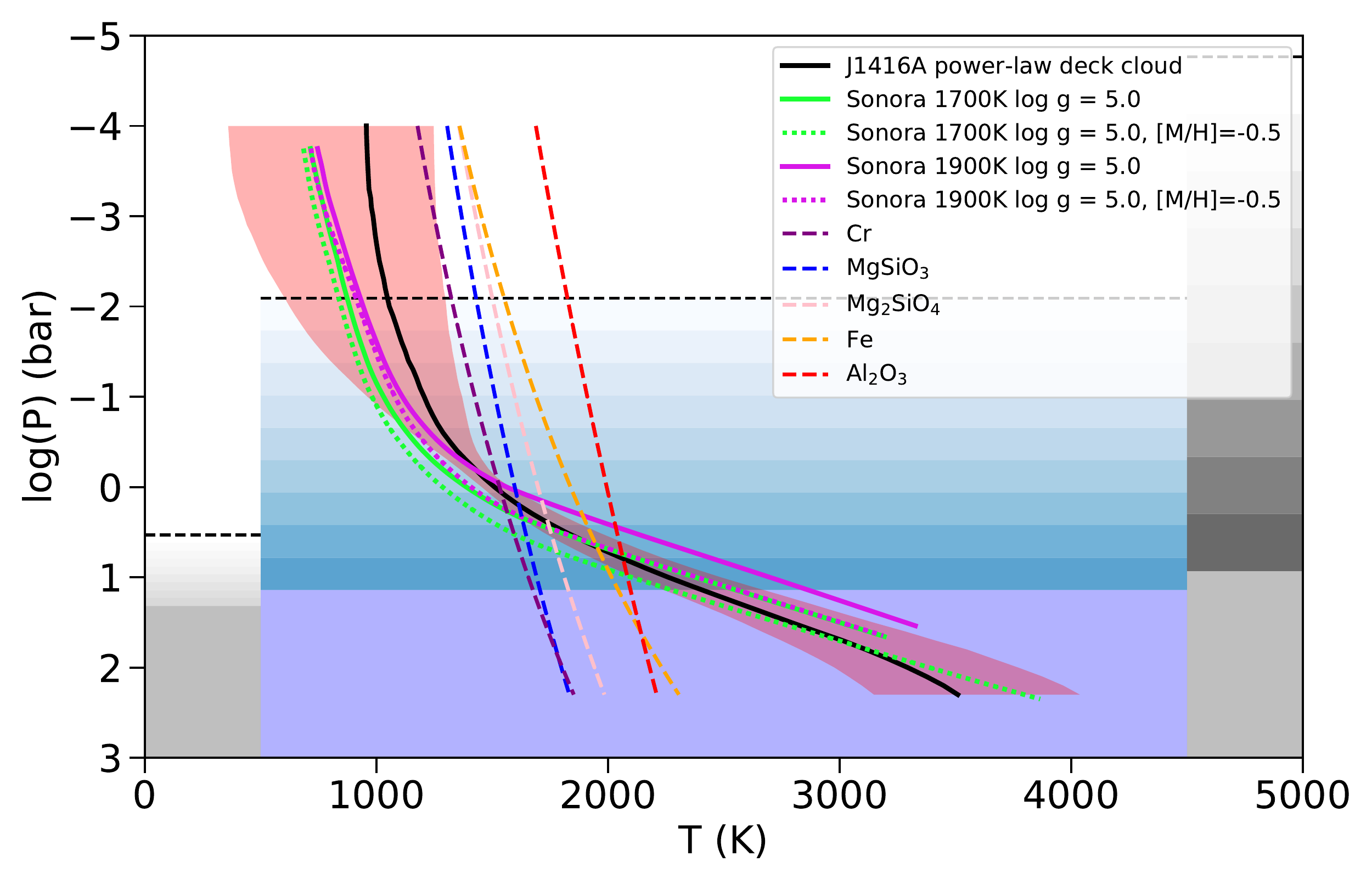}{0.5\textwidth}{\large(a)}
          \fig{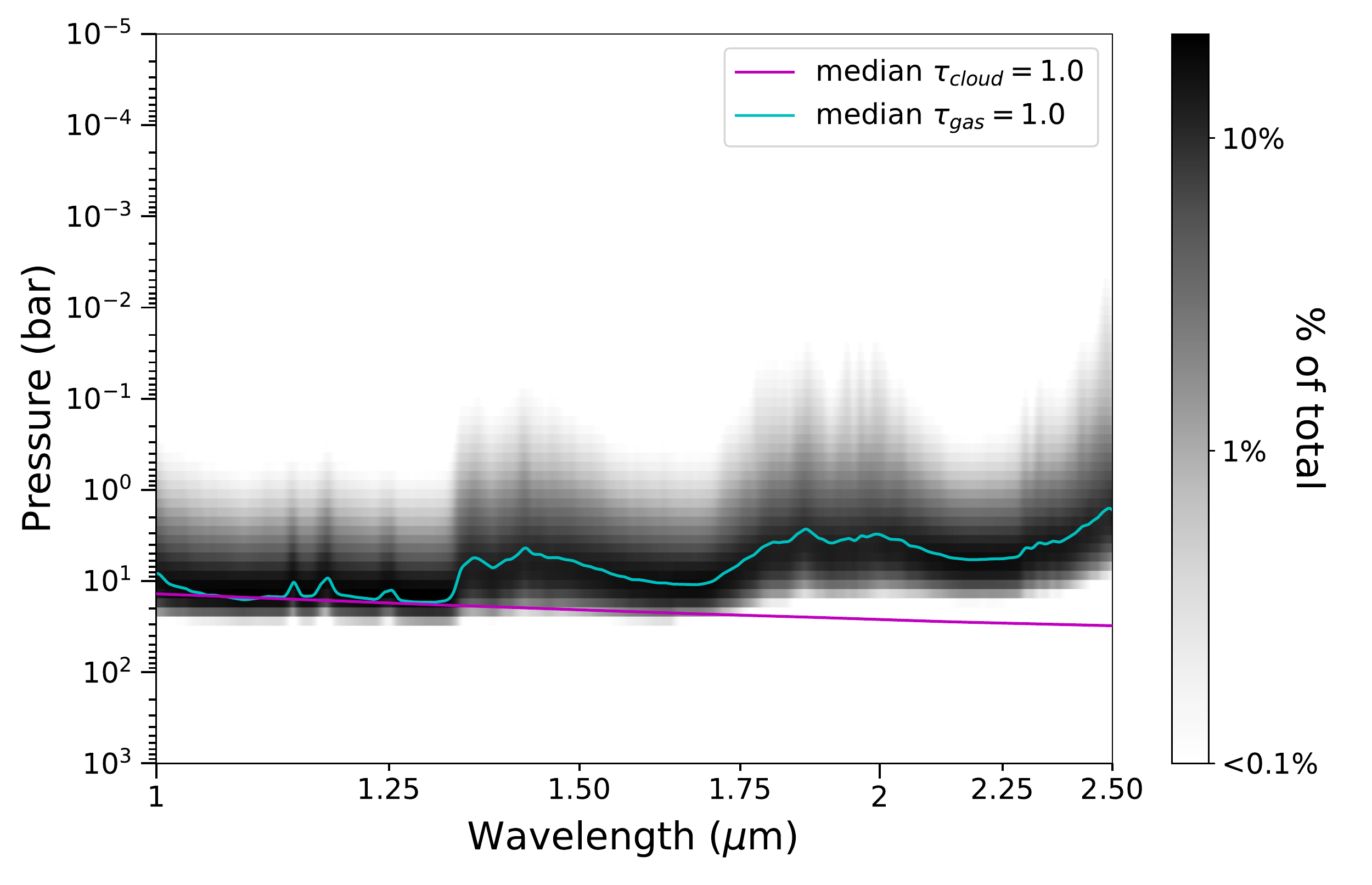}{0.5\textwidth}{\large(b)}} 
\caption{(a) Retrieved Pressure-Temperature Profile (black) compared to Sonora cloudless solar and low-metallicity model profiles similar to the semi-empirical and retrieval-derived \Teff (neon green and purple). The median cloud deck is shown in shades of blue. The median deck reaches an optical depth of $\tau=1$ at the boundary between the darkest blue and purple located at log $P=1.42$ bar. The purple region is where the cloud is optically thick and the blue shading indicates the vertical distribution where the cloud opacity drops to $\tau=0.5$ at the dash line. The grey bars on either side show the 1 $\sigma$ cloud deck location and vertical height distribution. The colored dashed lines are condensation curves for the listed species. (b) The contribution function associated with this cloud model, with the median cloud (magenta) and gas (aqua) at an optical depth of $\tau=1$ over plotted.}
\label{fig:1416A_deck_PT_profiles}
\vspace{0.5cm} 
\end{figure*}

Figure~\ref{fig:1416A_deck_PT_profiles}(a) shows the retrieved PT profile and location of the winning deck cloud model for J1416A. The Sonora (Marley et al. in prep) solar-metallicity, log $g=5.0$, 1700K model and the [M/H]$=-0.5$, log $g=5.0$, 1900K model agree with the retrieved profile 1$\sigma$ bounds, throughout the main photospheric pressure range ($\sim0.5-18$~bars, see panel b) and deeper. However, one should note that our deep PT profile (below photosphere) is an extrapolation of the shape at lower pressures as there is little contribution to the observed flux. At pressures lower than 1 bar (higher up in the atmosphere), the median PT profile is hotter than the Sonora models, which was also seen for two L dwarfs in \cite{Burn17}. The median deck cloud, shown in the center of Figure~\ref{fig:1416A_deck_PT_profiles}(a), becomes optically thick deeper than $\sim$10 bar with the cloud top location in pressure space quite tightly constrained to log\,$P=1.14\substack{+0.18\\-0.21}$ bars. However, the extent of the cloud (gradient region) where the optical depth falls to $\tau=1/2$ (dashed black line) is poorly constrained.

Figure \ref{fig:1416A_deck_PT_profiles}(b) shows the contribution function for this model along with the $\tau=1$ gas and cloud contributions. The contribution function in a layer is defined as

\begin{equation}
\begin{aligned}
    C(\lambda, P) =\frac{B(\lambda,T(P))\int_{P_1}^{P_2}d\tau}{\exp{\int_0^{P_2}d\tau}}
\end{aligned}
\label{eqn:contributionlayers}
\end{equation}

where $B(\lambda,T(P))$ is the Planck function, zero is the pressure at the top of the atmosphere, $P_1$ is the pressure at the top of the layer, and $P_2$ is the pressure at the bottom of the layer. The majority of the flux contributing to the observed spectrum of J1416A comes from the approximately 1 to 18 bar region, corresponding to the photosphere. The observed flux in the $Y$ band is dominated by the gas at shorter wavelengths ($\lesssim 1.11\,\mu$m) while the cloud opacity dominates from $\sim1.06-1.11\,\mu$m. The $J$ band is shaped by the gas opacity, with the cloud opacity sitting just below the $\tau=1$ gas line, potentially contributing minor amounts of opacity. In the $H$ and $K$ bands, the gas opacity dominates our observed flux as it becomes optically thick well before (higher up) the cloud contribution. 

\begin{figure}
\gridline{\fig{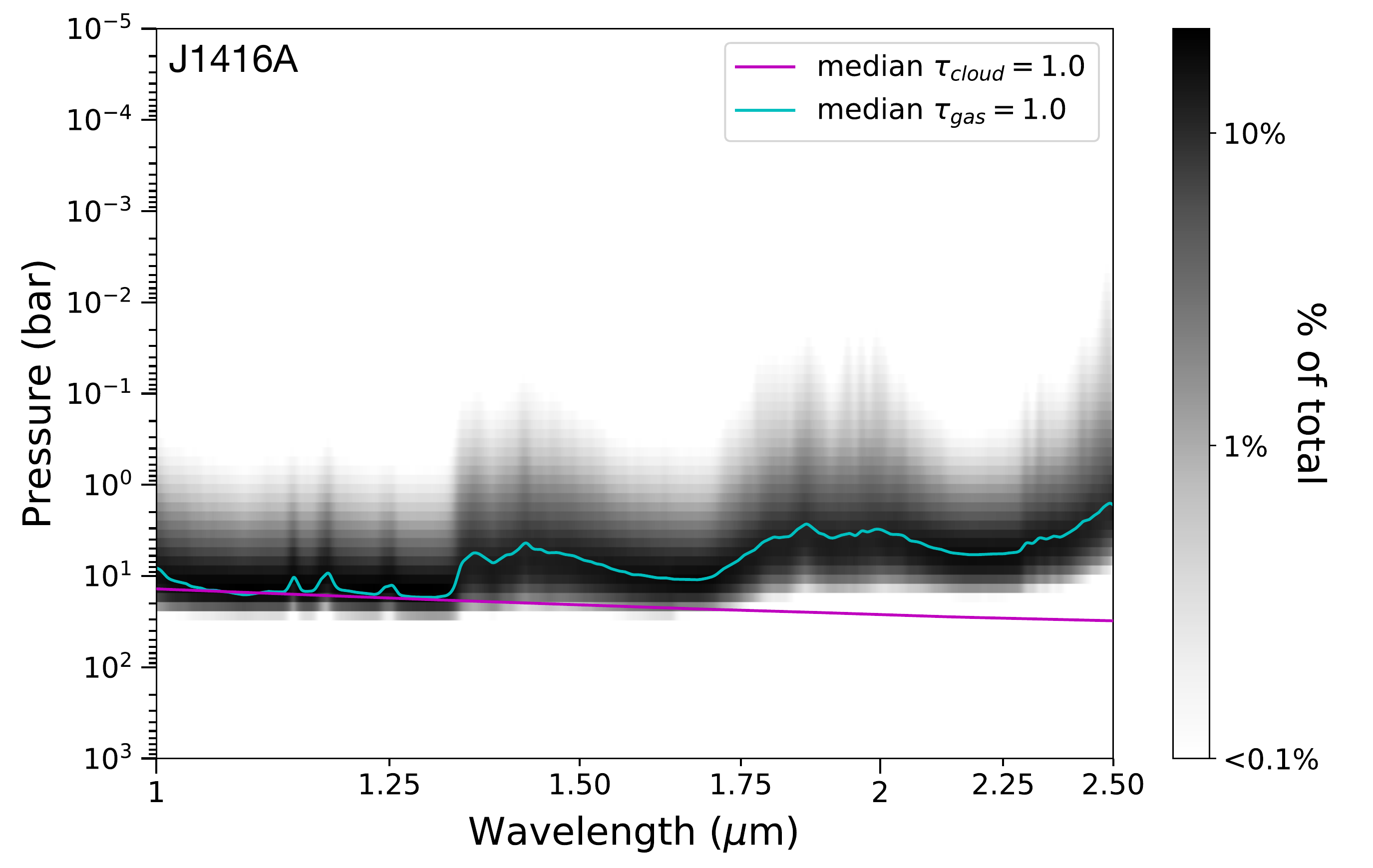}{0.48\textwidth}{\large(a)}} 
\gridline{\fig{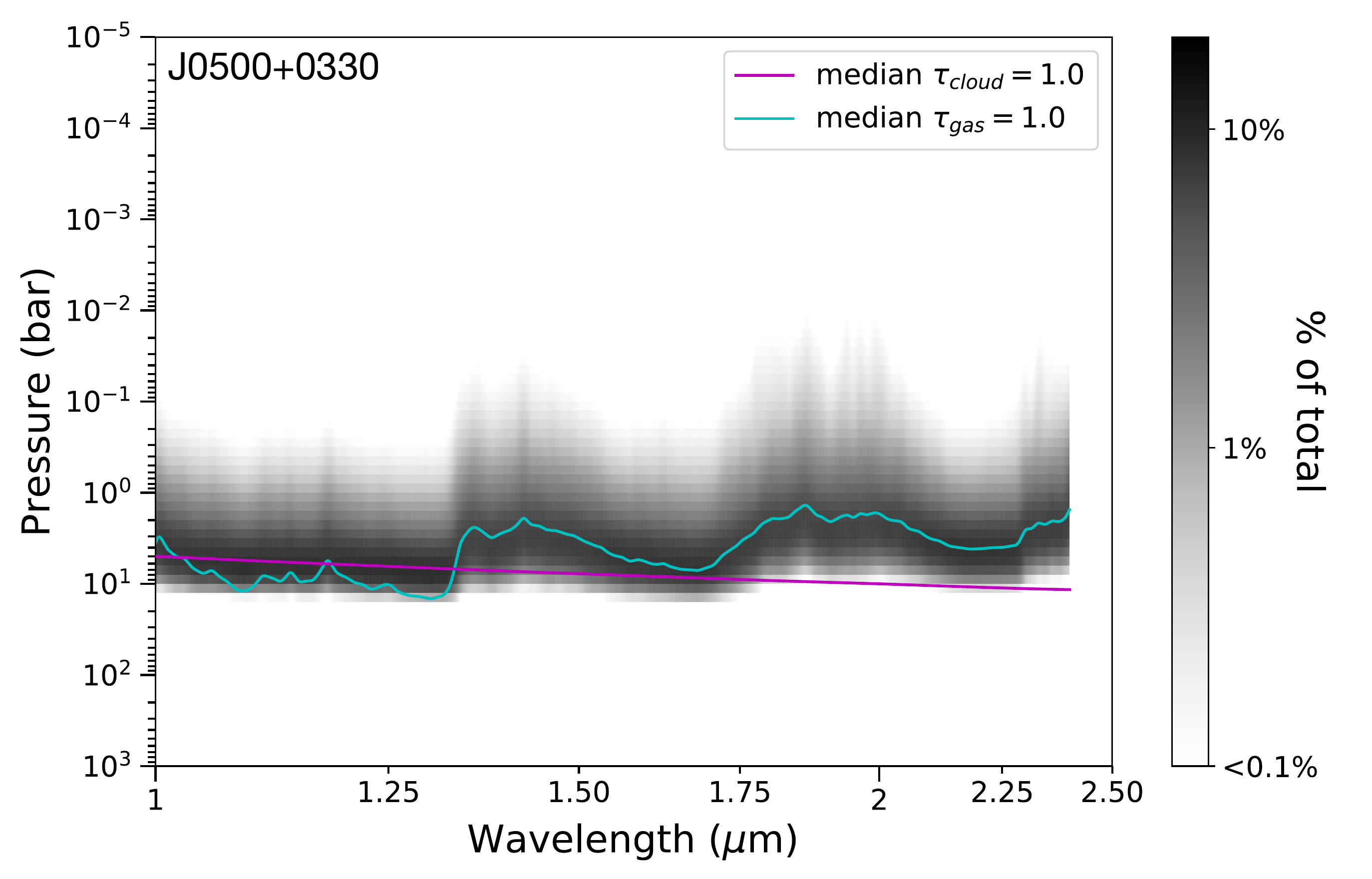}{0.48\textwidth}{\large(b)}}
\gridline{\fig{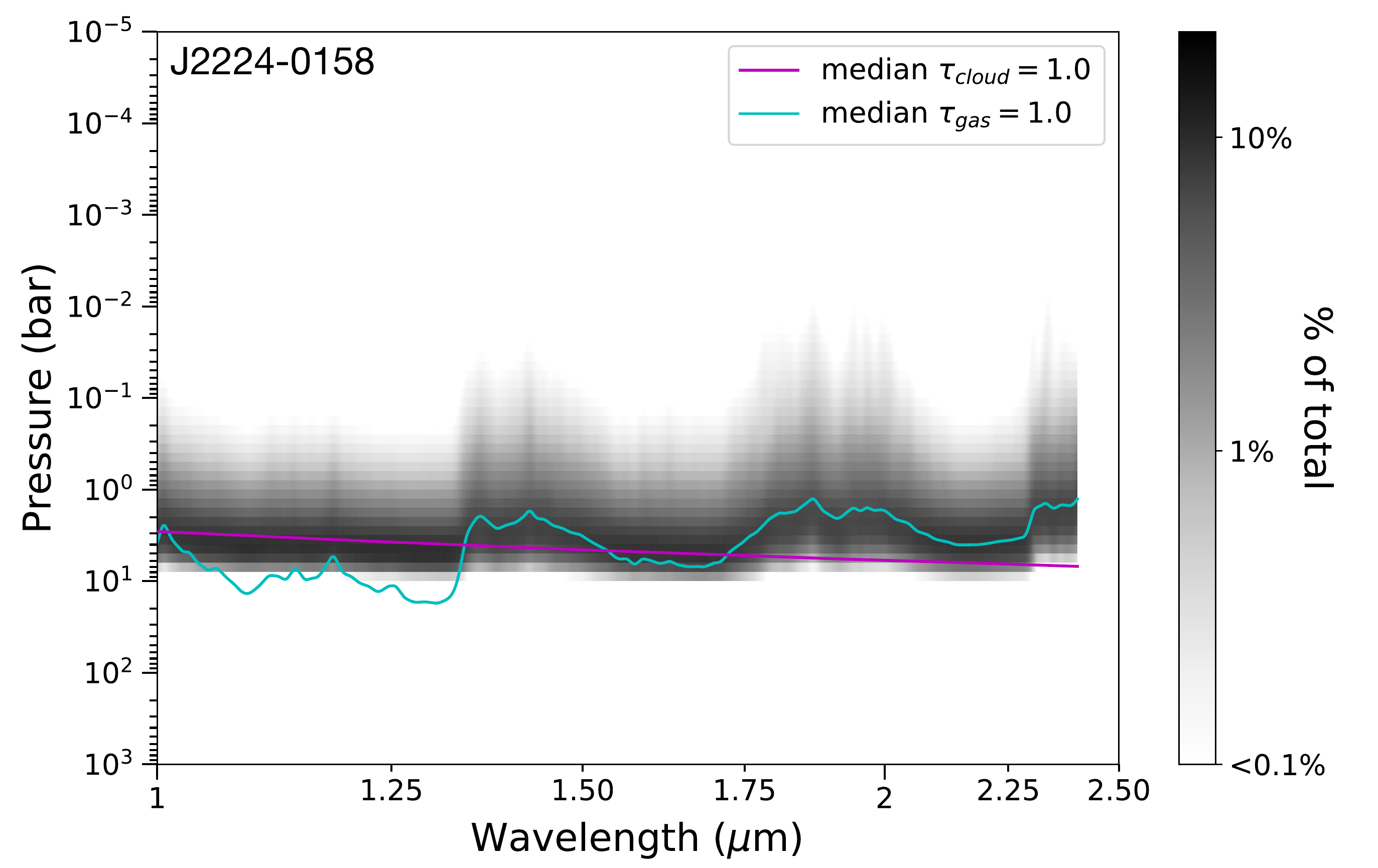}{0.48\textwidth}{\large(c)}}
\caption{Contribution functions for J1416A winning model compared to the power-law deck cloud models for J2224$-$0158 and J0050$+$0330 from \cite{Burn17}, in order form bluest to reddest $J-K$ color. (a) J1416A, (b) J0050$+$0330, (c) J2224$-$0158.}
\label{fig:ContributionFuncComp2017}
\end{figure}

The lack of the cloud's contribution to the $J$ band is a possible factor for J1416A's observed unusually blue $J-K$ color of $1.03\pm0.03$. In Figure~\ref{fig:ContributionFuncComp2017} we compare the contribution functions for J1416A with the two L dwarfs in \cite{Burn17}, 2MASS J05002100$+$033050 (hereafter J0500$+$0330) and 2MASSW J2224438$-$015852 (hereafter J2224$-$0158). We can see that the median $\tau_{cloud}=1$ level is reached at shallower pressures for both comparison targets and lies above the median $\tau_{gas}=1$ level in most of the $Y$ and the entire $J$ band for J0500$+$0330, and the entire $Y$ and $J$ bands in the case of J2224$-$0158. This points towards seeing deeper into the atmosphere at the $J$ band of J1416A potentially due to its lower metallicity, than the field source J0500$+$0330 and the red L dwarf J2224$-$0158, as the possible cause of the observed blue $J-K$ color.

\subsubsection{Retrieved gas abundances and derived properties}
\begin{figure*}
  \centering
   \includegraphics[scale=.218]{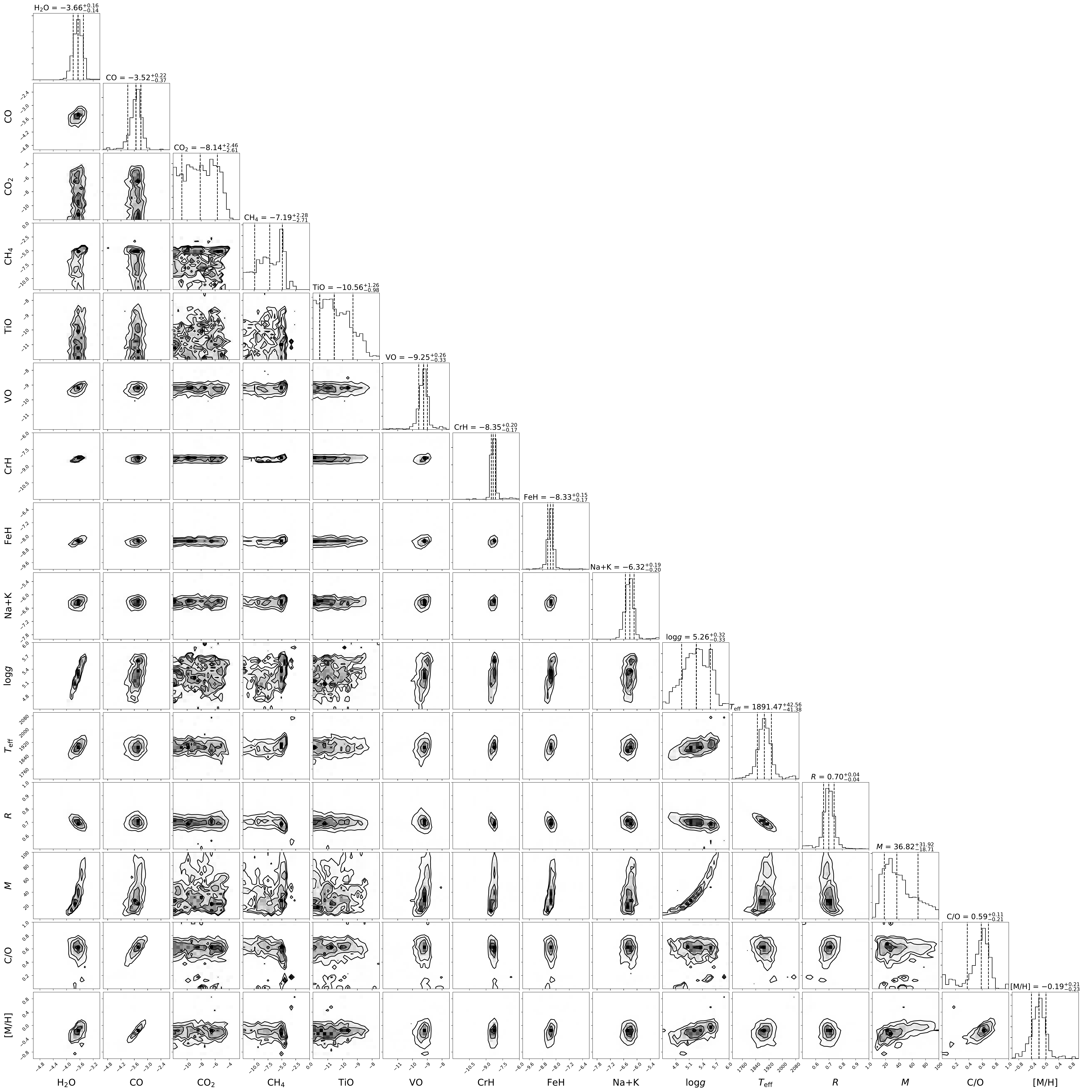}
\caption{J1416A power-law deck cloud posterior probability distributions for the retrieved parameters and extrapolated parameters. 1D histograms of the marginalized posteriors are shown along the diagonals with 2D histograms showing the correlations between the parameters. The dashed lines in the 1D histograms represent the 16\textsuperscript{th}, 50\textsuperscript{th}, and 84\textsuperscript{th} percentiles, with the 68\% confidence interval as the width between the 16\textsuperscript{th} and 84\textsuperscript{th} percentiles. Parameter values listed above are shown as the median~$\pm1\sigma$. Gas abundances are displayed as log$_{10}$(X) values, where X is the gas. \Teff, radius, mass, C/O ratio, and {[M/H]} are not directly retrieved parameters, but are calculated using the retrieved $R^2/D^2$ and log($g$) values along with the predicted spectrum. Our derived C/O ratio is absolute, where Solar C/O is 0.55, while our [M/H] is relative to Solar. Values for CO$_2$, CH$_4$, and TiO are not constrained and thus only provide upper limits.}
\label{fig:1416A_d2_89_gascorner}
\vspace{0.5cm} 
\end{figure*}

\begin{deluxetable}{l c}
\tablecaption{Retrieved Gas Abundances and Derived Properties for J1416A deck cloud model\label{tab:Deck_Corner_values}} 
\tablehead{\colhead{Parameter}\phm{stringssssssssssssssssss} & \colhead{Value}}
  \startdata
  \multicolumn{2}{c}{Retrieved} \\\hline
  H$_2$O & $-3.66\substack{+0.16 \\ -0.14}$\\
  CO & $-3.52\substack{+0.22 \\ -0.37}$\\
  CO$_2$ & <$-5.68$\\
  CH$_4$ & <$-4.91$\\
  TiO & $<-9.3$\\
  VO & $-9.25\substack{+0.26 \\ -0.33}$\\
  CrH & $-8.35\substack{+0.20 \\ -0.17}$\\
  FeH & $-8.33\substack{+0.15 \\ -0.17}$\\
  Na+K & $-6.32\substack{+0.19 \\ -0.20}$\\
  log $g$ (dex) & \phm{+}$5.26\substack{+0.32 \\ -0.33}$ \\ \hline
  \multicolumn{2}{c}{Derived}  \\ \hline
  \Lbol & $-4.23 \pm 0.01$ \\ 
  \Teff (K) & $1891.47\substack{+42.56 \\ -41.38}$ \\ 
  Radius ($R_\mathrm{Jup}$) & \phm{+}$0.7\pm0.04$ \\ 
  Mass ($M_\mathrm{Jup}$) & \phm{+}$36.82\substack{+31.92 \\ -18.71}$ \\
  {C/O}\tablenotemark{a,b} & \phm{+}$0.59\substack{+0.11 \\ -0.21}$ \\ 
  {[M/H]}\tablenotemark{a,b} & $-0.19\substack{+0.21 \\ -0.23}$ \\
  \enddata
  \tablenotetext{a}{Additional comparatives are listed in Table~\ref{tab:1416data}.}
  \tablenotetext{b}{Atmospheric values.}
  \tablecomments{Molecular abundances are fractions listed as log values. For unconstrained gases, 1$\sigma$ confidence is used to determine upper limit.}
\end{deluxetable}

Figure~\ref{fig:1416A_d2_89_gascorner} shows the posterior probability distributions for the retrieved gas abundances and surface gravity, as well as \Teff, radius, mass, C/O ratio, and [M/H] which are determined based on retrieved quantities. The values in  Figure~\ref{fig:1416A_d2_89_gascorner} are listed in Table~\ref{tab:Deck_Corner_values} for ease of reading. An extrapolated value for \Lbol is not shown in Figure~\ref{fig:1416A_d2_89_gascorner} as \Lbol showed no interesting correlations with any parameter. Our retrieved gas abundances are compared to values expected from chemical equilibrium grids in Section \ref{sec:spectrum_VMR_deck}.

The derived radius and mass are calculated from the retrieved scaling factor ($R^2/D^2$) and log\,$g$ values, along with the parallax measurement. To derive the \Teff, we use the radius and integrate the flux in the resultant forward model spectrum across $0.6-20$~$\mu$m. Our retrieval-derived \Teff is $\sim200$K hotter than our semi-empirical \Teff ($T_\mathrm{eff_{Retrieval}}=1891\substack{+42.56 \\ -41.38}$~K versus $T_\mathrm{eff_{SED}}=1694\pm 74$~K). This is due to the retrieval-based radius being 0.12~$R_\mathrm{Jup}$ smaller than the model radius from the SED method. Our retrieved gravity and extrapolated mass agree within 1$\sigma$ to the gravity and mass we derive from evolutionary models when generating the SED (Retrieval: log~$g= 5.26\substack{+0.32 \\ -0.33}$; $M$=$36.82\substack{+31.92 \\ -18.71}$~$M_\mathrm{Jup}$, SED: log~$g$=$5.22\pm0.22$; $M=60\pm18$~$M_\mathrm{Jup}$).

To derive the C/O ratio we exclude all carbon and oxygen bearing molecules that are not constrained for both cloud models of J1416A, thus assuming all of the carbon exists in CO and CH$_4$ and all of the oxygen is in H$_2$O, CO, and VO. To derive [M/H], we use the following equations:
\begin{equation}
    f_{H_2}=0.84(1-f_{gases})
\end{equation}
\begin{equation}
    N_{H}=2f_{H_2}N_{tot}
\end{equation}
\begin{equation}
    N_{element}=\sum_{molecules}  n_{atom}f_{molecule}N_{tot}
\end{equation}
\begin{equation}
    N_{M}=\sum_{elements} \frac{N_{element}}{N_H}
\end{equation}
where $f_{H_2}$ is the H$_2$ fraction, $N_{H}$ is the number of neutral hydrogen atoms, $N_{element}$ is the number atoms for the element of interest (C, O, V, Cr, Fe, and Na+K), $n_{atom}$ is the number of atoms of that element in a molecule (e.g. 2 for oxygen in CO$_2$), $f_{gases}$ is the total gas fraction containing only the constrained gases, and $N_{tot}$ is the total number of gas molecules. Thus the final value of [M/H] is
\begin{equation}
    [M/H]=log\frac{N_M}{N_{Solar}}
\end{equation}
where $f_{solar}$ is calculated as the sum of the solar abundances relative to H. Examining our derived C/O and [M/H], we find that J1416A has a roughly solar C/O and a slightly subsolar metallicity (C/O$=0.59\substack{+0.11 \\ -0.21}$; [M/H]$=-0.19\substack{+0.21 \\ -0.23}$). We note that for both C/O and [M/H] it does not matter if we include or exclude VO, which is done when comparing to J1416B, the ratios agree within $1\sigma$ of each other. This C/O ratio does not account for oxygen lost to silicate formation, which we address in further detail in Section~\ref{sec:co_ratio1416}.

\subsubsection{Cloud Properties}
\begin{figure}
  \centering
  \hspace{-0.25cm}
   \includegraphics[scale=0.36]{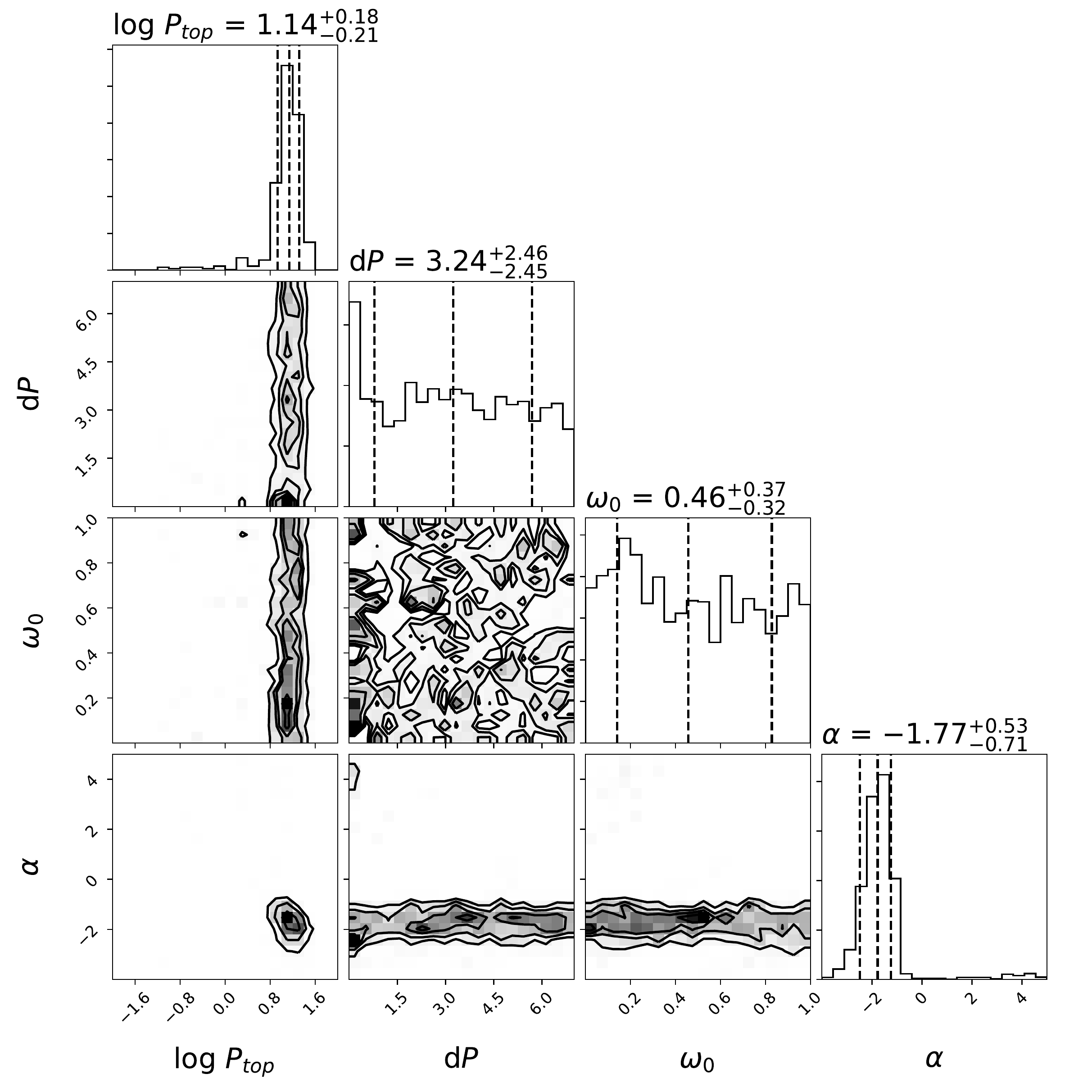}
\caption{J1416A power-law deck cloud posterior probability distributions for the cloud parameters. The cloud top pressure (log $P_{top}$) and the cloud height (d$P$) are shown in bars, and $\alpha$ is from the optical depth equation $\tau = \tau_0\lambda^\alpha$.}
\label{fig:1416A_d2_89_cloudcorner}
\vspace{0.5cm} 
\end{figure}

Figure~\ref{fig:1416A_d2_89_cloudcorner} shows the four retrieved deck cloud properties for J1416A: (1) the pressure at which the optical depth of the cloud passes one (the cloud top), (2) the decay height of the cloud in $\Delta$log$P$ (vertical extent above the cloud top, see Section~\ref{sec:cloudmodel}), (3) the single scattering albedo, and (4) the wavelength exponent $\alpha$ for the optical depth function $\tau = \tau_0\lambda^\alpha$ characterizing how ``non-grey'' the cloud is. We find the cloud top location is well constrained, while the vertical extent of the cloud and the albedo are unconstrained. With $\alpha$ being a negative value ($\alpha= -1.77\substack{+0.53\\-0.32}$), \cite{Burn17} investigated what could give rise to similar cloud opacity seen in two L dwarfs and found $\alpha = -2$ to be most consistent with a Hansen distribution \citep{Hans71} dominated by small sub-micron particles. 

By examining the over-plotted condensation curves on the PT profile (Figure~\ref{fig:1416A_deck_PT_profiles}) to identify the possible cloud deck species, we find no condensation curves intersect the profile at the cloud top location.  \cite{Burn17} found iron or corundum as likely cloud compositions for their two L dwarfs as these condensation curves intersected the PT profile at the top of the deck cloud. Thus for J1416A, iron or corundum (Al$_2$O$_3$) could be possible deck cloud candidates; however, the cloud optical depth continues to increase beneath the phase-equilibrium condensation point on our thermal profile. This could be due to cloud opacity deriving from the condensation of other species at deeper layers, or opacity arising from a process such as virga:  when condensed material (rain) falls through the atmosphere before vaporizing again.  

Interestingly, we find a slight positive correlation between the retrieval-derived radius of J1416A and the $\alpha$ parameter. With a more negative $\alpha$, the cloud has a lower optical depth at longer wavelengths, allowing for flux to escape from hotter, brighter layers. The retrieval compensates for this to provide a good fit by reducing the scale factor ($R^2/D^2$), resulting in a smaller radius estimate.

\subsubsection{Retrieved Spectrum and Composition\label{sec:spectrum_VMR_deck}}
\begin{figure*}
\centering
 \gridline{\fig{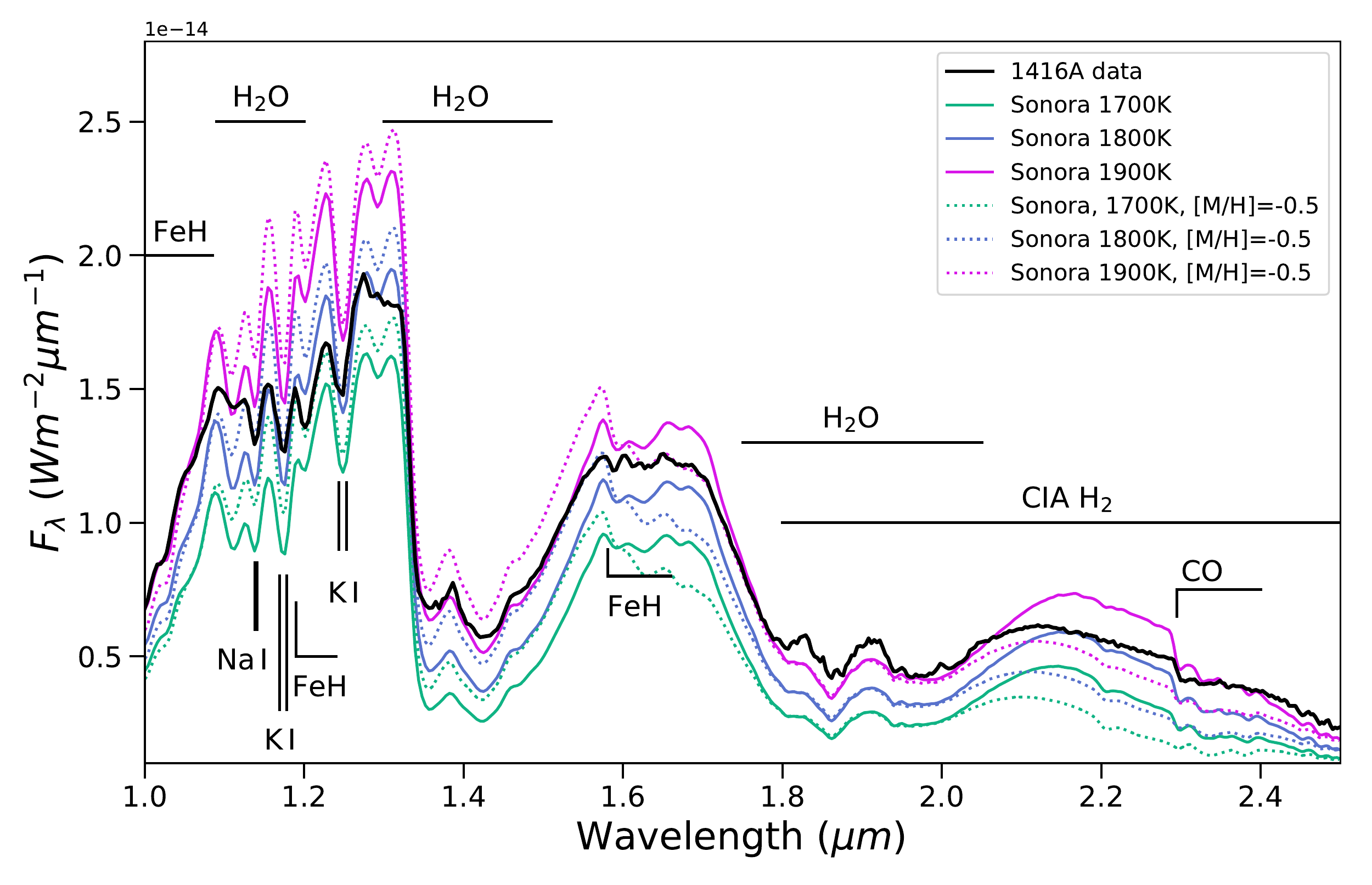}{0.5\textwidth}{\large(a)}}
 \gridline{\fig{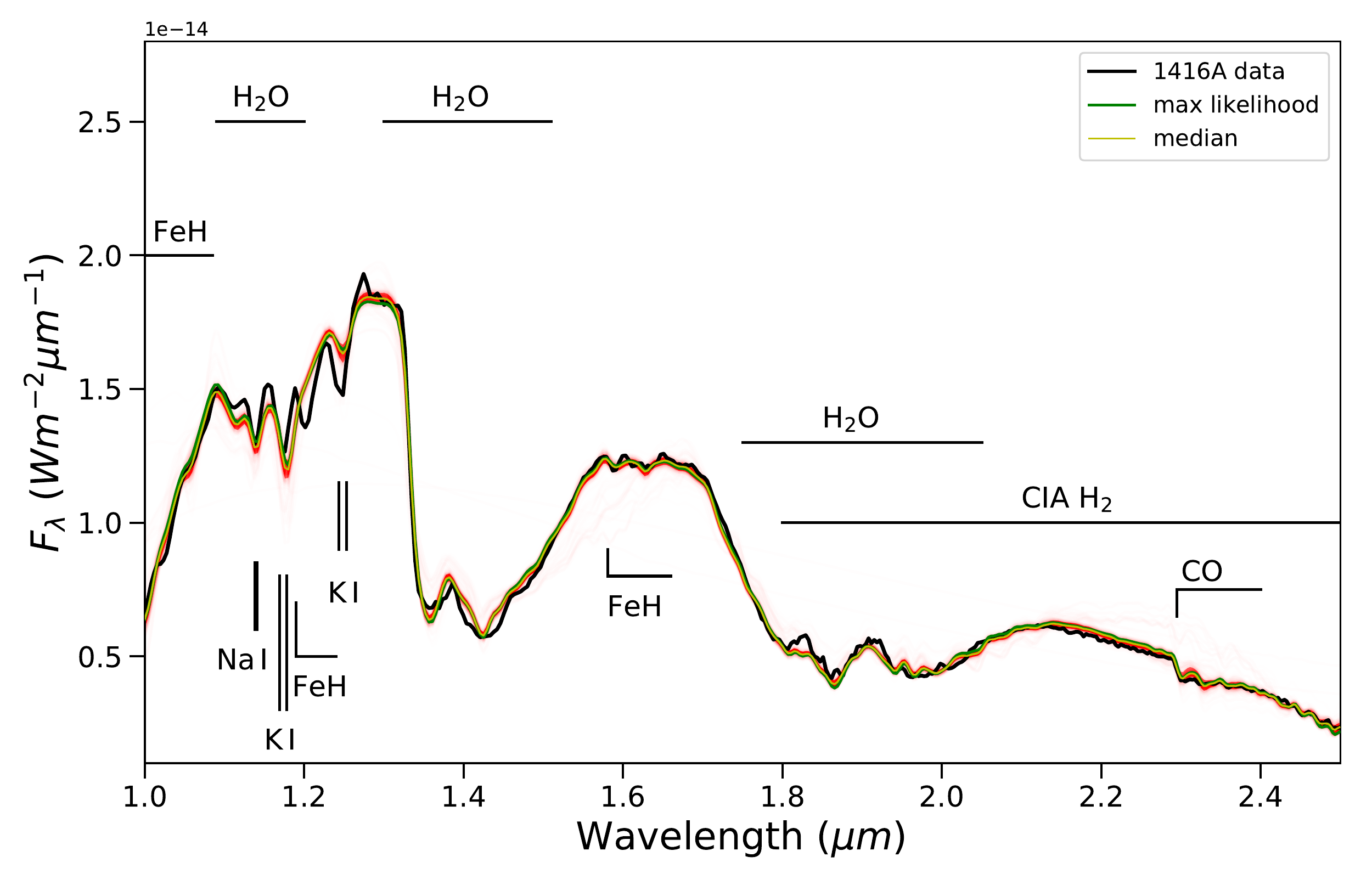}{0.5\textwidth}{\large(b)}
           \fig{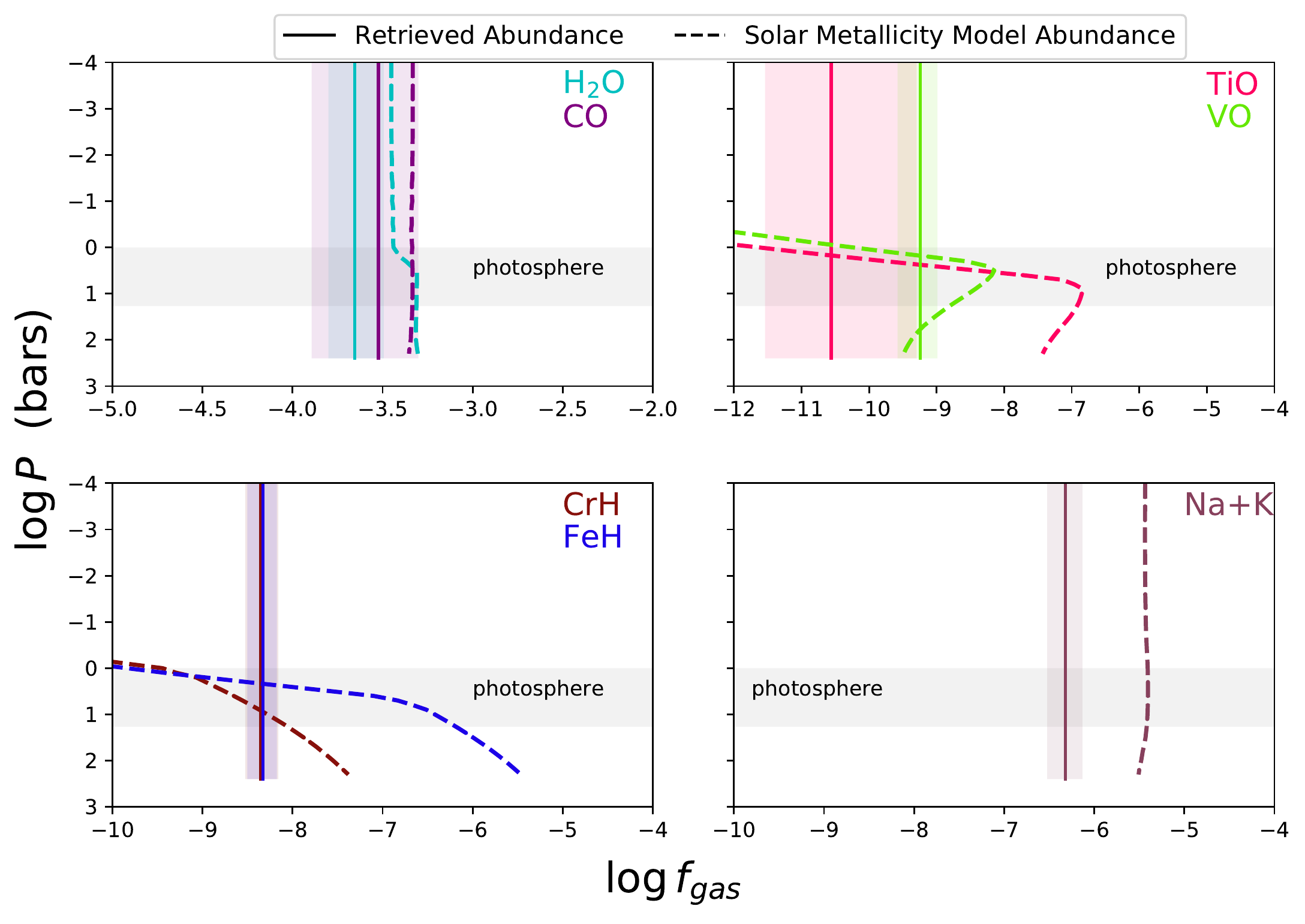}{0.5\textwidth}{\large(c)}}
\caption{(a) Retrieved forward model spectra for the deck cloud model of J1416A. The maximum likelihood spectrum is shown in dark green, the median spectrum in yellow, and 500 random draws from the final 2000 samples of the EMCEE chain in red. The SpeX prism data is shown in black. For comparison the cloud-free Sonora grid model solar metallicity spectra for log $g= 5.0$ and \Teff~$=1600$K, 1700K and 1800K (solid teal, blue, and purple), as well as [M/H]~$=-0.5$ for log $g= 5.0$ and \Teff~$=1800$K and 1900K (dotted teal, blue, and purple). These \Teff values bracket the range of the SED-derived and retrieval-derived \Teff. (b) Retrieved uniform-with-altitude mixing abundances for constrained gases compared to Solar metallictiy and C/O model abundances. The approximate location of the photosphere is shown in gray.}
\label{fig:1416A_deck_spectrum_vmr}
\vspace{0.5cm} 
\end{figure*}

Figure \ref{fig:1416A_deck_spectrum_vmr}(a) compares the observed SpeX prism data and Sonora model spectra, which are cloud-free and consistent with the retrieved PT profile (see Figure~\ref{fig:1416A_deck_PT_profiles}). Figure \ref{fig:1416A_deck_spectrum_vmr}(b) the retrieved forward model spectrum for the deck cloud model to the observed SpeX prism data. To compare the Sonora spectra to our retrieved forward model spectrum, the Sonora models were scaled to the median retrieved scale factor. Even though J1416A is best fit with a power-law deck cloud, the fit to the cloudless Sonora models are not very far off. This is likely due to the deck cloud affecting only a small portion of J1416A's spectrum, thus these models can do a fair job at fitting the observed data. 

When comparing the observed spectrum of J1416A to the Sonora model spectra, we find the 1900K solar metallicity model provides the best fit overall but struggles to fit features in the $J$ band and the $H$ band plateau. The $J$ band is best fit by the 1800~K solar model, while the peak of the $H$ band is best fit by the 1900~K low metallicity model, and while the 1900K solar model fits some of the $K$ band pseudo continuum it is a poor match to the CO feature.

In Figure \ref{fig:1416A_deck_spectrum_vmr}(b), the retrieval spectrum fits the overall shape of the observed spectrum quite well, but has difficulties fitting the \ion{Na}{1} doublet, \ion{K}{1} doublets, and the FeH feature between the \ion{K}{1} doublets in the $J$ band. Issues in fitting the \ion{Na}{1} and \ion{K}{1} doublets are likely due to how the pressure broadening is treated in the opacity models for these lines. With pressure broadening from the 0.77$\mu$m \ion{K}{1} doublet impacting the slope in $J$ band through about $1.1\,\mu$m, the retrieved spectrum is likely unable to fit both the broad slope of the $J$ band in this region as well as the narrow \ion{K}{1} and \ion{Na}{1} doublet features. We find that the Allard alkali opacities provide a better fit to J1416A than the Burrows alkali opacities, discussed in further detail in Section~\ref{sec:Alkalies}. In the $H$ band, the retrieval does a much better job of fitting the FeH band to the data. This is likely driven by the $H$ band feature being broader than the $J$-band FeH feature an thus has a larger impact on the goodness of fit. The FeH fitting issue is an example of a problem introduced by the assumption of uniform-with-altitude mixing ratios, as the $J$ and $H$ band features are at different pressures and should have different abundances at those pressure layers.

Figure \ref{fig:1416A_deck_spectrum_vmr}(c) shows the retrieved abundances for constrained gases compared to the solar metallicity and solar C/O thermochemical equilibrium model values from the grid introduced in Section~\ref{sec:gas_abundances}. Here we see the Na+K and H$_2$O abundances are less than expected from models, pointing towards a subsolar metallicity for J1416A. The median retrieved CO abundance is also less than the solar model value but is just within the 1$\sigma$ confidence interval. The photosphere is shown as a gray strip to guide reasonable abundance ranges for metal-oxides and metal-hydrides. As these are not close to uniform-with-altitude, it is difficult to compare our retrieved values to the models. We do find that our abundances for TiO, VO, CrH, and FeH all fall within the very wide range of possible model abundances in the photosphere. Examining our FeH abundance, we see that the retrieved value is less than the maximum abundance of $\approx-6$ that is possible in the photosphere. This maximum abundance corresponds to deeper into the atmosphere where we see the $J$ band FeH feature. With our lower than expected abundance, this points towards Fe being condensed in the atmosphere and agrees with Fe as our predicted cloud species.

Interestingly, we find that the uniform-with-altitude model is preferred over the thermochemical equilibrium model. At these temperatures, J1416A is expected to be in thermochemical equilibrium as the thermochemical timescale should be faster than the mixing timescale (\citealt{Viss06}, Section 5.1). Based on Figure \ref{fig:1416A_deck_spectrum_vmr}(b), the alkalies are likely to be driving this preference as their abundance is the only one that is discrepant with the thermochemical grid abundance. Therefore, the uniform-with-altitude method is able to capture this discrepancy while still allowing for the other gas abundances to be in agreement with the thermochemical grid.

\subsection{It's a Different Cloud, which is Indistinguishable: The power-law slab cloud tells the same story \label{sec:1416ASlab89}}
As listed in Table~\ref{tab:1416AModels}, the power-law slab cloud is indistinguishable from the power-law deck cloud model and thus should tell a similar story about the atmosphere of J1416A. Here we present the retrieval results of the power-law slab cloud retrieval.

\subsubsection{Pressure-Temperature Profile and Contribution Function}
\begin{figure*}
 \gridline{\fig{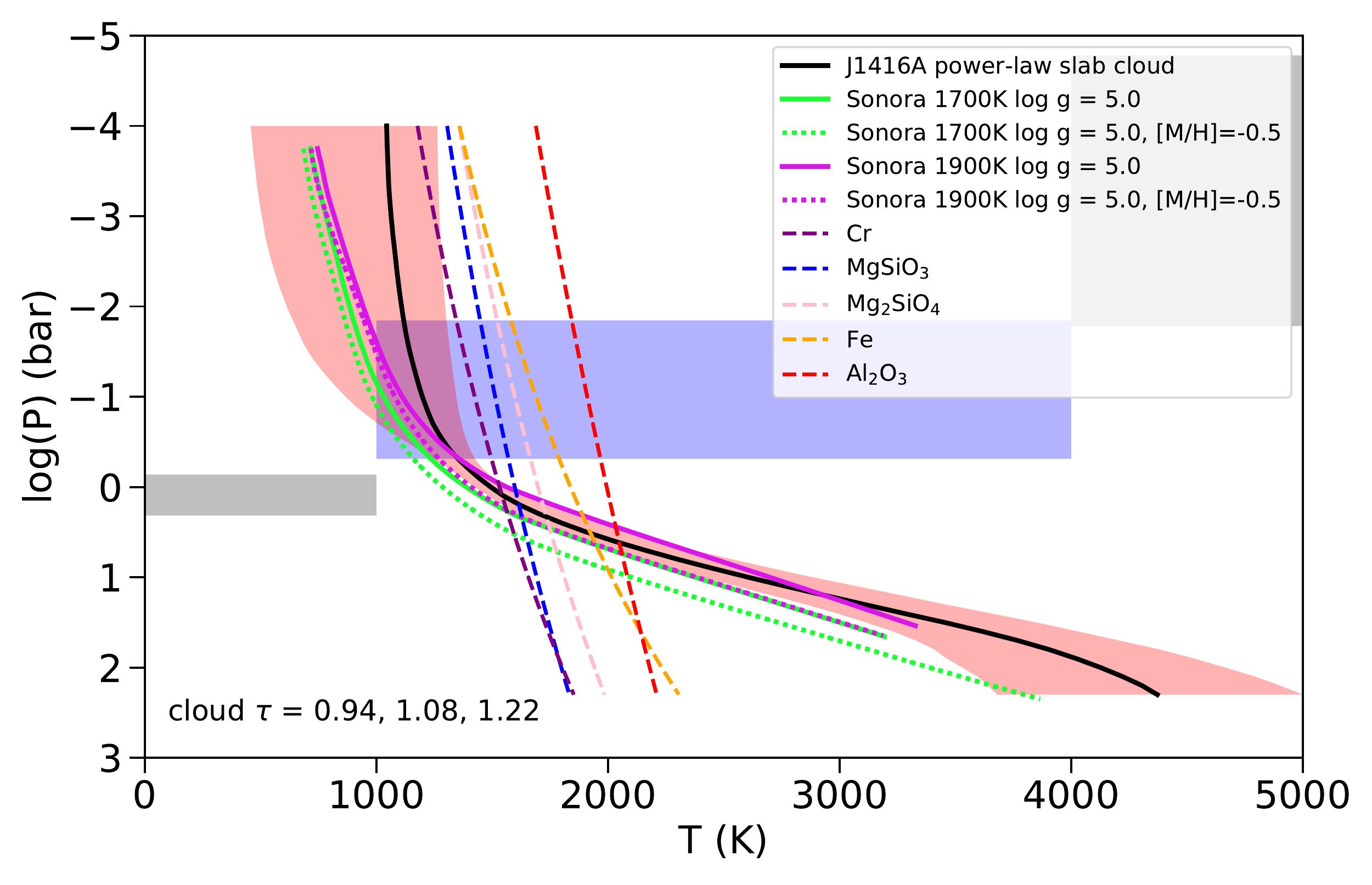}{0.5\textwidth}{\large(a)}
          \fig{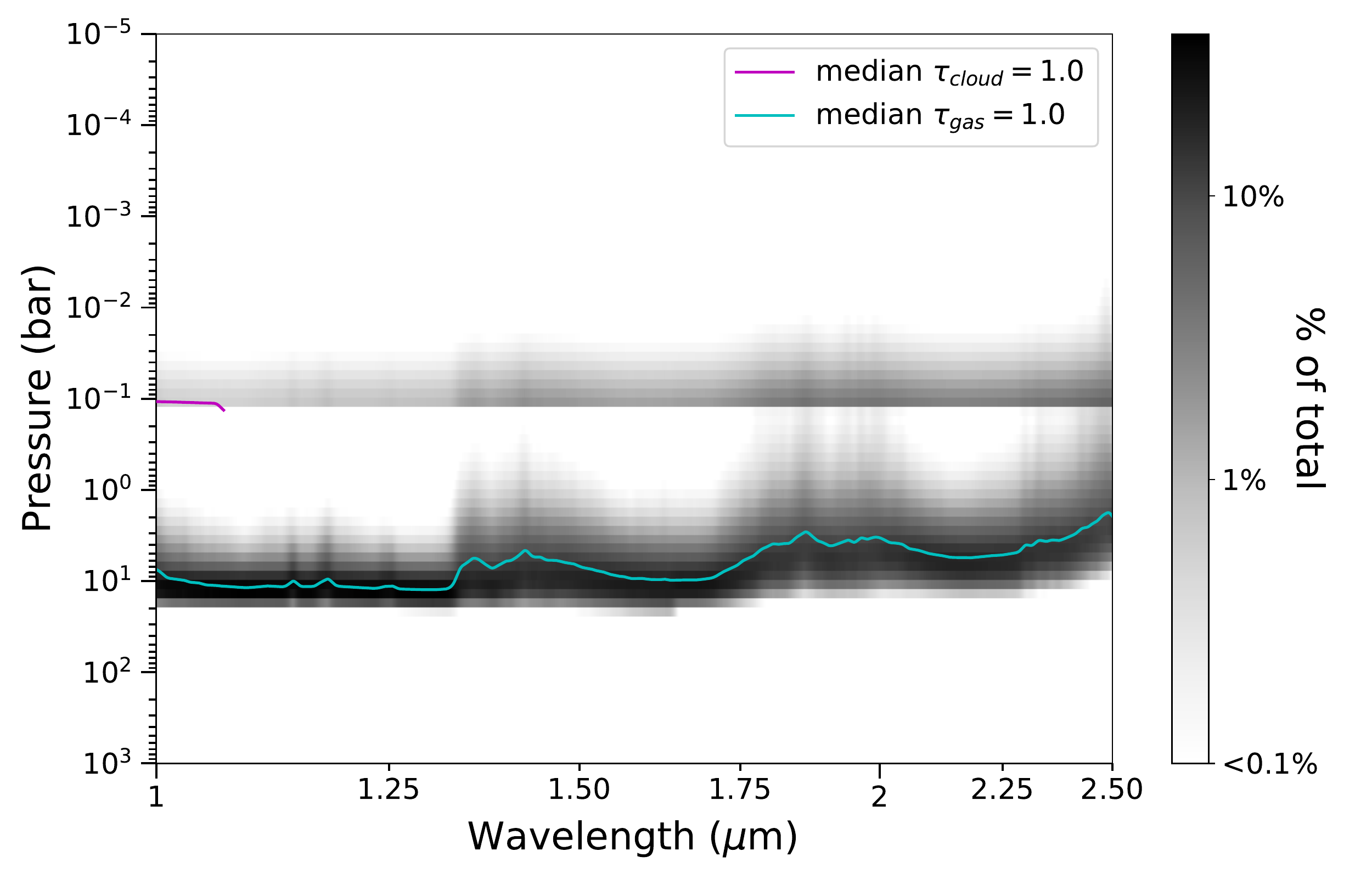}{0.5\textwidth}{\large(b)}} 
\caption{(a) Retrieved Pressure-Temperature Profile (black) compared to cloudless Sonora solar and low-metallicity model profiles similar to the SED-derived and retrieval-derived effective temperatures (neon green and purple). The median cloud slab height and location is shown purple with the 1 $\sigma$ shown in grey, indicating the ranges of height and base locations. Optical depth for the cloud is shown in the bottom left corner. The colored dashed lines are condensation curves for the listed species. (b) The contribution function associated with this cloud model, with the median cloud (magenta) and gas (aqua) at an optical depth of $\tau=1$ over plotted.}
\label{fig:1416A_slab_PT_profiles}
\vspace{0.5cm} 
\end{figure*}

Figure~\ref{fig:1416A_slab_PT_profiles}(a) shows the retrieved PT profile, slab cloud location, and total optical depth of the cloud. For this model, we find the bulk of the flux roughly between 1 and 18 bars like the deck cloud. The median retrieved profile in this region agrees within the 1$\sigma$ confidence interval with the Sonora solar metallicity, log\,$g=5.0$, 1900K, and 1700K models and the [M/H]~$=-0.5$, log\,$g=5.0$, 1900K model. Compared to the 1700K/5.0/solar and 1900K/5.0/$-$0.5 models, the retrieved profile is slightly hotter at the same pressure, while it is slightly cooler than the 1900K/5.0/solar model at the same pressure. At higher pressures, deeper in the atmosphere, the retrieved profile has a similar slope to that of the Sonora models, while at pressures lower than the photosphere the retrieved profile is more isothermal than the models. This is similar to the behavior of the power-law deck cloud profile compared to the Sonora models. The location in pressure space of the slab cloud, as well as its vertical height, are both poorly constrained due to the cloud being primarily optically thin with a total median optical depth across the cloud thickness of $\tau=1.08$at 1 $\mu$m, with a $\lambda^{-1.27}$ drop off to longer wavelengths.

Figure~\ref{fig:1416A_slab_PT_profiles}(b) shows the contribution function for this model, which shows the opacity from the slab cloud having a small effect on the overall flux emitted. The optically thick portion of the slab cloud is only between $\sim1-1.06\,\mu$m, whereafter it becomes optically thin and no longer significantly contributes to the observed flux. In the optically thick $Y$ band region we see that even though the cloud contributes $\sim1\%$ of the total flux in this region, the optical depth of $\tau_\mathrm{median}=1.08$ is primarily from here. Unlike the deck cloud, we see that the slab contributes to the flux at higher altitudes; however, this only contributes $\sim1-10\%$ of the total flux observed. The lack of cloud opacity in the $J$ band contributes to the unusually blue $J-K$ color in the same way as the power-law deck cloud model. With the cloud only affecting part of the $Y$ band, the flux from the $J$ band likely coming from a deeper pressure layer than that of field L dwarfs causing the bluer $J-K$ color (see the comparison in Section~\ref{sec:PT_cont_deck}).

\subsubsection{Retrieved gas abundances and derived properties}
\begin{figure*}
  \centering
   \includegraphics[scale=.218]{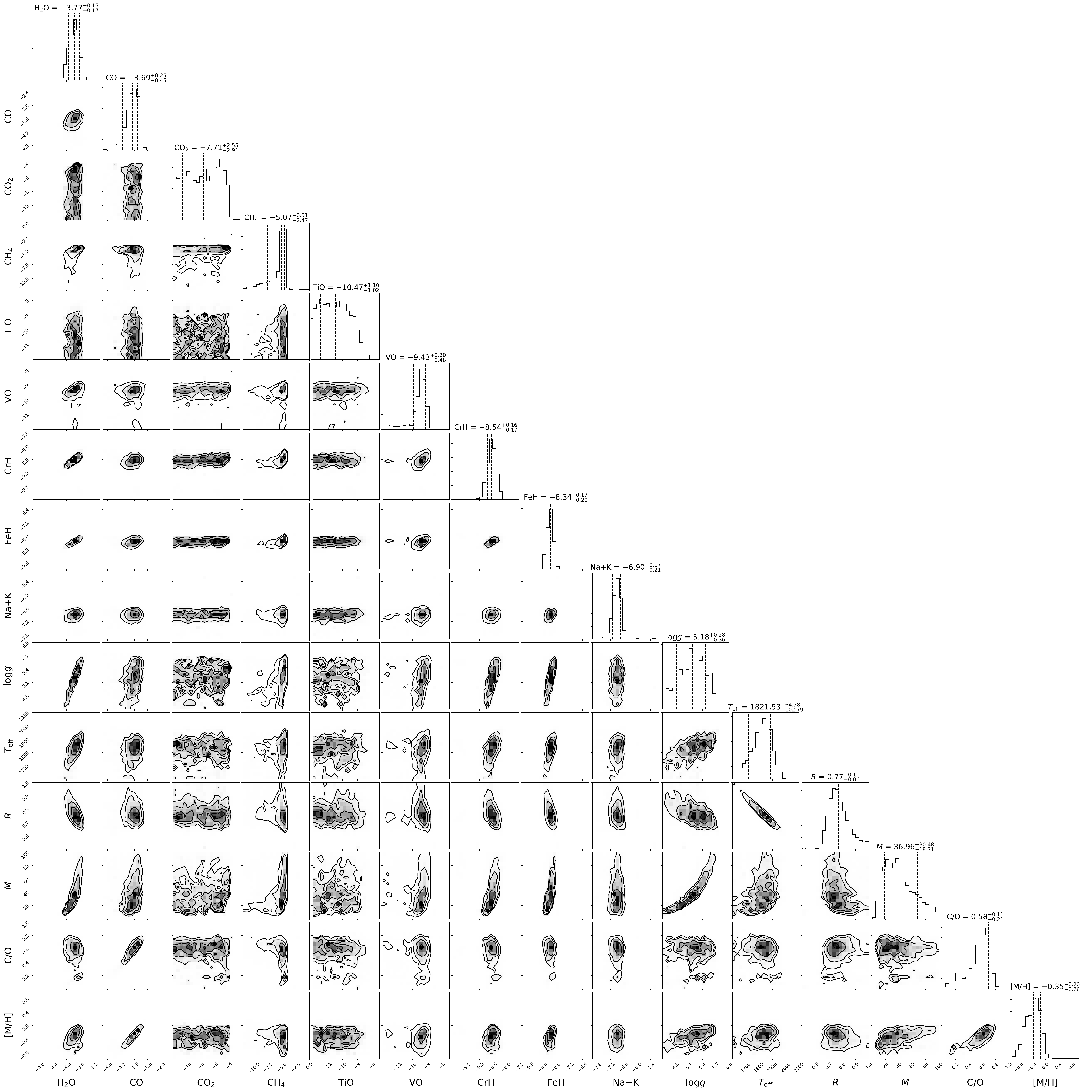}
\caption{J1416A power-law slab cloud posterior probability distributions for the retrieved parameters and extrapolated parameters. 1D histograms of the marginalized posteriors are shown along the diagonals with 2D histograms showing the correlations between the parameters. The dashed lines in the 1D histograms represent the 16\textsuperscript{th}, 50\textsuperscript{th}, and 84t\textsuperscript{th} percentiles, with the 68\% confidence interval as the width between the 16\textsuperscript{th} and 84\textsuperscript{th} percentiles. Parameter values listed above are shown as the median~$\pm1\sigma$. Gas abundances are displayed as log$_{10}$(X) values, where X is the gas. \Teff, radius, mass, C/O ratio, and {[M/H]} are not directly retrieved parameters, but are calculated using the retrieved $R^2/D^2$ and log($g$) values along with the predicted spectrum. Our derived C/O ratio is absolute, where Solar C/O is 0.55, while our [M/H] is relative to Solar. Values for CO$_2$ and TiO are not constrained and thus only provide upper limits.}
\label{fig:1416A_s1_89_postcorner}
\end{figure*}

\begin{deluxetable}{l c}
\tablecaption{Retrieved Gas Abundances and Derived Properties for J1416A slab cloud model\label{tab:Slab_Corner_values}} 
\tablehead{\colhead{Parameter}\phm{stringssssssssssssssssss} & \colhead{Value}}
  \startdata
  \multicolumn{2}{c}{Retrieved} \\\hline
  H$_2$O & $-3.77\substack{+0.15 \\ -0.17}$\\
  CO & $-3.69\substack{+0.25 \\ -0.45}$\\
  CO$_2$ & <$-5.16$\\
  CH$_4$ & $-5.07\substack{+0.51 \\ -2.47}$\\
  TiO & <$-9.37$\\
  VO & $-9.43\substack{+0.30 \\ -0.48}$\\
  CrH & $-8.54\substack{+0.16 \\ -0.17}$\\
  FeH & $-8.34\substack{+0.17 \\ -0.20}$\\
  Na+K & $-6.90\substack{+0.17 \\ -0.21}$\\
  log $g$ (dex) & \phm{+}$5.18\substack{+0.28 \\ -0.36}$ \\ \hline
  \multicolumn{2}{c}{Derived}  \\ \hline
  \Lbol & $-4.21 \pm 0.01$ \\ 
  \Teff (K) & $1821.53\substack{+64.58 \\ -102.49}$ \\ 
  Radius ($R_\mathrm{Jup}$) & \phm{+}$0.77\substack{+0.10 \\-0.06}$ \\  
  Mass ($M_\mathrm{Jup}$) & \phm{+}$36.96\substack{+30.48 \\ -18.71}$ \\
  {C/O}\tablenotemark{a,b} & \phm{+}$0.58\substack{+0.11 \\ -0.21}$ \\
  {[M/H]}\tablenotemark{a,b} & $-0.35\substack{+0.20 \\ -0.26}$ \\
  \enddata
  \tablenotetext{a}{Additional comparatives are listed in Table~\ref{tab:1416data}.}
  \tablenotetext{b}{Atmospheric values.}
  \tablecomments{Molecular abundances are fractions listed as log values. For unconstrained gases, 1$\sigma$ confidence is used to determine upper limit.}
\end{deluxetable}

Figure~\ref{fig:1416A_s1_89_postcorner} shows the posterior probability distributions for the gases, surface gravity, \Teff, radius, mass, C/O, and [M/H] for the slab cloud model, with the values also listed in Table~\ref{tab:Slab_Corner_values} for ease of readability. Comparisons to the chemical equilibrium grid values of the gases are discussed in Section~\ref{sec:spectrum_VMR_slab}. The majority of the gas abundances, \Teff, radius, mass, C/O, and [M/H] values for the slab model agree with those from the deck cloud model. The exception is the Na+K abundance, which differs from the deck cloud abundance by 1.4$\sigma$. This key difference in alkali abundance will be discussed in more detail in Section~\ref{sec:Alkalies}, when we compare the alkali abundances between the retrievals for J1416A and J1416B.

\subsubsection{Cloud Properties}
\begin{figure}
  \centering
   \includegraphics[scale=.29]{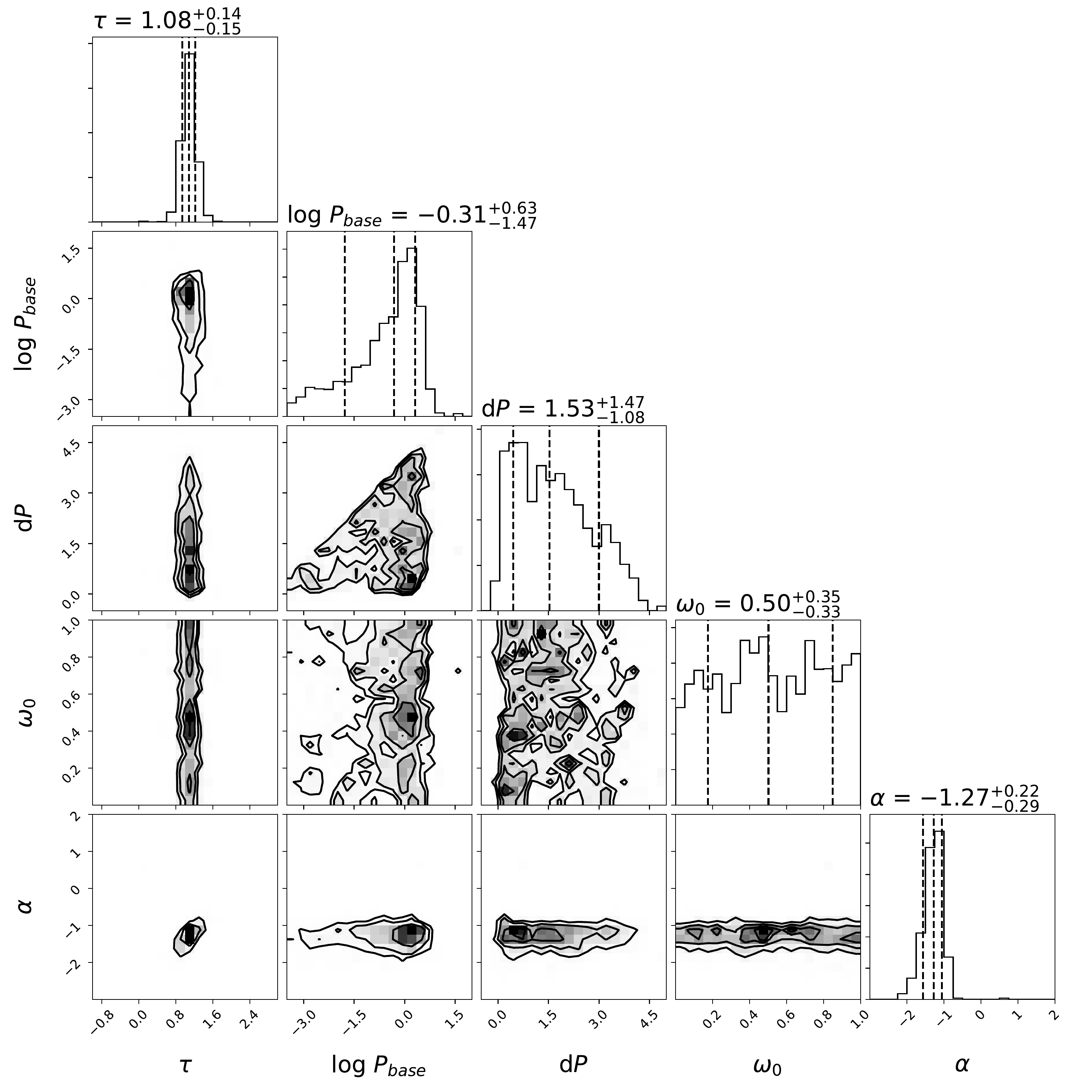}
\caption{J1416A power-law slab cloud posterior probability distributions for the cloud parameters. The cloud top pressure (log $P_{top}$) and the cloud height (d$P$) are shown in bars, and $\alpha$ is from the optical depth equation $\tau = \tau_0\lambda^\alpha$.}
\label{fig:1416A_s1_89_cloudcorner}
\vspace{0.5cm} 
\end{figure}

Retrieved cloud properties for the total optical depth, the pressure level for the base of the cloud (log\,$P_\mathrm{base}$), the height of the cloud, the single scattering albedo, and the wavelength exponent $\alpha$ that describes how ``non-grey'' the cloud is for the slab model are shown in Figure~\ref{fig:1416A_s1_89_cloudcorner}. The cloud base, height, and albedo are unconstrained for this model. The power $\alpha$ is more tightly constrained than in the deck cloud model and agrees within 1$\sigma$. The slab cloud also has a negative power, corresponding to a reddening cloud with sub-micron sized particles likely described by a Hansen distribution. As the slab cloud is higher in the atmosphere, multiple condensates are stable at its location. As the slab and deck cloud models are indistinguishable, distinguishing between the condensates is critical to atmospheric understanding and will be the subject of future work. Like the deck cloud, the slab cloud also has a positive correlation between the radius and $\alpha$, causing a smaller opacity at longer wavelengths, thus allowing for a smaller radius.

\subsubsection{Retrieved Spectrum and Composition\label{sec:spectrum_VMR_slab}}
\begin{figure*}
\centering
 \gridline{\fig{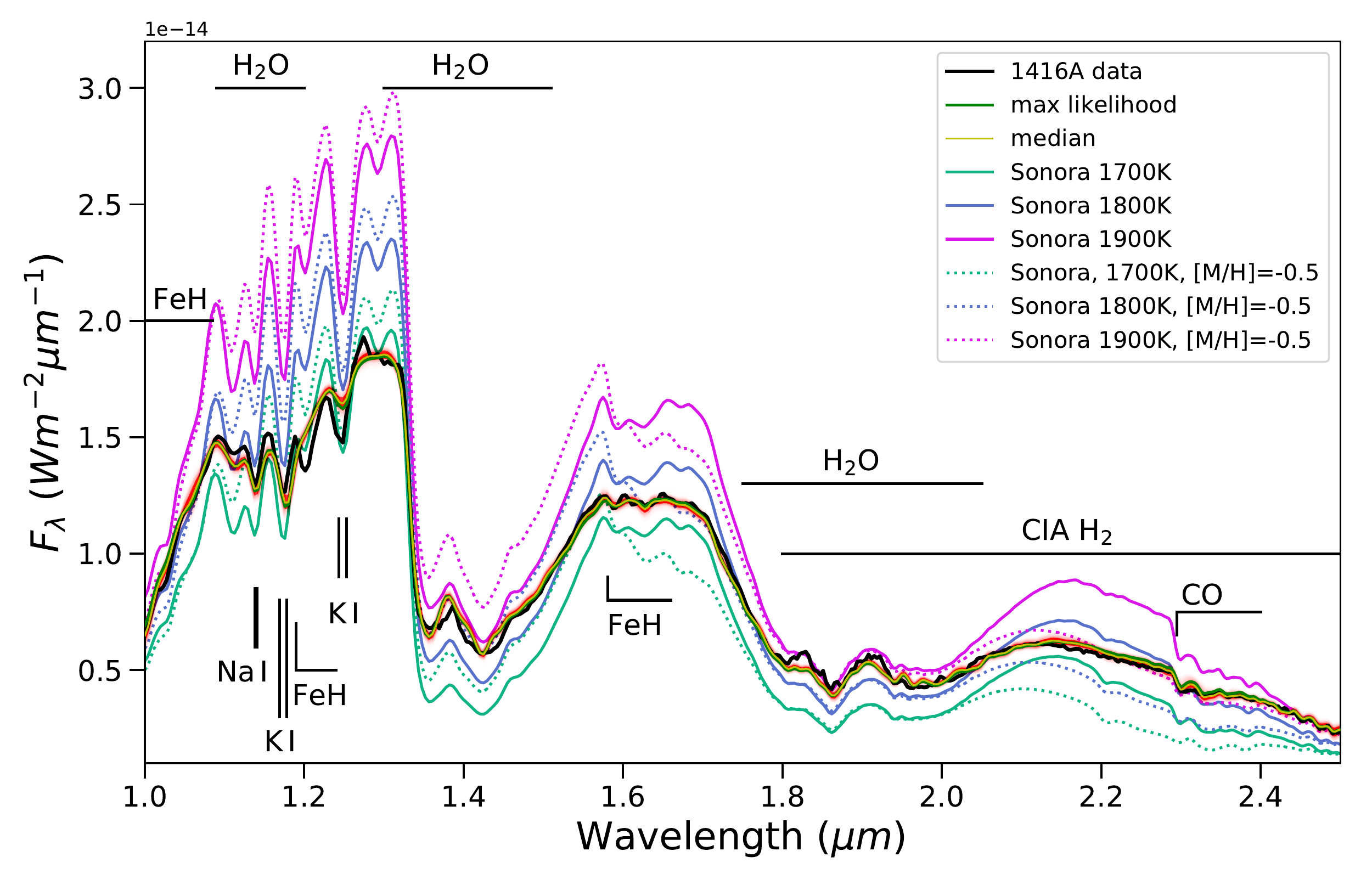}{0.5\textwidth}{\large(a)}}
\gridline{ \fig{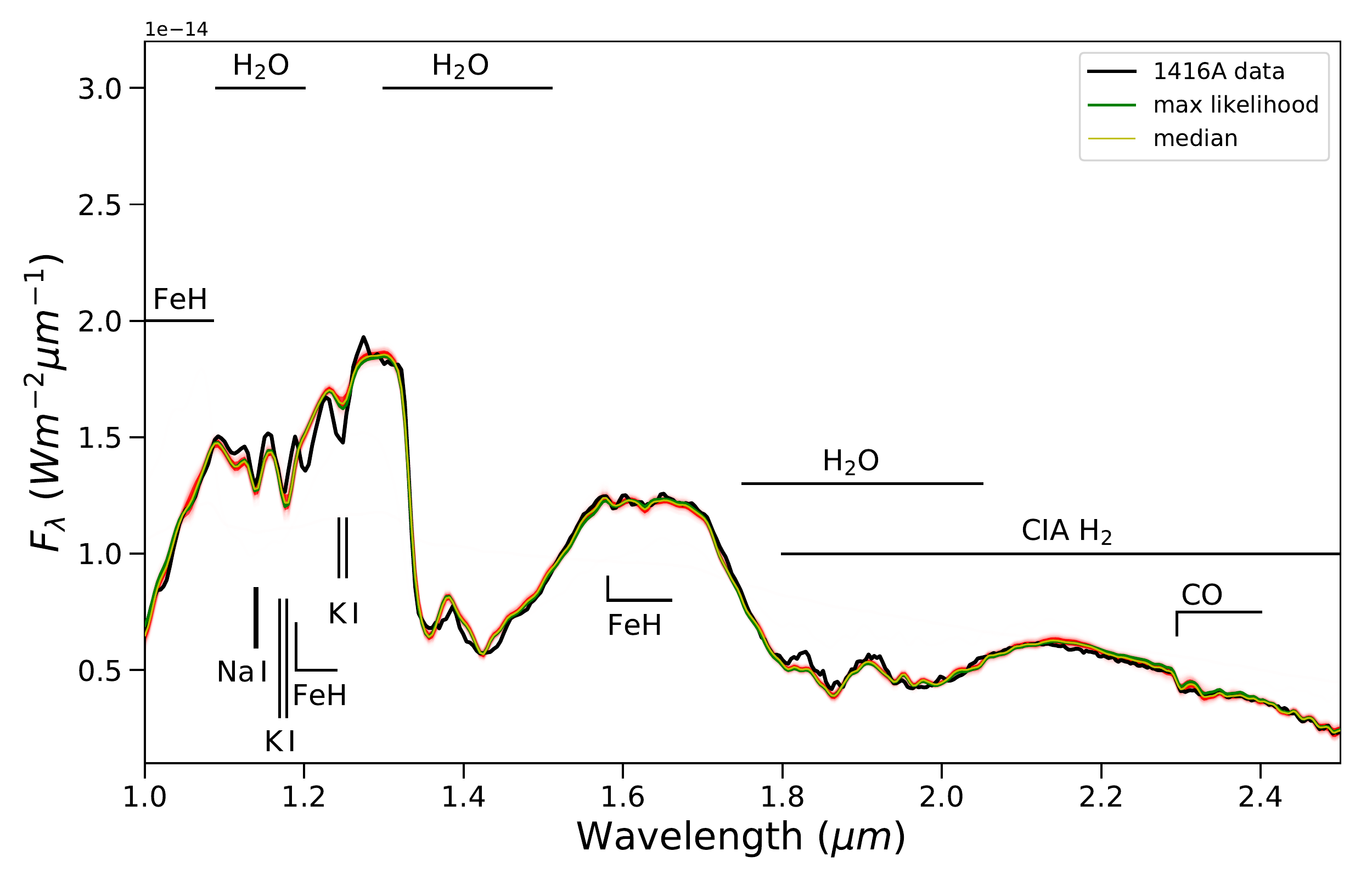}{0.5\textwidth}{\large(b)}
           \fig{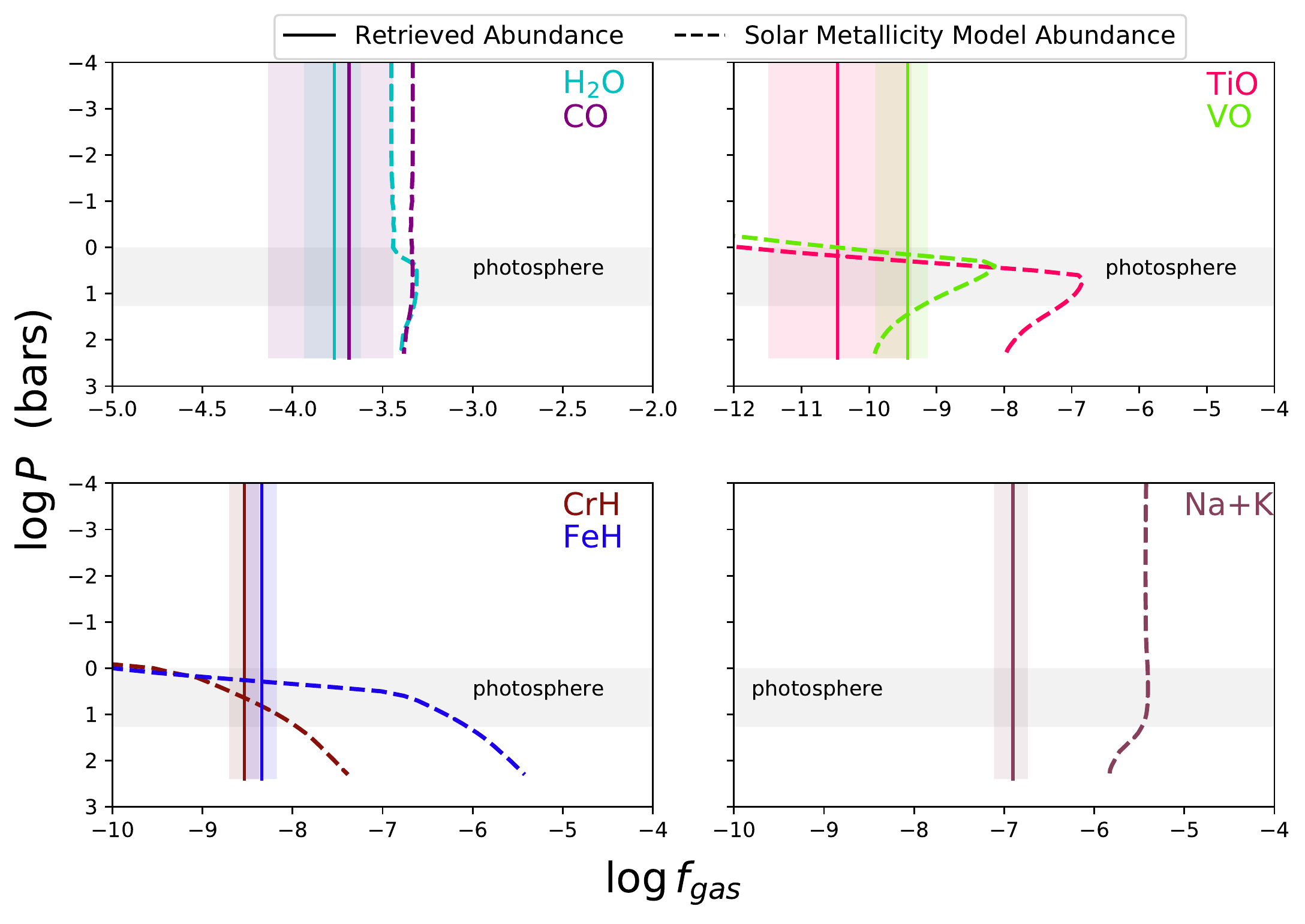}{0.5\textwidth}{\large(c)}} 
\caption{(a) Retrieved forward model spectra for the slab cloud model of J1416A. The maximum likelihood spectrum is shown in dark green, the median spectrum in yellow, and 500 random draws from the final 2000 samples of the EMCEE chain in red. The SpeX prism data is shown in black. For comparison the Sonora grid model solar metallicity spectra for log $g= 5.0$ and \Teff~$=1600$K, 1700K and 1800K (solid teal, blue, and purple), as well as [M/H]~$=-0.5$ for log $g= 5.0$ and \Teff~$=1800$K and 1900K (dotted teal, blue, and purple). These \Teff values bracket the range of the SED-derived and retrieval-derived \Teff. (b) Retrieved uniform-with-altitude mixing abundances for constrained gases compared to Solar metallictiy and C/O model abundances. The approximate location of the photosphere is shown in gray.}
\label{fig:1416A_slab_spectrum_vmr}
\vspace{0.5cm} 
\end{figure*}

The forward model spectrum for the slab cloud model is shown in Figure~\ref{fig:1416A_slab_spectrum_vmr}(a) compared to the observed SpeX spectrum, and various temperature and metallicity Sonora models that bracket the retrieved \Teff. For the slab cloud forward maximum-likelihood model spectrum, we find it is best fit by the 1700K solar metallicity model in the $J$ band, while the 1800K solar metallicity or 1800K [M/H]$=0.5$ models fit better in the $H$ and $K$ bands. In Figure~\ref{fig:1416A_slab_spectrum_vmr}(b), we compare the retrieved spectrum and the observed spectrum. The spectrum from the slab cloud model is quite similar to that of the deck cloud, fitting both the FeH feature and the $1.25\,\mu$m \ion{K}{1} doublet in the $J$ band poorly, for similar reasons as discussed in Section~\ref{sec:spectrum_VMR_deck}. Figure~\ref{fig:1416A_slab_spectrum_vmr}(c) compares the retrieved gas abundances for the constrained gases to the solar metallicity values expected from the thermochemical equilibrium model values from the grid introduced in Section~\ref{sec:gas_abundances}. Unlike the deck cloud model, the retrieved CO abundance is below the solar model expected values. All of our retrieved gas fractions for this model are consistent with the deck cloud model, with the exception of the Na+K abundance. These low abundances of H$_2$O, CO, and the tied Na+K again confirm the low metallicity atmosphere that we derive.

\section{Retrieved Model of 1416B}\label{sec:Retrieval_Models_B}

We initially used the Burrows alkali opacities as done in \cite{Line17} for J1416B which produced the best fit model. However, we find that the Allard alkali opacities give consistent abundances between J1416A and J1416B, and thus we effectively treat the cloud-free Allard alkali model as the best model for J1416B. Thus in this section, we present the results of the second best fitting model (our winning model, $\Delta$BIC=10) the cloud-free, uniform-with-altitude mixing ratio, Allard alkali opacity model for J1416B. The $\Delta$BIC for all tested models for J1416B are listed in Table~\ref{tab:1416BModels} and the cloud-free Burrows alkali opacity model results are shown in Section~\ref{sec:1416B_Burrows_alk}. Detailed examination of our choice of alkali line models is discussed in Section~\ref{sec:Alkalies}.  We will compare our J1416B results to retrieval results from \cite{Line17} throughout this section. Inter-comparison of retrieved and extrapolated parameters between J1416B and J1416A, as well as comparisons to the literature will be discussed in Section~\ref{sec:discussion1416}.

\begin{deluxetable*}{l c c }
\tablecaption{$\Delta$BIC for J1416B retrieval models\label{tab:1416BModels}} 
\tablehead{\colhead{Model} & \colhead{Number of Parameters} & \colhead{$\Delta$BIC}}
  \startdata
  Cloud Free & 14 & 0 \\
  Cloud Free Chemical Equilibrium & 11 & 14 \\
  Cloud Free, Allard Alkali & 14 & 10\\
  Grey Slab cloud & 18 & 14 \\
  Power-law Slab cloud & 19 & 25 \\
  Grey Deck cloud & 19 & 17 \\
  Power-law Deck cloud & 20 & 18 \\  
  \enddata
  \tablecomments{Unless otherwise listed default alkali opacities are Burrows.}
\end{deluxetable*}

\subsubsection{Pressure-Temperature Profile and Contribution Function}
\begin{figure*}
 \gridline{\fig{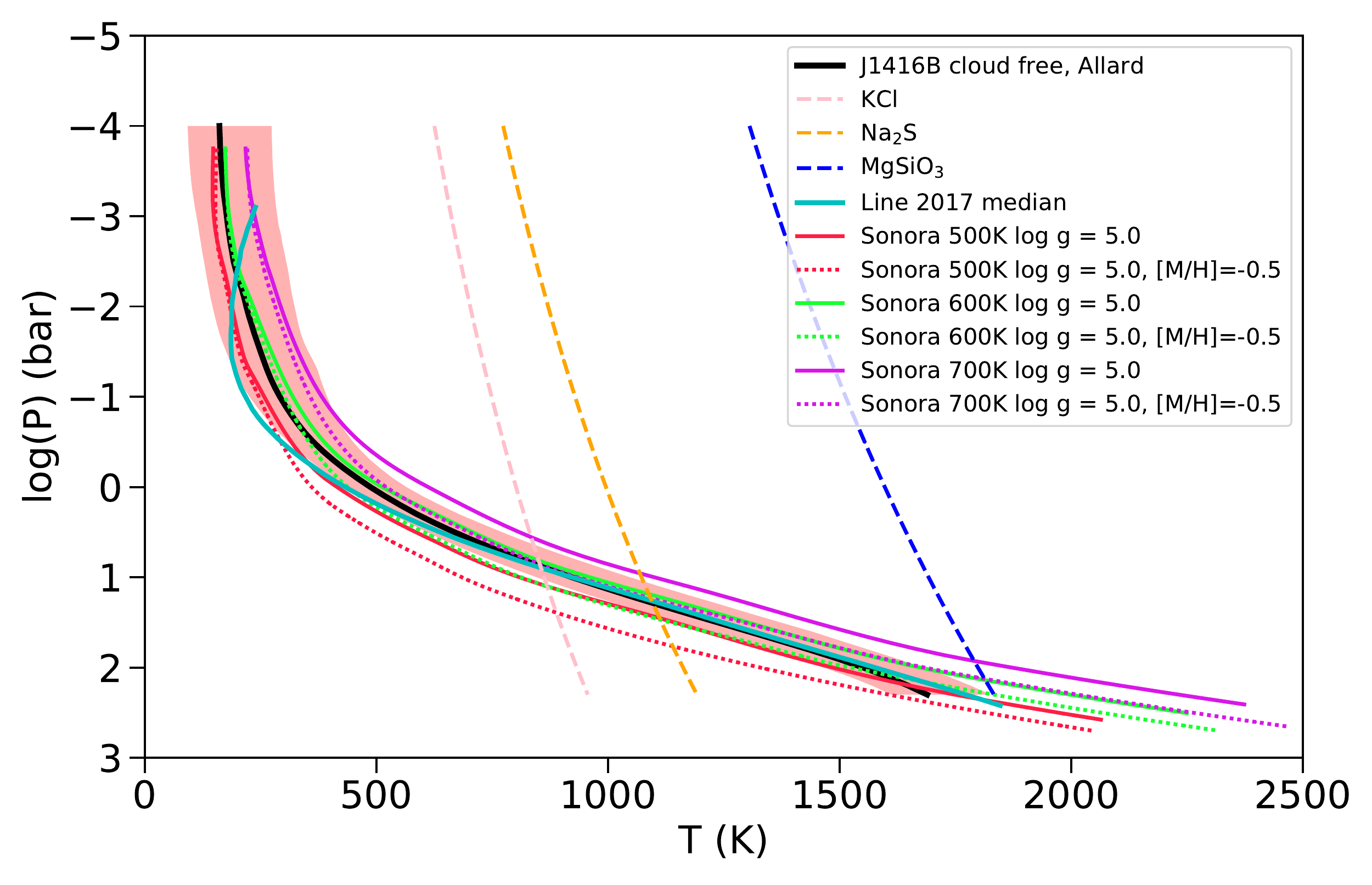}{0.5\textwidth}{\large(a)}
          \fig{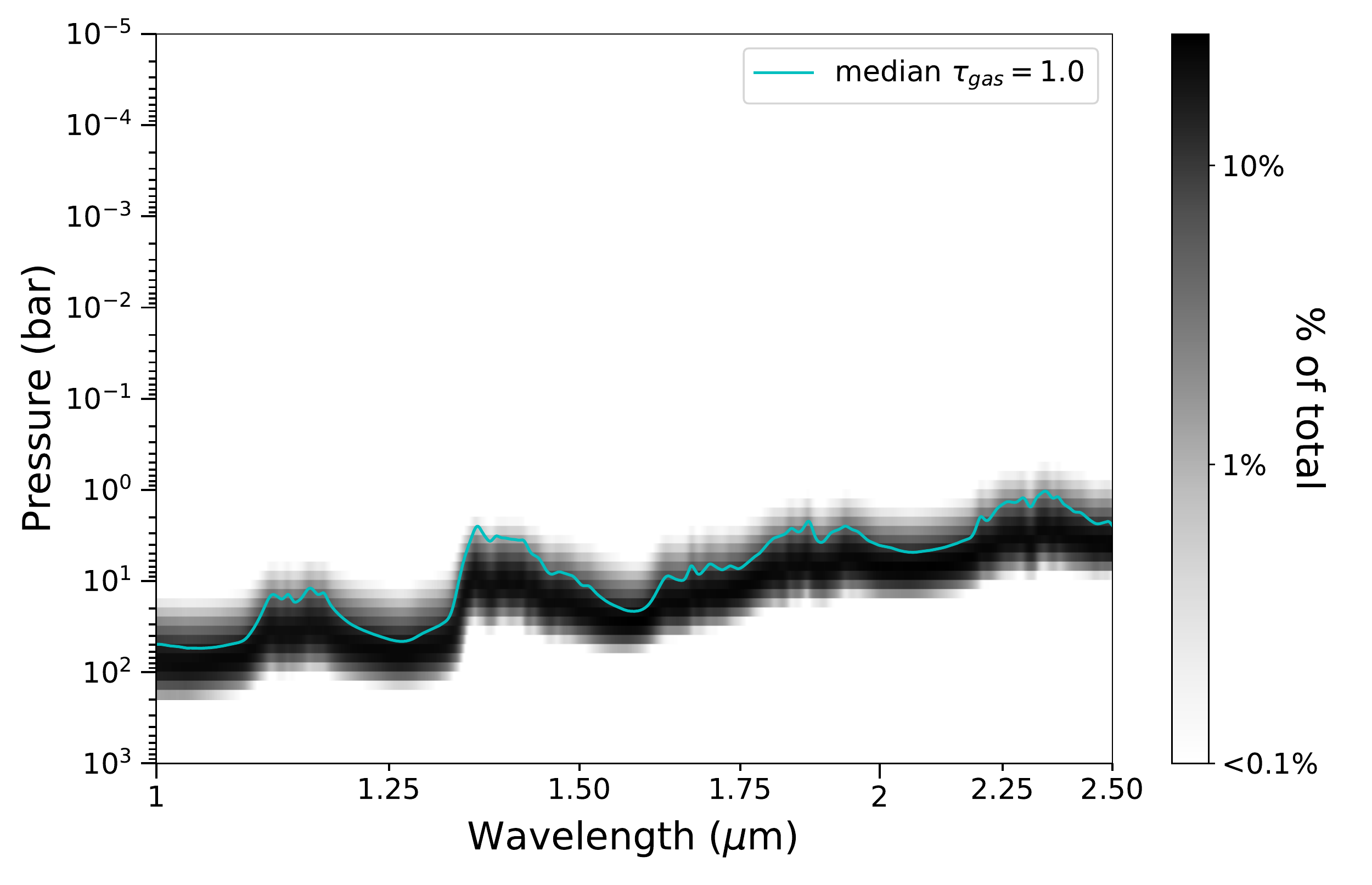}{0.5\textwidth}{\large(b)}} 
\caption{(a)Retrieved Pressure-Temperature Profile (black) compared to cloudless Sonora solar and low-metallicity model profiles (neon green, purple and bright pink) and the \cite{Line17} median profile in aqua. (b) Contribution plot with median gas at $\tau= 1$.}
\label{fig:1416b_PT_profile_allard}
\vspace{0.5cm} 
\end{figure*} 

Figure~\ref{fig:1416b_PT_profile_allard}(a) compares our retrieved median profile to Sonora 500K, 600K, and 700K solar and [M/H]=-0.5 models, and the median retrieved profile from \cite{Line17}. We see that our retrieved profile has a similar slope and is consistent within 1$\sigma$ across the entire profile with all models except the solar 700K and low-metallicity 500K Sonora models. Compared to the median profile from \cite{Line17}, we find our profile consistent within $1\sigma$; however, the shape of our profile differs from \cite{Line17} at pressures below $\sim-0.5$ bar. Many of the retrieved T dwarf profiles in \cite{Line17} were more isothermal than the models and they suggested it could be due to additional heating; however, temperature constraints are unreliable in this region of the profile. Figure~\ref{fig:1416b_PT_profile_allard}(b) shows the contribution function for this model with the photosphere ranging from about 1-100 bars.

\subsubsection{Retrieved Gas Abundances and Derived Properties}
\begin{figure*}
  \hspace{-0.25cm}
   \includegraphics[scale=.3]{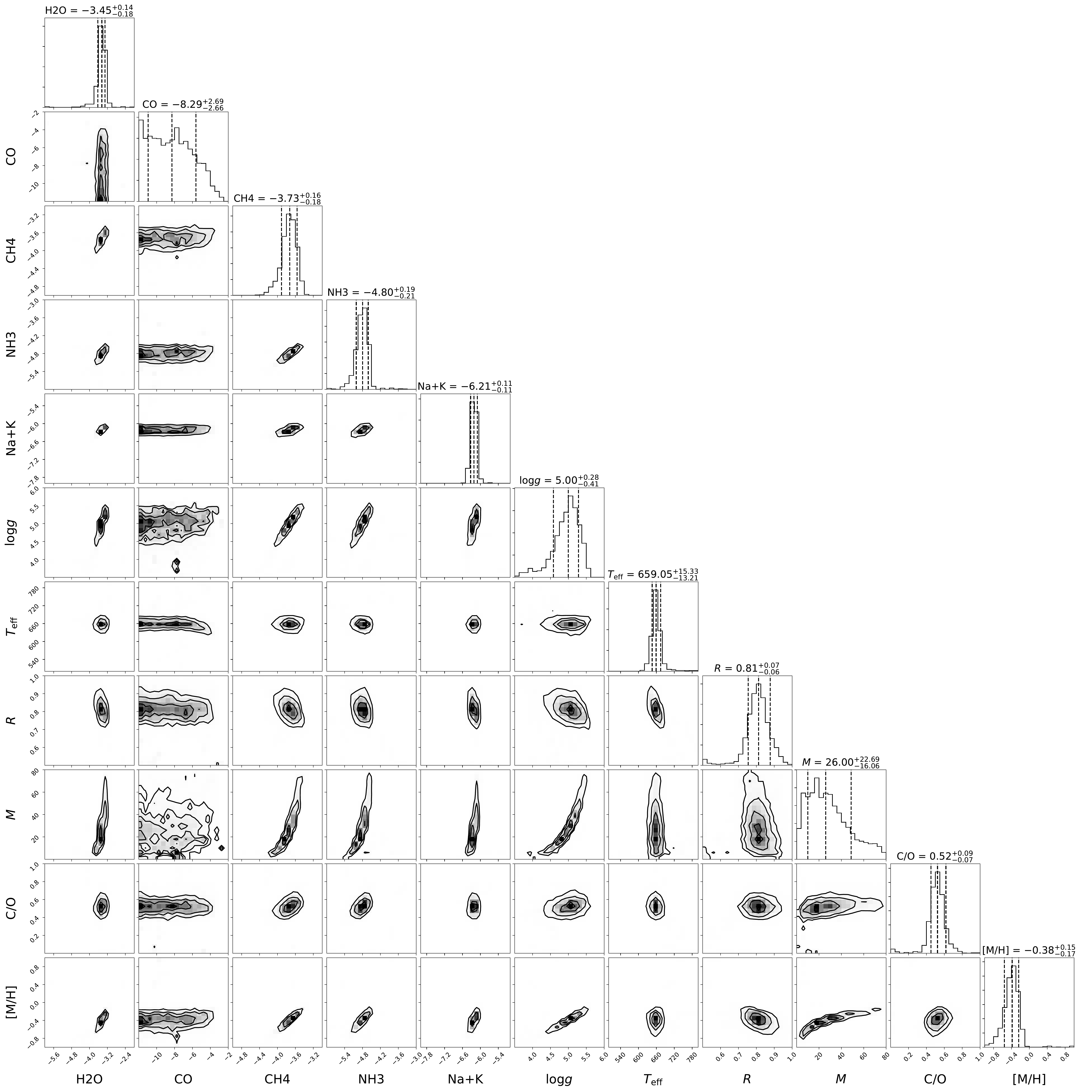} 
\caption{J1416B cloud-free posterior probability distributions for the retrieved and derived parameters using the Allard alkalis. 1D histograms of the marginalized posteriors are shown along the diagonals with 2D histograms showing the correlations between the parameters. The dashed lines in the 1D histograms represent the 16\textsuperscript{th}, 50\textsuperscript{th}, and 84\textsuperscript{th} percentiles, with the 68\% confidence interval as the width between the 16\textsuperscript{th} and 84\textsuperscript{th} percentiles. Parameter values listed above are shown as the median~$\pm1\sigma$. Gas abundances are displayed as log$_{10}$(X) values, where X is the gas. \Teff, radius, mass, C/O ratio, and {[M/H]} are not directly retrieved parameters, but are calculated using the retrieved $R^2/D^2$ and log($g$) values along with the predicted spectrum. Our derived C/O ratio is absolute, where Solar C/O is 0.55, while our [M/H] is relative to Solar. \Teff, radius, mass, C/O ratio, and {[M/H]} are not directly retrieved parameters, but are calculated using the retrieved $R^2/D^2$ and log $g$ values along with the predicted spectrum. CO abundance is not constrained and thus only provides an upper limit.}
\label{fig:1416b_nc_allard_postcorner}
\end{figure*}

\begin{deluxetable}{l c}
\tablecaption{Retrieved Gas Abundances and Derived Properties for J1416B \label{tab:1416B_Corner_values}} 
\tablehead{\colhead{Parameter}\phm{stringssssssssssssssssss} & \colhead{Value}}
  \startdata
  \multicolumn{2}{c}{Retrieved} \\\hline
  H$_2$O & $-3.45\substack{+0.14 \\ -0.18}$\\
  CO & <$-5.68$\\
  CH$_4$ & $-3.73\substack{+0.16 \\ -0.18}$\\
  NH$_3$ & $-4.80\substack{+0.19 \\ -0.21}$\\
  Na+K & $-6.21\pm0.11$\\
  log $g$ (dex) & \phm{+}$5.00\substack{+0.28 \\ -0.41}$ \\ \hline
  \multicolumn{2}{c}{Derived}  \\ \hline
  \Lbol & $-5.93\substack{+0.05 \\ -0.04}$ \\ 
  \Teff (K) & $659.05\substack{+15.33 \\ -13.21}$ \\ 
  Radius ($R_\mathrm{Jup}$) & \phm{+}$0.81\substack{+0.07 \\ -0.06}$ \\ 
  Mass ($M_\mathrm{Jup}$) & \phm{+}$26.01\substack{+22.68 \\ -16.07}$ \\
  {C/O}\tablenotemark{a,b,c} & \phm{+}$0.53\substack{+0.10 \\ -0.08}$ \\
  {[M/H]}\tablenotemark{a,c} & $-0.35\substack{+0.15 \\ -0.17}$ \\
  \enddata
  \tablenotetext{a}{Ratios determine from same gases in both the A and B components. Additional comparatives are listed in Table~\ref{tab:1416data}.}
  \tablenotetext{b}{$\mathrm{C/O}_\mathrm{Corr.}=0.39\substack{+0.07\\-0.05}$ when corrected ratio using the 25\% correction from \cite{Line17} to account for rainout.}
  \tablenotetext{c}{Atmospheric values.}
  \tablecomments{Molecular abundances are fractions listed as log values. For unconstrained gases, 1$\sigma$ confidence is used to determine upper limit.}
\end{deluxetable}

Posterior probability distributions for gases, surface gravity, \Teff, radius, mass, C/O, and [M/H] are shown in Figure~\ref{fig:1416b_nc_allard_postcorner} with their values along with the derived \Lbol listed in Table~\ref{tab:1416B_Corner_values}. Compared to results from \cite{Line17}, our derived \Teff is hotter and is not consistent within 1$\sigma$ (\Teff$=659.05\substack{+15.33 \\ -13.21}$ versus \Teff$=605\substack{+29 \\ -35}$), while our radius, surface gravity, and metallicity agree within $1\sigma$. Comparing our retrieved gas abundances to \cite{Line17} we find all the gases we have in common are consistent except for the Na+K alkali abundance. \cite{Line17} used the Burrows alkali opacities, while we use the Allard opacities in this model. When comparing to our model that used the Burrows opacities we find the Na+K abundance is consistent with \cite{Line17}. Similar to \cite{Line17}, we detect ammonia with our constraints equally as tight.

Our retrieved abundances yield a C/O ratio of $\mathrm{C/O}=0.52\substack{+0.09\\-0.07}$. To consider the effect of oxygen sequestration by silicate condensation in the atmosphere of J1416B, \cite{Line17} made a correction of 25\% to their retrieved C/O ratio, resulting in $\mathrm{C/O}_\mathrm{Corr.}=0.45\substack{-0.16\\+0.26}$. If we apply this same correction, we have $\mathrm{C/O}_\mathrm{Corr.}=0.39\substack{+0.07\\-0.05}$ which is consistent within $1\sigma$ of the pre-corrected value, the \cite{Line17} value, and is subsolar relative to the solar C/O ratio of C/O=0.55 \citep{Aspl09}. It should be noted that the correction used from \cite{Line17} (where 3.28 oxygen atoms are removed per silicon atom) is under the assumption of uniform metallicity variations in elemental abundance ratios (e.g.,Si/H$\sim$M/H; cf. \citealt{Viss10a}), as variations in the abundances of rock-forming elements (such as Mg and Si) will affect the proportion of oxygen removed by silicate condensation. However, as J1416B is subsolar, corrections to the C/O ratio may differ as subdwarf atmospheres have weak or absent metal oxides. If there is a relative depletion or lack of rock-forming elements, less oxygen would be sequestered, yielding a smaller correction in the C/O ratio. In Figure 13 of \cite{Niss14}, they show that as metallicity ([Fe/H]) decreases, the C/O is expected to decrease for thin-disk stars. Using our uncorrected metallicity, we find that our C/O ratio lies within the scatter of their expected metallicity prediction. The \cite{Line17} C/O ratio also falls within the scatter of the \cite{Niss14} metallicity prediction.

\subsubsection{Retrieved Spectrum and Composition\label{sec:spectrum_VMR_1416B}}

\begin{figure*}[!ht]
\centering
 \gridline{\fig{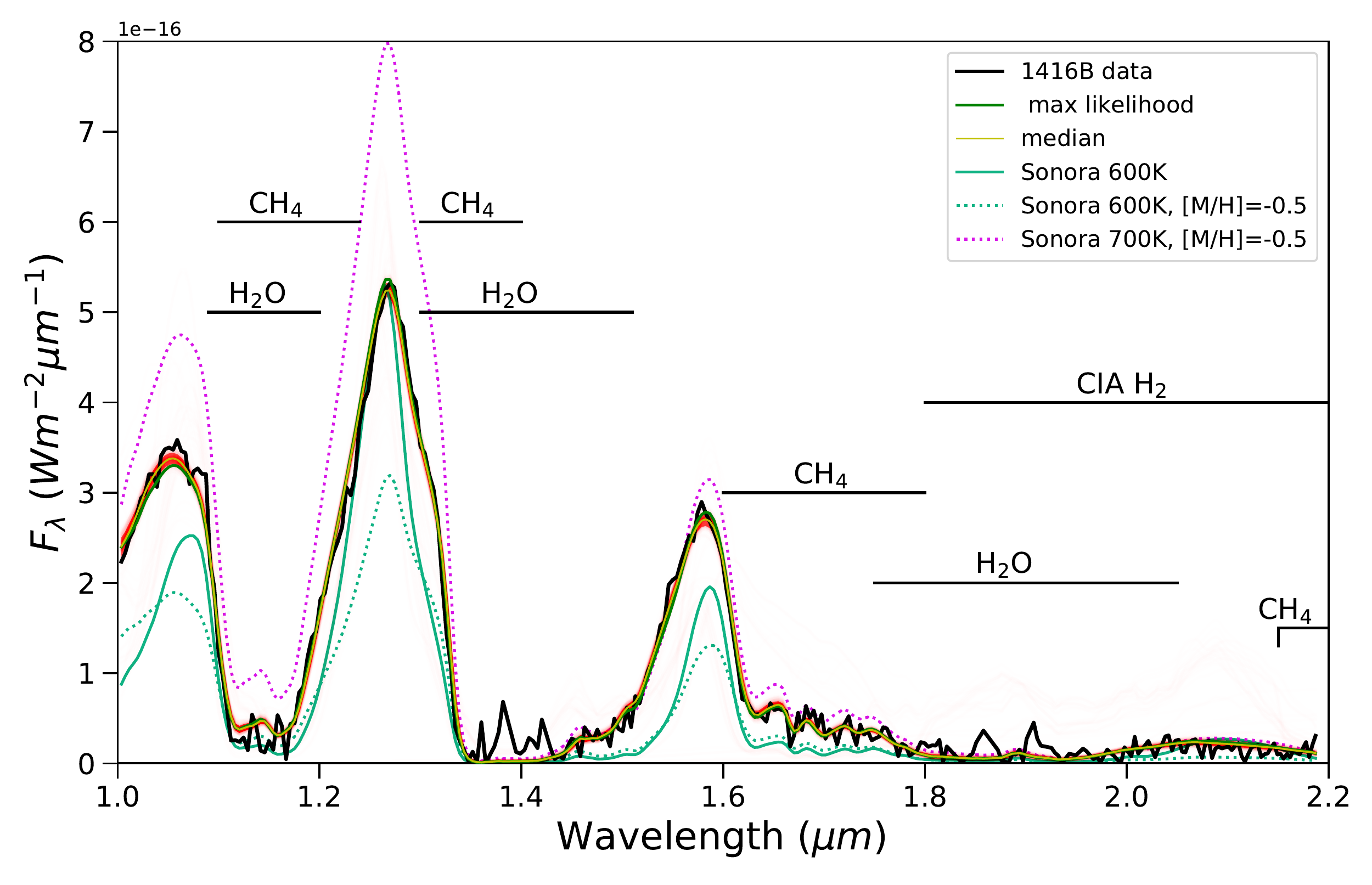}{0.5\textwidth}{\large(a)}
          \fig{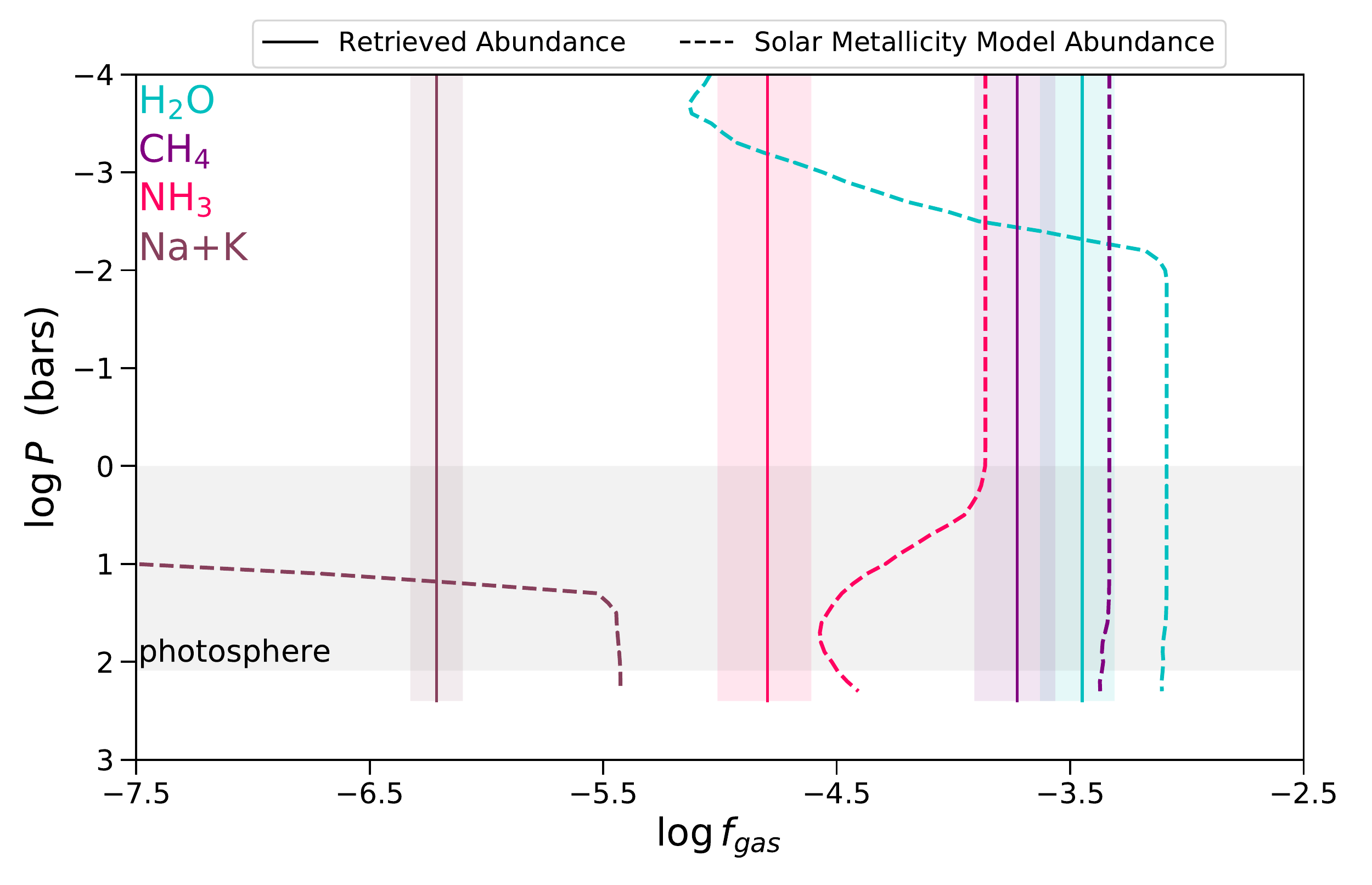}{0.5\textwidth}{\large(b)}} 
\caption{(a) Data (in black) compared to the retrieved maximum likelihood (in green) and median (in yellow) spectra. In red we show 500 random draws from the final 5,000 walkers of the converged MCMC chain. Sonora solar and low-metallicity model spectra are shown in teal and purple, respectively. (b) Retrieved uniform-with-altitude gas abundances for the cloudless Allard alkali model compared to solar abundances.}
\label{fig:1416B_nc_SPEC_VMR_allard}
\vspace{0.5cm} 
\end{figure*}

Figure~\ref{fig:1416B_nc_SPEC_VMR_allard}(a) compares our retrieved median and maximum likelihood spectra to the SpeX prism J1416B data and the best-fitting Sonora solar and [M/H]=-0.5 grid model spectra. We find that our retrieval spectrum fits quite well, with the exception of the $Y$-band peak being slightly below the data. In comparison to the Sonora model spectra, we find none of the models fit the $Y$-band peak, the 600K solar metallicity model does a good job fitting the $J$-band peak but is unable to fit the slope on either side quite well, and the $H$ and $K$ band features are best fit by the 700K low-metallicity model. We find our retrieved gas abundances for H$_2$O, CH$_4$, and NH$_3$ are subsolar in the photosphere, while the alkalies are broadly consistent with the solar value in Figure~\ref{fig:1416B_nc_SPEC_VMR_allard}(b). These values are consistent with those in \cite{Line17}. We find that relatively small changes in composition can drive major observable differences in the spectrum, particularly at lower temperatures. Therefore, with a slightly subsolar metallicity for J1416B, its spectrum differs quite drastically from field T dwarfs.

\section{Fundamental Parameter Discussion \label{sec:FundParmComp1416}} 

\subsection{J1416A Fundamental Parameter Comparison}
Table~\ref{tab:LitFunParamsA} compares our SED- and retrieval-based fundamental parameters to the literature. Additionally, we list new $UVW$ values using \cite{GaiaDR2} proper motions and parallax along with the radial velocity from \cite{Schm10a}. Our empirical \Lbol is 2.5$\sigma$ and 1.5$\sigma$ discrepant from our deck and slab retrieval-based bolometric luminosities respectively; however, all three measurements have very small uncertainties. The largest discrepancy between our SED and retrieval derived parameters are the \Teff and radius, with our \Teff for the deck cloud at minimum, 81K hotter and the slab 50K hotter than the semi-empirical \Teff of 1694K. This is due to our small retrieved radius of $R_{deck}=0.7\pm0.04$, $R_{slab}=0.77\substack{+0.10 \\-0.06}$, which is about 20\% smaller than the evolutionary model radius from the SED method (see Section~\ref{sec:evo_comp} for further discussion). Compared to the literature, our retrieval-based \Teff is hotter than all, except the model-based \Teff from \cite{Bowl10} (which also calculates a \Lbol, but using an atmospheric spectra model), while the retrieval-based masses are consistent with our SED method value and \cite{Bowl10}. The log\,$g$ we derive agrees between the SED and the retrieval methods. As this work is the first to derive a metallicity for J1416A, we find that the metallicity is consistent between both cloud models.

\begin{deluxetable*}{l c c c c c c c c c}
\tablecaption{Comparison of fundamental parameters from the literature for 1416A\label{tab:LitFunParamsA}} 
\tablehead{\colhead{Parameter} & \colhead{This Paper} & \colhead{This Paper} & \colhead{This Paper} &\colhead{Burn10} & \colhead{Schm10} & \colhead{Bowl10} & \colhead{Scho10} & \colhead{Cush10} \\
\colhead{} & \colhead{SED} & \colhead{Retrieval-Deck} &\colhead{Retrieval-Slab} &\colhead{} & \colhead{} & \colhead{} & \colhead{} & \colhead{}& \colhead{}}
  \startdata
  log $L_*/L_{\odot}$       & $-4.18\pm0.011$          & $-4.23 \pm 0.01$                   &$-4.21 \pm 0.01$                      &$\cdots$ & $\cdots$     & $-4.36\pm0.21$                & $\cdots$    & $\cdots$\\
  \Teff (K)                 & $1694\pm74$              & $1891\substack{+42.56 \\ -41.38}$  &$1821.53\substack{+64.58 \\ -102.79}$ &1500     & 1722         & 2200                          &$\cdots$     &1700 \\
  Radius ($R_\mathrm{Jup}$) & $0.92\pm0.08$            & $0.7 \pm 0.04$                     &$0.77\substack{+0.10 \\ -0.06}$       &$\cdots$ &$\cdots$      & $\cdots$                      &$\cdots$     & 0.81\\
  Mass ($M_\mathrm{Jup}$)   & $60\pm18$                & $36.82\substack{+31.92 \\ -18.71}$ &$36.96\substack{+30.48 \\ -18.71}$    & 75      &$\cdots$      & $61\pm9$\tablenotemark{a}     &$\cdots$     &$\cdots$\\
  log $g$                   & $5.22\pm0.22$            & $5.26\substack{+0.32 \\ -0.33}$    &$5.18\substack{+0.28 \\ -0.36}$       & 5.5     &$\cdots$      & 5.5                           &$\cdots$     &5.5\\
  Age (Gyr)                 & $0.5-10$                 &$\cdots$                            &$\cdots$                              & 10      & >0.8         & 1\tablenotemark{a}            & $\cdots$    & $\cdots$\\
  {[M/H]}               & $-0.3$\tablenotemark{b}   & $-0.17\substack{+0.21 \\ -0.23}$\tablenotemark{f}   & $-0.33\substack{+0.20 \\ -0.26}$\tablenotemark{f}     &$\cdots$ &$\cdots$      &$\cdots$                       &$\cdots$     & $\cdots$ \\
  C/O                         &$\cdots$                  &$0.59\substack{+0.11\\-0.21}$\tablenotemark{f}       & $0.58\substack{+0.11\\-0.21}$\tablenotemark{f}        & $\cdots$& $\cdots$     & $\cdots$                      &$\cdots$     & $\cdots$\\ 
  distance (pc)     & $9.3\pm 0.03$\tablenotemark{c}   &10\tablenotemark{d}                 &10\tablenotemark{d}                   &5-15     & $8\pm 1.6$   & $8.4 \pm 1.9$\tablenotemark{e}& $7.9\pm1.7$ &$\cdots$\\ 
  $U$\tablenotemark{g}      & $-17.48\pm0.5$           & $\cdots$                           &$\cdots$                              &$\cdots$ &$-17.9\pm0.5$  & $6\pm4$                       & $\cdots$    & $\cdots$ \\ 
  $V$\tablenotemark{g}     & $5.81\pm0.04$             &$\cdots$                            &$\cdots$                              &$\cdots$ &$10.2\pm1.2$  & $10.2\pm1.2$                  & $\cdots$    & $\cdots$\\
  $W$\tablenotemark{g}     & $-38.4\pm1.1$             &$\cdots$                            &$\cdots$                              &$\cdots$ &$-31.4\pm4.7$ & $-27\pm9$                     & $\cdots$    & $\cdots$\\
  \enddata
  \tablenotetext{a}{Additional masses based on assumed ages of 3 Gyr: $78\pm 3$ $M_\mathrm{Jup}$ and 10 Gyrs: $80.9\pm1.2$ $M_\mathrm{Jup}$.}
  \tablenotetext{b}{Due to the low metallicity in the literature, we use the \cite{Saum08} low-metallicity (-0.3 dex) cloudless evolutionary models to determine the radius range.}
  \tablenotetext{c}{From \cite{GaiaDR2}.}
  \tablenotetext{d}{For the retrieval, the distance-calibrated spectrum from the SED was used, thus it was set to a distance of 10 pc. Distance uncertainty is included for determining the extrapolated parameters using the measured distance uncertainty.}
  \tablenotetext{e}{An estimated distance of $9.4\pm 1.3$ pc is given assuming a low metallicity and using the \cite{Cush09} relations.}
  \tablenotetext{f}{Same gas set between J1416AB used for deriving value.}
  \tablenotetext{g}{We derive new $UVW$ values in this work and do not correct for LSR. $UVW$ values from \cite{Schm10a} and \cite{Bowl10} were both corrected for LSR using \cite{Dehn98}. Thus as \cite{Schm10a} uses LSR$_\mathrm{Corr}=(10, 5, 7)$ making $UVW_\mathrm{No LSR}=(-17.9,2.2,-38.4)$, while \cite{Bowl10} uses LSR$_\mathrm{Corr}=(-10, 5.25, 7.17)$ making $UVW_\mathrm{No LSR}=(16,4.95,-34.17)$.} 
  \tablecomments{Column shortnames correspond to: Burn10: \cite{Burn10}, Schm10: \cite{Schm10a}, Bowl10: \cite{Bowl10}, Scho10: \cite{Scho10}, Cush10: \cite{Cush10}.}
\end{deluxetable*}

\subsection{J1416B Fundamental Parameter Comparison}
\begin{deluxetable*}{l c c c c c c c c c}
\tablecaption{Comparison of fundamental parameters from the literature for 1416B\label{tab:LitFunParamsB}} 
\tablehead{\colhead{Parameter} & \colhead{This Paper} & \colhead{This Paper} &\colhead{Burn10} & \colhead{Scho10} & \colhead{Burg10b}& \colhead{Burg10c\tablenotemark{a}}& \colhead{Burg10c\tablenotemark{a}} & \colhead{Fili15} & \colhead{Line17}  \\
\colhead{} & \colhead{SED} & \colhead{Retrieval\tablenotemark{b}} &\colhead{} & \colhead{} & \colhead{}& \colhead{cloudless}& \colhead{cloudy} & \colhead{} & \colhead{}}

  \startdata
  log $L_*/L_{\odot}$      & $-5.80\pm0.07$ & $-5.93\substack{+0.05\\-0.04}$     & $\cdots$ & $\cdots$    &$\cdots$                       &$\cdots$                       &$\cdots$                    & $-5.813\pm0.013$ &$\cdots$                        \\
  \Teff (K)                & $660\pm62$     & $659.05\substack{+15.33\\-13.21}$  & $500$    & 600         & $650\pm60$                    &$685\substack{+55\\-65}$       & $595\substack{+25\\-45}$   & $656\pm54$       & $605\substack{+29\\-35}$       \\
  Radius ($R_\mathrm{Jup}$)& $0.94\pm0.16$  &$0.81\substack{+0.07\\-0.06}$       & $\cdots$ & $\cdots$    & $0.83\substack{+0.14\\-0.10}$ &$0.84\pm0.06$                  & 0.86                       &$0.96\pm0.16$     &$0.8\substack{+0.07\\-0.06}$    \\
  Mass ($M_\mathrm{Jup}$)  & $33\pm22$      &$26.01\substack{+22.68\\-16.07}$    & $30$     & 30          & $22-47$                       &$43.0\substack{+11.5\\-10.5}$  &$36.7\substack{+1.0\\-3.1}$ &$30.23\pm19.86$   & $\cdots$                       \\
  log $g$                  & $4.83\pm0.51$  &$5.00\substack{+0.28\\-0.41}$       & 5.0      & $\cdots$    & $5.2\pm0.4$                   & $5.2\pm 0.3$                  &$5.5$                       &$4.80\pm0.52$     & $4.93\pm 0.4$                  \\
  Age (Gyr)                & $0.5-10$       &$\cdots$                            & 10       & 5           & $2-10$.                       & $8\pm 4$                      &$6-12$                      & $0.5-10$         &$\cdots$                        \\
  C/O                      &$\cdots$        &$0.52\substack{+0.09\\-0.07}$\tablenotemark{c,d}       & $\cdots$ & $\cdots$    & $\cdots$                      & $\cdots$                      &$\cdots$                    & $\cdots$         &$0.45\substack{+0.26\\-0.16}$\tablenotemark{e} \\ 
  {[M/H]}               & $-0.3$\tablenotemark{f} &$-0.35\substack{+0.10\\-0.08}$\tablenotemark{c}       & -0.3     &$\cdots$     & $<-0.3$                       &$-0.17\substack{+0.17\\-0.13}$ & 0.0                        & 0.0              & $-0.35\substack{+0.10\\-0.11}$ \\
  distance (pc) & $9.3\pm0.03$\tablenotemark{g}  &10\tablenotemark{h}            &$5-15$    & $7.9\pm1.7$ & $10.6\substack{+3.0\\-2.8}$   & $11.1\pm3.2$                  &$11.4\pm3.4$                & $9.12\pm0.11$\tablenotemark{i}         & $9.12\pm0.11$\tablenotemark{j} \\
  \enddata
  \tablenotetext{a}{Mean values listed.}
  \tablenotetext{b}{Here we list values from the Allard alkalies for the winning model.}
  \tablenotetext{c}{Same gas set between J1416AB used for deriving value.}
  \tablenotetext{d}{If we using the rainout correction from \cite{Line17}, $\mathrm{C/O}_\mathrm{Corr.}=0.39\substack{+0.07\\-0.05}$.}
  \tablenotetext{e}{Rainout corrected value listed in \cite{Line17} in log$_{10}$C/O.}
  \tablenotetext{f}{Due to the low metallicity in the literature, we use the \cite{Saum08} low-metallicity (-0.3 dex) cloudless evolutionary models to determine the radius range.}
  \tablenotetext{g}{From \cite{GaiaDR2}.}
  \tablenotetext{h}{For the retrieval, the distance-calibrated spectrum from the SED was used, thus it was set to a distance of 10 pc. Distance uncertainty is included for determining the extrapolated parameters using the measured distance uncertainty.}
  \tablenotetext{i}{Used parallax from \cite{Fahe12}.}
  \tablenotetext{j}{Used parallax from \cite{Dupu12a}.}
  \tablecomments{Column shortnames correspond to: Burn10: \cite{Burn10}, Scho10: \cite{Scho10}, Burg10b: \cite{Burg10b}, Burg10c: \cite{Burg10c}, Fili15: \cite{Fili15}, Line17: \citealt{Line17}.} 
\end{deluxetable*}

Table~\ref{tab:LitFunParamsB} lists our SED and retrieval method fundamental parameters compared to the literature. Comparing \Lbol, we find that both our SED and retrieval method values agree with \cite{Fili15} within 1$\sigma$. Our semi-empirical and retrieval-based \Teff radius, mass, and log\,$g$ are consistent with one another and the literature within 1$\sigma$, with the exception of $T_\mathrm{eff}$ which is consistent within 2$\sigma$. Our retrieval C/O and [M/H] measurements are consistent with those in the literature.

\subsection{Comparison of characteristics to evolutionary diagrams\label{sec:evo_comp}}

\begin{figure*}
\gridline{\fig{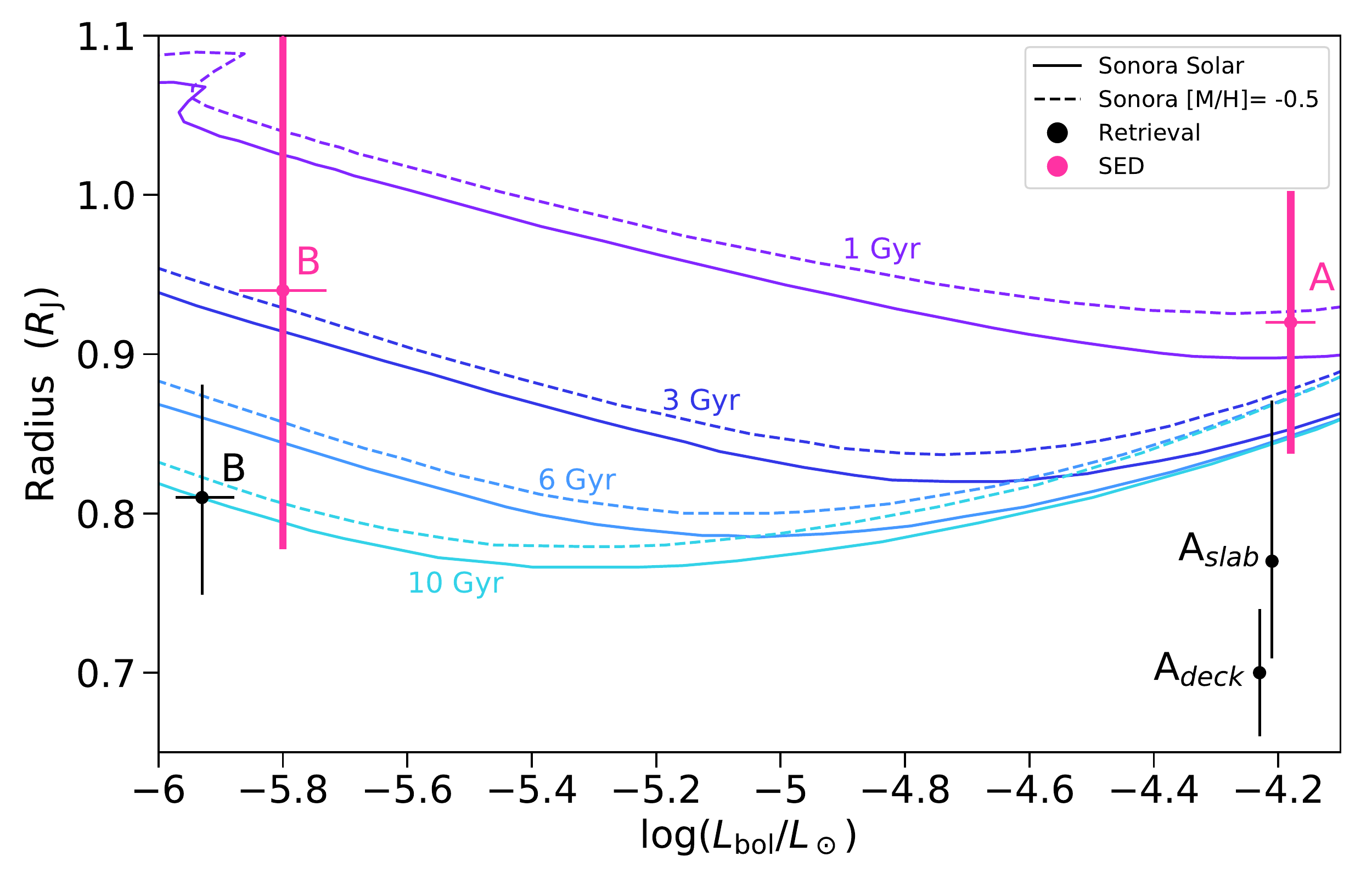}{0.5\textwidth}{\large(a)}
          \fig{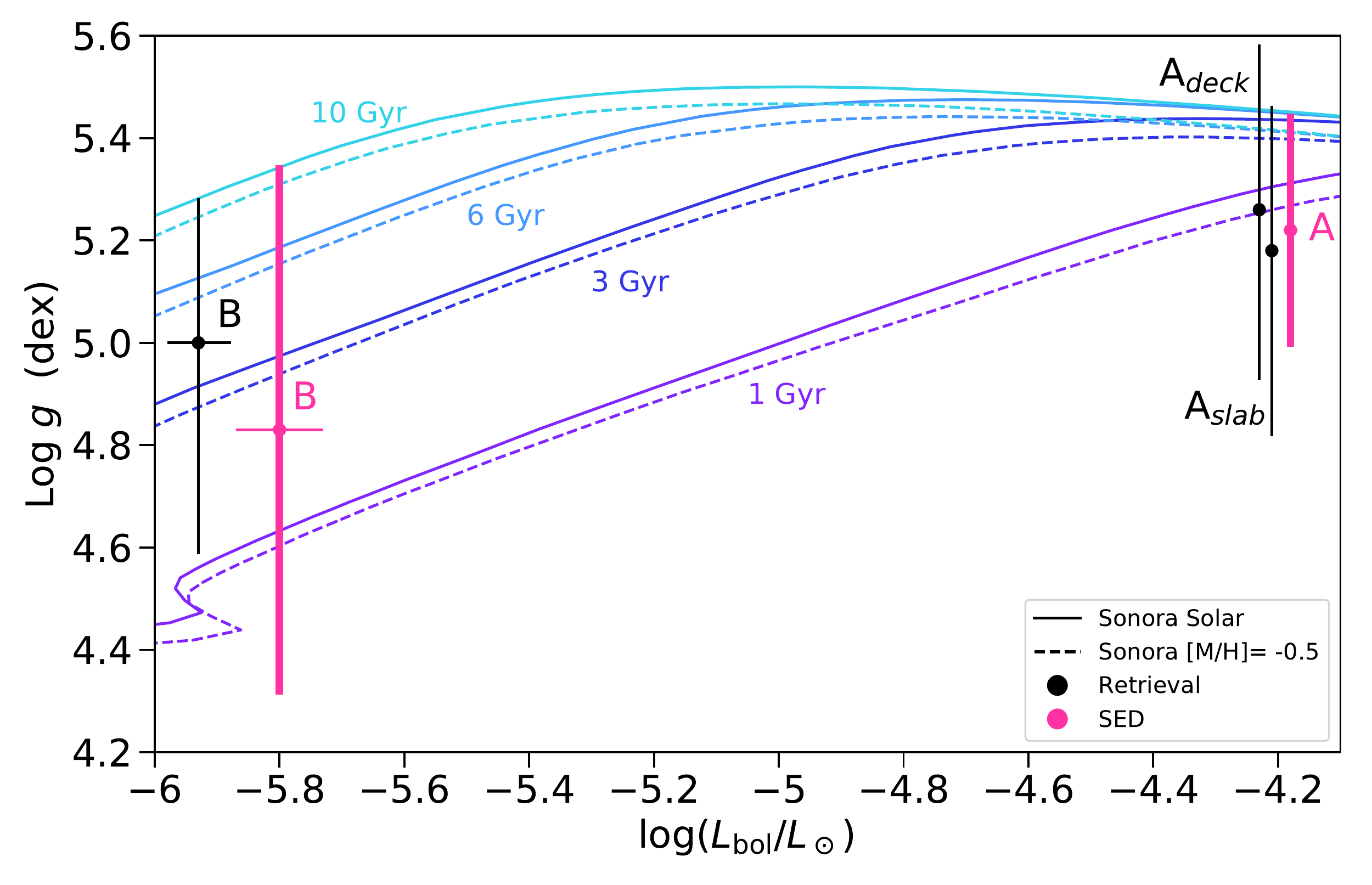}{0.5\textwidth}{\large(b)}}
 \gridline{\fig{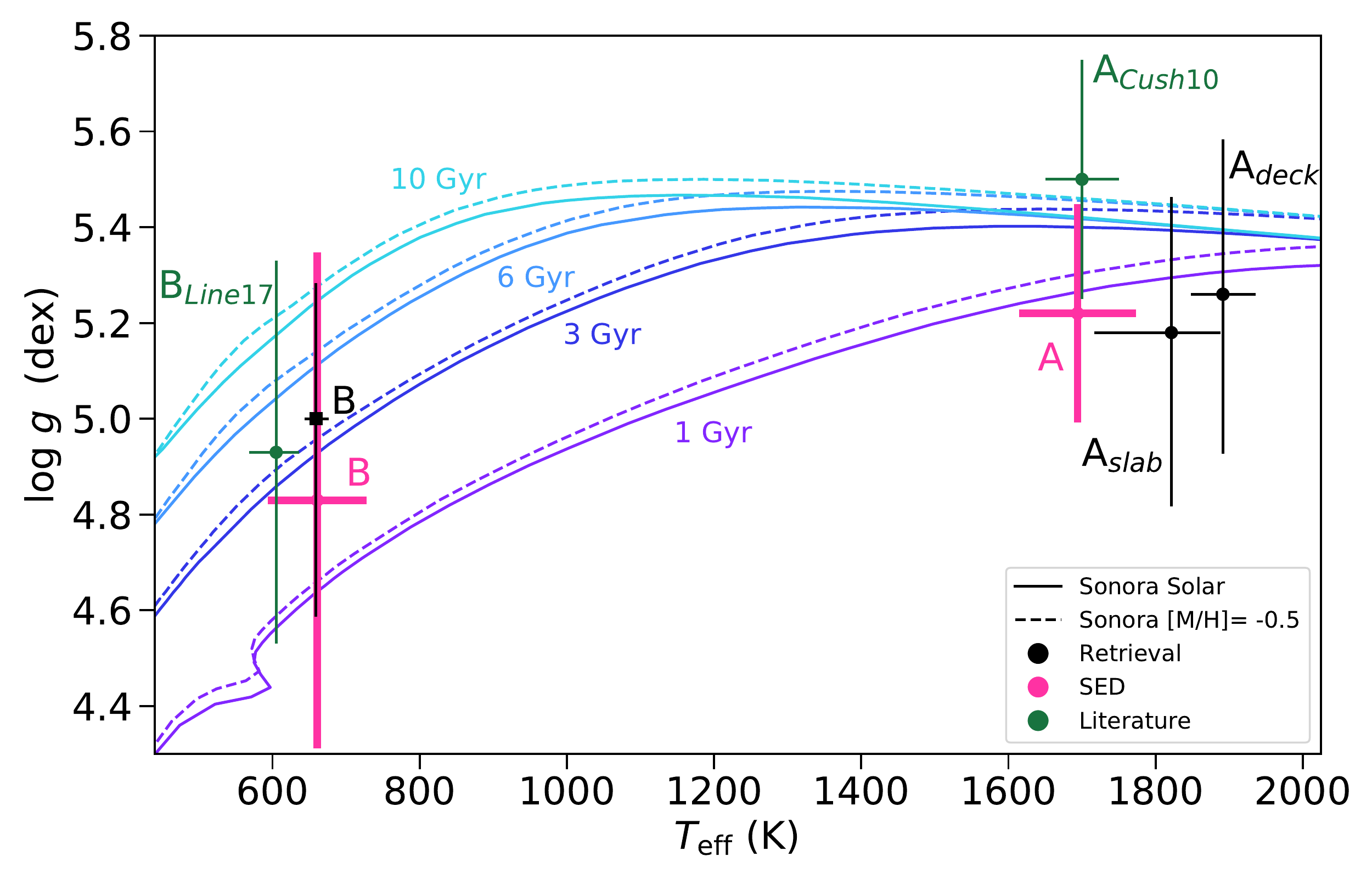}{0.5\textwidth}{\large(c)}
           \fig{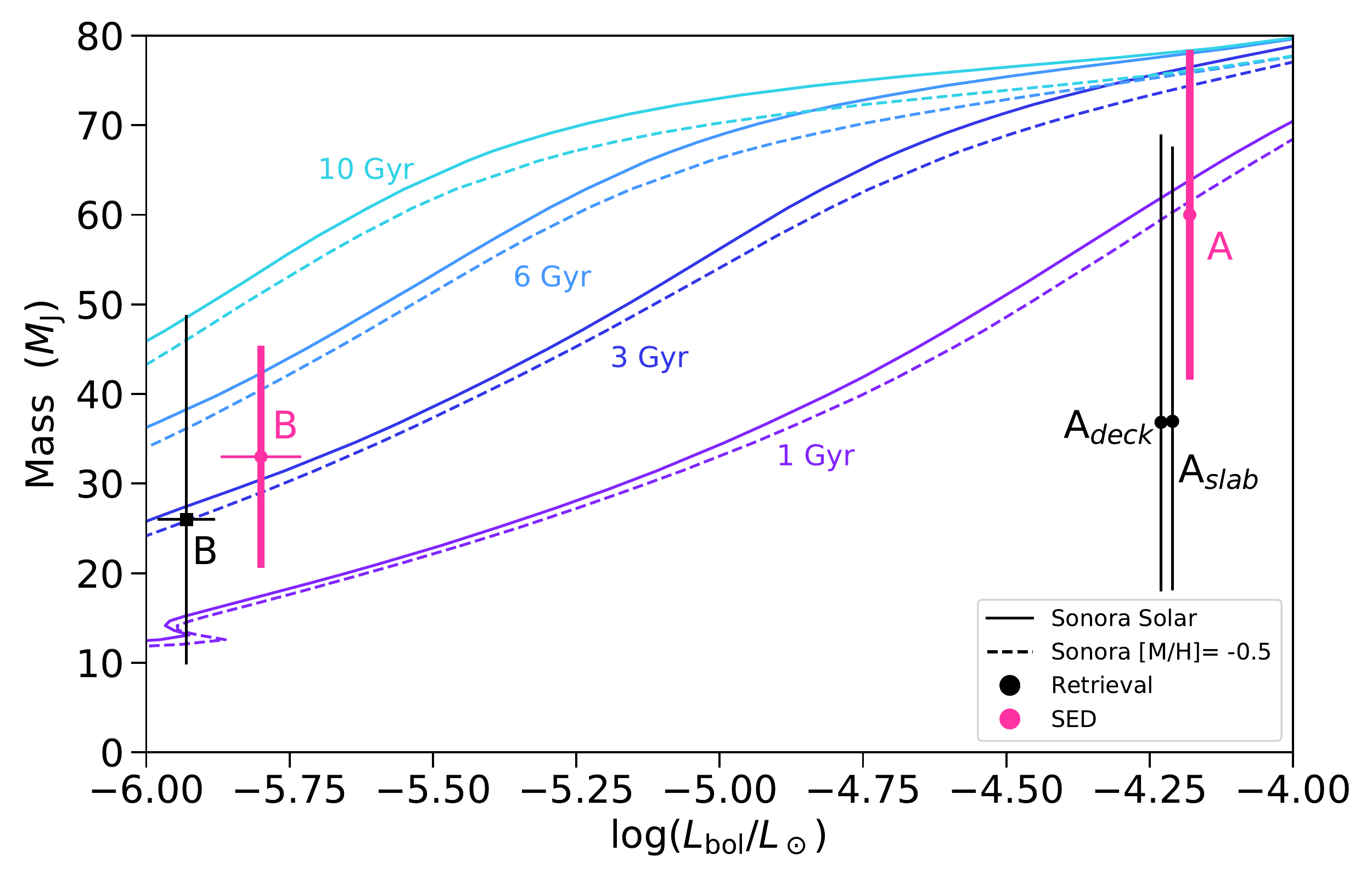}{0.5\textwidth}{\large(d)}} 
\caption{Comparison of retrieved bolometric luminosity, radius, surface gravity, and mass to the Sonora Bobcat evolutionary solar and low metallicity models. {[M/H]}$=0.0$ are displayed as solid lines, while {[M/H]}$=-0.5$ are dashed lines, with ages ranging from $1-10$~Gyr in shades of blue and purple. Black symbols show the retrieved values using the Allard alkalies, while pink points show the SED derived values, with values that are ranges as thick pink lines as they use the radius range in the determination of the value, while thin lines are uncertainties. Non-visible uncertainties are smaller than the point size. (a) Radius versus \Lbol. (b) Log\,$g$ versus \Lbol. (c) Log\,$g$ versus \Teff. Green points show the \cite{Cush10} and \cite{Line17} values for J1416A and J1416B respectively. (d) Mass versus \Lbol.}
\label{fig:Comparison_evo_models}
\vspace{0.5cm} 
\end{figure*}

Figures~\ref{fig:Comparison_evo_models}(a)--(d) compare our SED- and retrieval-based fundamental parameters to Sonora evolutionary model grids for solar and low metallicity ([M/H]=-0.5). As the SED-based parameters \Teff, mass, and radius are drawn from different evolutionary models (see Section~\ref{sec:1416data} for evolutionary models used), these are plotted for comparison to the retrieval-inferred values and not to the evolutionary models themselves. 

A comparison of radius versus \Lbol is shown in Figure~\ref{fig:Comparison_evo_models}(a), with the retrieval shown in black and the SED in pink. It is quite clear that the derived retrieval radius for the deck cloud model of J1416A is smaller than predicted by the evolutionary models, while the slab cloud is consistent with the low-metallicity $6-10$~Gyr and the solar-metallicity $3-6$~Gyr models. While the radius of the deck cloud model may appear to be unphysically small, \cite{Sora13} estimated radii of brown dwarfs from the scale factor, similar to our method, using \textit{AKARI} spectra and found that most of their mid- to late-L dwarfs had radii smaller than predicted from evolutionary models. We find both the deck and slab cloud radii for J1416A fall within the radius range of $0.64-0.81$\, $R_\mathrm{Jup}$ for the mid- to late-L dwarfs in \cite{Sora13} with a \Teff between 1500 and 2000K. The problem of unphysically small radii is an ongoing problem for atmospheric retrievals (e.g. \citealt{Zale19}) and has been seen as an issue for the directly-imaged exoplanets as well. We caution the reader against using the retrieved radii for J1416A.

Compared to our SED method radius, we see that it is only consistent with J1416A's slab cloud model radius. For J1416B, the retrieval radius is consistent with the Sonora evolutionary models and the SED method radius. As seen with J1416A, J1416B's SED method radius is larger than the retrieval-derived radius. J1416A and B's empirical \Lbol from the SED are fainter than the retrieval-derived \Lbol, which is inferred from integration under the retrieved forward model spectrum. The retrieval derived radius for J1416B constrains the age to be $>6$~Gyrs.

Comparison of the retrieved and evolutionary model based (from the SED method) surface gravity versus \Lbol compared to the Sonora Bobcat evolutionary models are shown in Figure~\ref{fig:Comparison_evo_models}(b). The surface gravity for both retrieval models of J1416A are consistent with the SED value, and the same is seen between J1416B's retrieved and SED surface gravity. Here, we see that both J1416A's slab and deck retrieval, as well as the SED, log\,$g$ give an age range of $1-10$~Gyr. For J1416B, we find the retrieved log\,$g$ produces an age range of $1-10$~Gyr, which is broader than the range given from the radius. 

Figure~\ref{fig:Comparison_evo_models}(c) compares the log\,$g$ versus \Teff, where here we also compare J1416A to literature results from model values in \cite{Cush10}, and the retrieval results for J1416B from \cite{Line17}. While the log\,$g$ for J1416A and B are consistent across the SED, retrievals, and the literature values plotted, the \Teff measurements vary over a wider range, particularly for J1416A. When comparing mass versus \Lbol in Figure~\ref{fig:Comparison_evo_models}(d), we find that the retrieval places J1416A with a very young age of likely less than $1$~Gyr, which is strikingly different from the very old age estimate from the radius. This age disagreement is likely due to the mass being tied to the radius and with a larger radius the derived mass range would be higher.

\section{Discussion\label{sec:discussion1416}}
Table \ref{tab:Comparison_disscussion} lists all parameters we will discuss when comparing between J1416A and J1416B, particularly those of interest when determining if the system formed and evolved together. We list the retrieved alkali abundances, C/O, and [M/H] determined when using both the Allard and Burrows alkali opacities for both J1416A and J1416B. Here we use the C/O and [M/H] ratios determined from using only gases that both J1416A and J1416B have in common (H$_2$O, CH$_4$, and CO). Agreement in the expected behavior of the alkali abundances was the primary deciding factor on the preferred cross-sections.

\begin{deluxetable*}{l c c c c c c}
\tablecaption{Properties for Comparison between J1416A and J1416B\label{tab:Comparison_disscussion}} 
\tablehead{\colhead{Object} & \multicolumn{2}{c} {log(Na+K)} & \multicolumn{2}{c} {C/O\tablenotemark{a}} & \multicolumn{2}{c} {[M/H]\tablenotemark{a}} \\
\cmidrule(lr){2-3}\cmidrule(lr){4-5}\cmidrule(lr){6-7}\colhead{} & \colhead{Allard} & \colhead{Burrows} & \colhead{Allard} & \colhead{Burrows} &\colhead{Allard} & \colhead{Burrows} }
  \startdata
  1416A deck & $-6.32\substack{+0.17 \\ -0.21}$ & $-6.62\substack{+0.18\\-0.24}$ & $0.59\substack{+0.11 \\ -0.21}$ & $0.60\substack{+0.11 \\ -0.16}$ & $-0.17\substack{+0.21 \\ -0.23}$ & $-0.11\substack{+0.18 \\ -0.21}$\\
  1416A slab & $-6.90\substack{+0.17 \\ -0.21}$ & $-7.15\substack{+0.41\\-0.29}$ & $0.58\substack{+0.11 \\ -0.21}$ & $0.57\substack{+0.11 \\ -0.26}$  & $-0.33\substack{+0.20 \\ -0.26}$ & $-0.29\substack{+0.21 \\ -0.27}$\\ 
  1416B & $-6.21\pm0.11$ & $-5.29\substack{+0.05\\-0.06}$ &$0.53\substack{+0.10\\-0.08}$ & $0.50\substack{+0.11\\-0.07}$ & $-0.35\substack{+0.15\\-0.17}$& $-0.47\substack{+0.16\\-0.14}$\\
  \enddata
  \tablenotetext{a}{The is the AB comparative for C/O and [M/H]. The other versions can be found in Table~\ref{tab:1416data}.}
  \tablecomments{C/O values are listed as absolute, where Solar C/O=0.55, and [M/H] is listed relative to Solar abundances. }
\end{deluxetable*} 

\subsection{Addressing the Differences in Alkalies \label{sec:Alkalies}}
The alkali abundances retrieved for J1416AB are listed in Table \ref{tab:Comparison_disscussion}, using both alkali opacity models. The Allard opacities are able to produce consistent alkali abundance between J1416A and J1416B, only when J1416A is parameterized with the deck cloud. Alkali abundances do not necessarily need to be consistent between J1416A and J1416B because they are condensing out at around the \Teff of J1416B \citep{Line17, Zale19}. However, the Burrows opacities result in J1416AB having a higher alkali abundance than J1416A, which is not expected to occur in T dwarfs due to rainout. To check for correlations or degeneracies between alkali abundance and the cloud parameters of both cloud models for J1416A, we created a corner plot using the Allard opacity retrieval results and found no correlations for either cloud model. Because the Allard alkalies produce the expected alkali abundance behavior between J1416A and B with the deck cloud and not the slab cloud, this is evidence that the deck cloud produces a more realistic fit to the data over the slab cloud for J1416A.

\subsection{C/O Ratio}  \label{sec:co_ratio1416}
To compare the C/O ratio between J1416A and J1416B, we have derived an atmospheric C/O ratio that only considers the gases in common between both sources (H$_2$O, CO, and CH$_4$), due to the differing gas assumptions in the L and T dwarf retrievals. For the L dwarf, there will be a small contribution from VO missing in the oxygen total. However, as VO has a very small abundance it does not make a large impact in the overall C/O ratio. Using this C/O ratio, we find that J1416AB are consistent within $1\sigma$, which points towards evidence in favor of formation and evolution as a pair. Both J1416A and J1416B are approximately solar in C/O and have slightly subsolar metallicities. Considering the various methods to determine the C/O for J1416AB, all methods are consistent within $1\sigma$ and do not differ based on which alkali opacities are used.

As a note, the C/O ratios in Table~\ref{tab:Comparison_disscussion} do not include the rainout correction, as we have not made any corrections to the C/O ratio of J1416A. The rainout correction applied to J1416B accounts for oxygen that should be in the atmosphere above any deep cloud not detected in the retrieval. For J1416A, because the retrieval prefers a cloudy model we have an entirely different situation to consider. If a correction is necessary for J1416A it would likely be a smaller amount, because a much smaller fraction of the total atmosphere is above the cloud (i.e. for the median slab or deck cloud we would be accounting for oxygen above $\sim0.1$~bar) than is in the case of J1416B. In addition, we should consider oxygen tied up in SiO gas in J1416A. Considering this the correction for J1416A could range from 0.5-12\%, which is well within our 68\% confidence interval of our C/O ratio.

\subsection{Metallicity Differences?}
To compare the metallicity between J1416A and J1416B, it is important to remember that the gases used to derive the individual atmospheric metallicities differed between the L and T dwarf atmospheres. To account for this, we take the same approach as for the C/O ratio and determine a metallicity using only the gases in common between the L and T dwarf to determine the elemental abundances in our metallicity calculation. This approach does not include elements that are expected to have a large portion taken up by unobservable sinks such as N$_2$ or condensation of iron in the L dwarf. However, both nitrogen (for J1416B) and iron (for J1416A) would affect the metallicity determination at the 10\% level, well within our 68\% confidence intervals.  

The $1\sigma$ confidence intervals are quite large for both alkali opacity variant metallicities, with the Allard opacities producing consistent [M/H] between J1416A and J1416B regardless of J1416A's cloud model. When using the Burrows opacities, the derived metallicities are inconsistent between J1416A's deck cloud model and J1416B. It should be noted that both alkali models produce a lower median metallicity for the deck cloud compared to the slab, however, only the Allard model is consistent. Additionally, the Burrows model produces a higher median metallicity for J1416A, while a lower median metallicity for J1416B. Only the Allard opacities produce a consistent picture of the co-moving pair.

\section{Conclusions}

In this work we present the first distance-calibrated SED of J1416A and an updated distance-calibrated SED of J1416B. We present the first retrieval of J1416A and the second retrieval of J1416B. J1416A is best parameterized by a power-law deck cloud model; however, it is indistinguishable from a power-law slab cloud model. While J1416B is best fit by a cloud-free model, agreeing with previous results from \cite{Line17}. For both cloud models of J1416A, we find our retrieval radius is smaller than the evolutionary model radius and inconsistent within 1$\sigma$. We also find that the retrieval produces a hotter \Teff than the SED to compensate for the smaller radius and to maintain the same flux we observe. We find that relatively small changes in the composition can drive major changes in observed features in the spectrum, particularly for low temperature sources.

Examining the retrieval results across the pair, we find that only the Allard alkali opacities produce alkali abundances expected for J1416AB (with the T dwarf abundance lower than the L dwarf) and only for the deck cloud model for A. Both J1416A and J1416B have slightly subsolar metallicities that are consistent with each other, no matter the chosen alkali opacity model. J1416AB is consistent with an approximately solar C/O ratio, with the median value slightly super-solar for A and slightly subsolar for B. These results point toward the pair having formed and evolved together. Retrieval results of this binary are the first look from a larger sample that aims to dive deeper into understanding subdwarf atmospheres by asking: (1) Are subdwarfs cloudless? and (2) How do their PT profiles compare to similar spectral type or \Teff sources (Gonzales et al. in prep). Having both cloudy and cloud-free results from this work provides a step in understanding the nuances of metallicity in L and T dwarf atmospheres.

\acknowledgments
This research was supported by the NSF under Grant No. AST-1614527, grant No. AST-1313278, and grant No. AST-1909776. This research was made possible thanks to the Royal Society International Exchange grant No. IES{\textbackslash}R3\textbackslash170266. E.G. thanks the LSSTC Data Science Fellowship Program, which is funded by LSSTC, NSF Cybertraining Grant No. 1829740, the Brinson Foundation, and the Moore Foundation; her participation in the program has benefited this work. BB acknowledges financial support from the European Commission in the form of a Marie Curie International Outgoing Fellowship (PIOF-GA-2013-629435). This publication makes use of data products from the Two Micron All Sky Survey, which is a joint project of the University of Massachusetts and the Infrared Processing and Analysis Center/California Institute of Technology, funded by the National Aeronautics and Space Administration and the National Science Foundation. This publication makes use of data products from the Wide-field Infrared Survey Explorer, which is a joint project of the University of California, Los Angeles, and the Jet Propulsion Laboratory/California Institute of Technology, funded by the National Aeronautics and Space Administration. This work has made use of data from the European Space Agency (ESA) mission {\it Gaia} (\url{https://www.cosmos.esa.int/gaia}), processed by the {\it Gaia} Data Processing and Analysis Consortium (DPAC, \url{https://www.cosmos.esa.int/web/gaia/dpac/consortium}). Funding for the DPAC has been provided by national institutions, in particular the institutions participating in the {\it Gaia} Multilateral Agreement.

\software{astropy \citep{Astropy},  
          SEDkit (\url{https://github.com/hover2pi/SEDkit},\textit{Eileen Branch}), 
          \textit{Brewster} \citep{Burn17},
          EMCEE \citep{emcee},
          Corner \citep{Corner}
          }

\appendix
\section{Alternative Alkalies for Winning Models}
\subsection{J1416A Burrows Models}
\begin{figure*}
\gridline{\fig{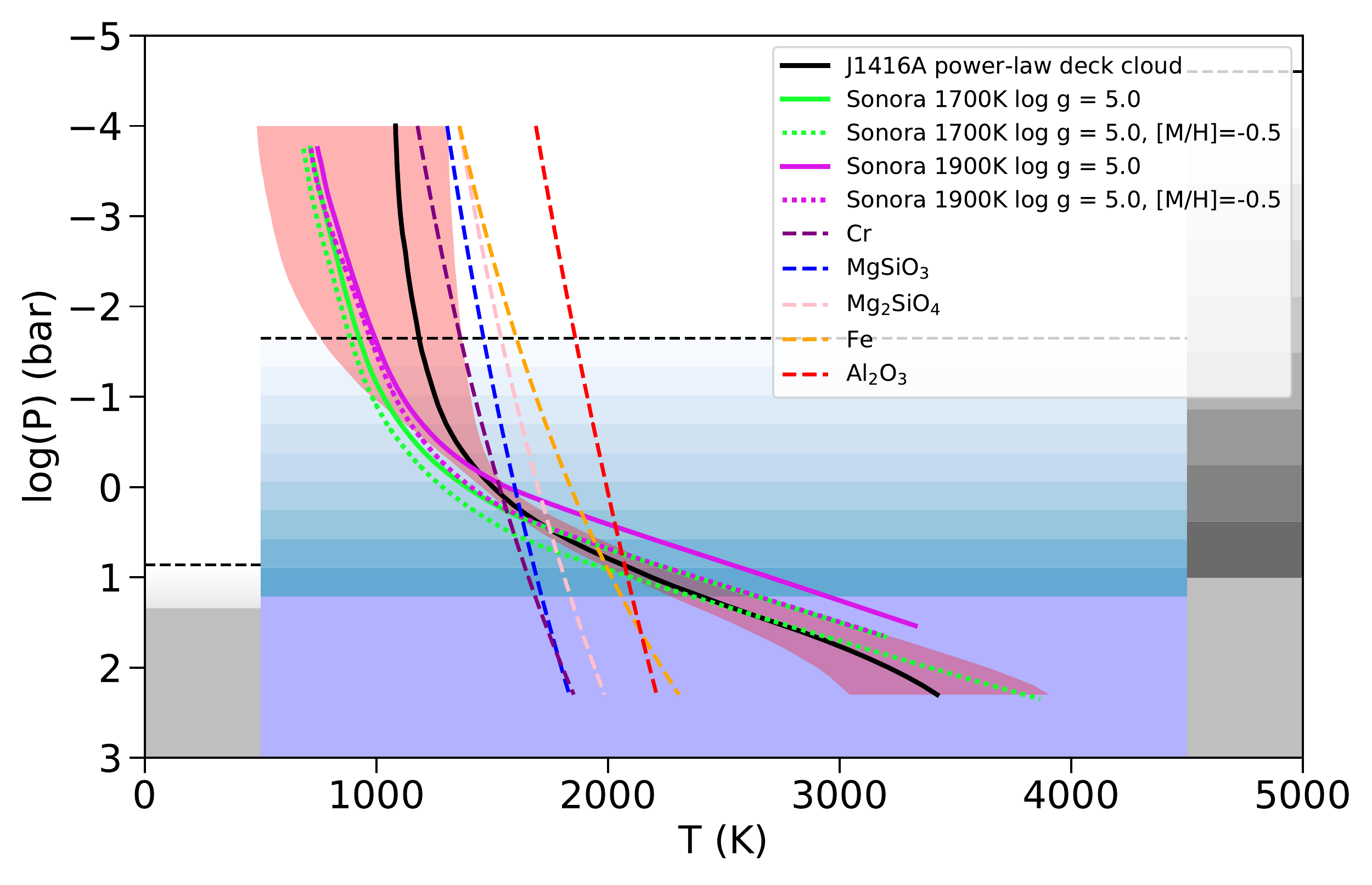}{0.5\textwidth}{\large(a)}
          \fig{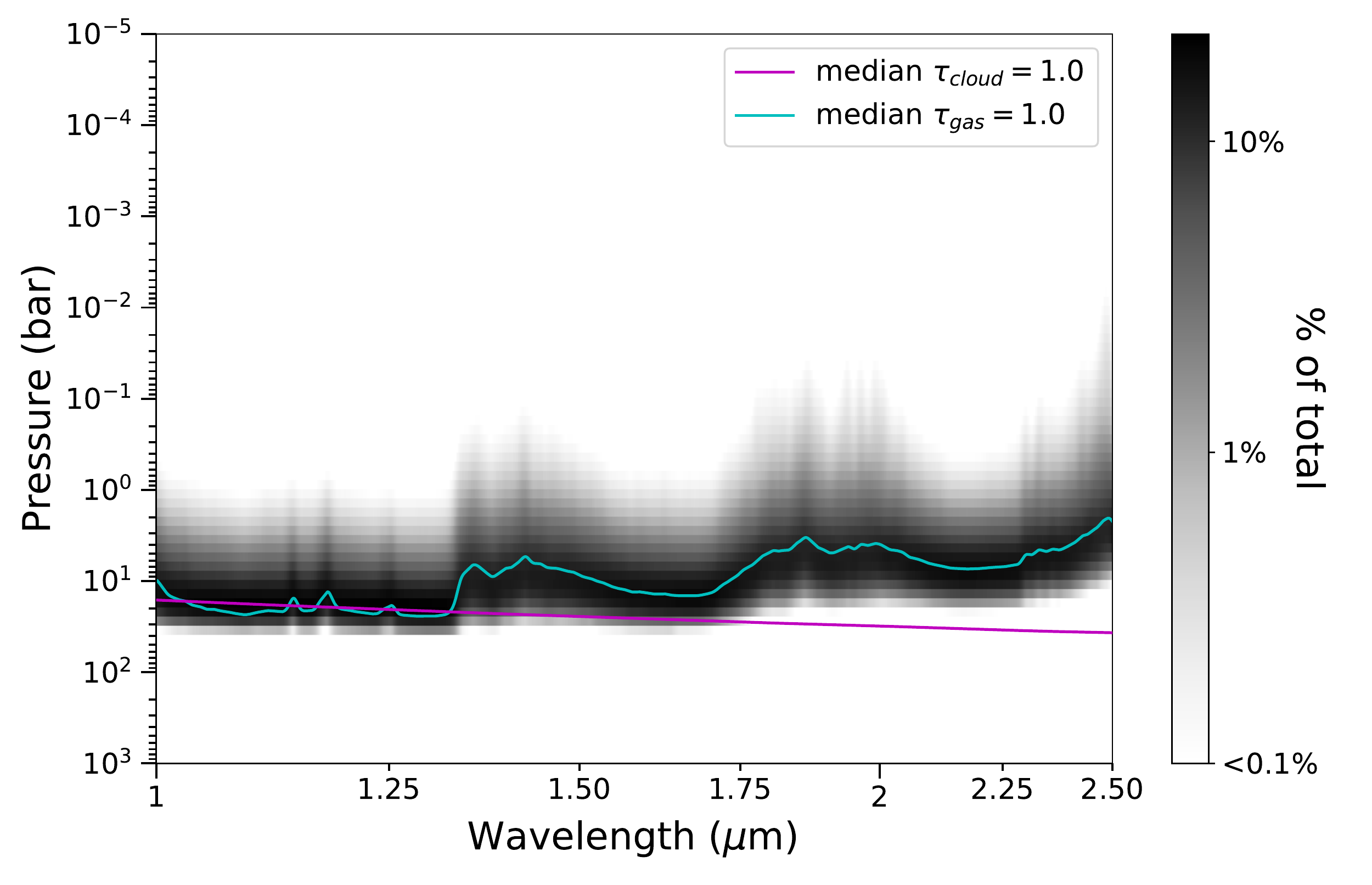}{0.5\textwidth}{\large(b)}} 
\caption{(a) Retrieved Pressure-Temperature Profile (black) compared to cloudless Sonora solar and low-metallicity model profiles similar to the SED-derived and retrieval-derived effective temperatures (neon green and purple). The median cloud deck is shown in shades of blue. The median deck reaches an optical depth of $\tau=1$ at the boundary between the darkest blue and purple. The purple region is where the cloud is optically thick and the blue shading indicates the vertical distribution where the cloud opacity drops to $\tau=0.5$ at the dash line. The grey bars on either side show the 1 $\sigma$ cloud deck location and vertical height distribution. The colored dashed lines are condensation curves for the listed species. (b) The contribution function associated with this cloud model, with the median cloud (magenta) and gas (aqua) at an optical depth of $\tau=1$ over plotted.}
\label{fig:1416A_deck_burrows_PT_profiles}
\vspace{0.5cm} 
\end{figure*}

\begin{figure*}
  \hspace{-0.25cm}
   \includegraphics[scale=.22]{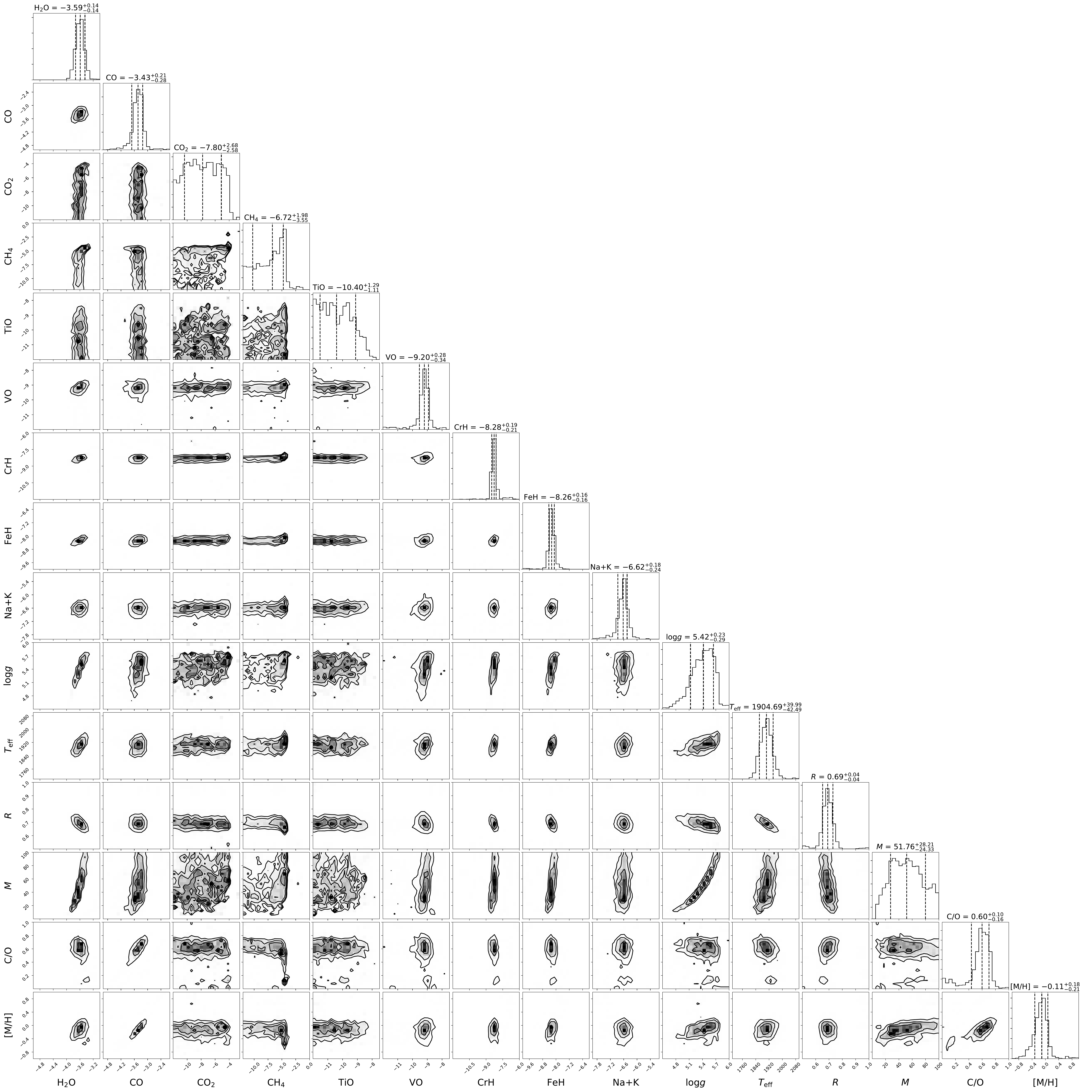}
\caption{J1416A power-law deck cloud posterior probability distributions for the retrieved parameters and extrapolated parameters. 1D histograms of the marginalized posteriors are shown along the diagonals with 2D histograms showing the correlations between the parameters. The dashed lines in the 1D histograms represent the 16\textsuperscript{th}, 50\textsuperscript{th}, and 84\textsuperscript{th} percentiles, with the 68\% confidence interval as the width between the 16\textsuperscript{th} and 84\textsuperscript{th} percentiles. Parameter values listed above are shown as the median~$\pm1\sigma$. Gas abundances are displayed as log$_{10}$(X) values, where X is the gas. \Teff, radius, mass, C/O ratio, and {[M/H]} are not directly retrieved parameters, but are calculated using the retrieved $R^2/D^2$ and log($g$) values along with the predicted spectrum. Our derived C/O ratio is absolute, where Solar C/O is 0.55, while our [M/H] is relative to Solar. CO$_2$, CH$_4$, and TiO abundances are not constrained and thus only provide upper limits.}
\label{fig:1416A_d2_89_burrows_postcorner}
\vspace{0.5cm} 
\end{figure*}

\begin{figure*}
  \centering
  \hspace{-0.25cm}
   \includegraphics[scale=.43]{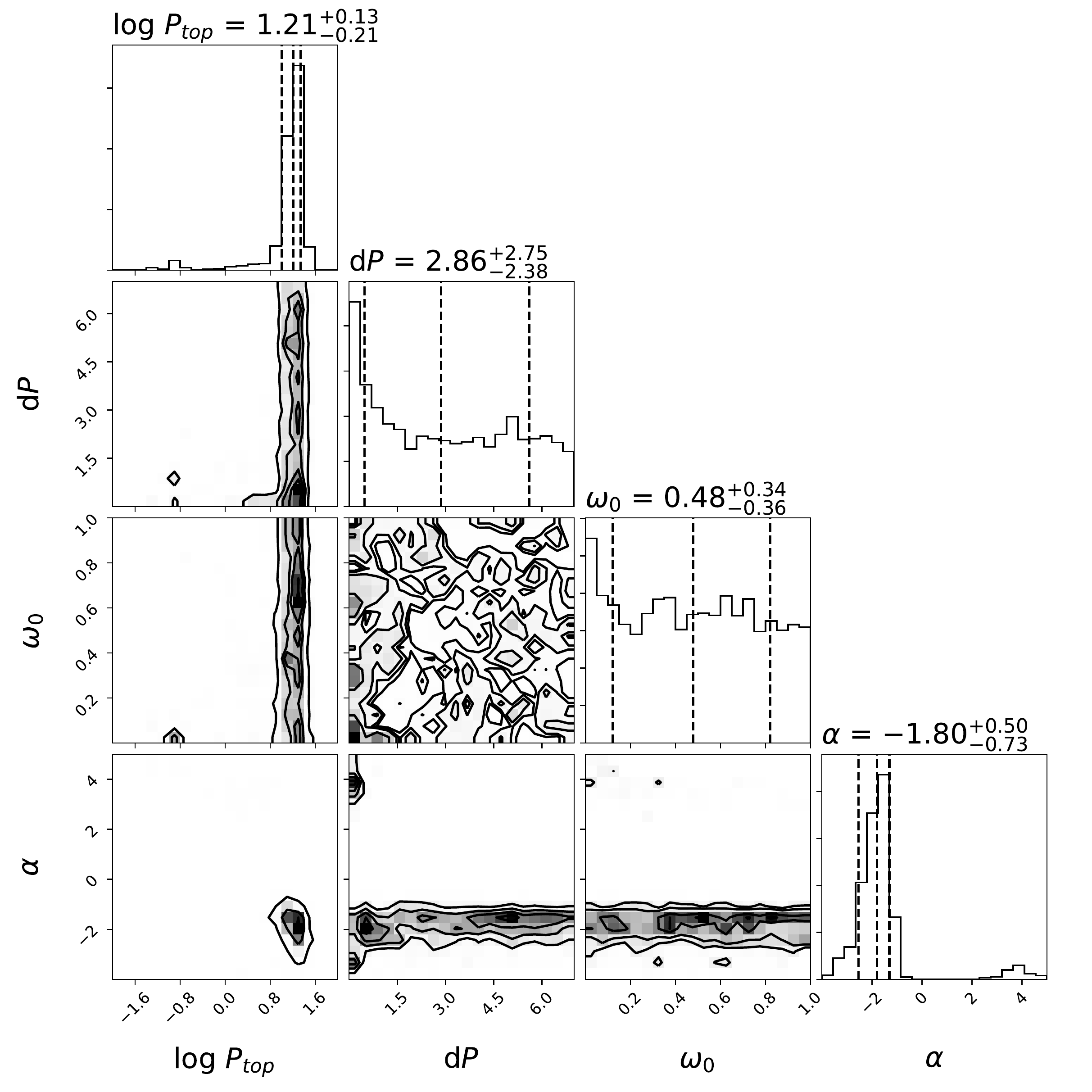}
\caption{J1416A power-law deck cloud posterior probability distributions for the cloud parameters. The cloud top pressure (log $P_{top}$) and the cloud height (d$P$) are shown in bars, and $\alpha$ is from the optical depth equation $\tau = \tau_0\lambda^\alpha$.}
\label{fig:1416A_d2_89_burrows_cloudcorner}
\end{figure*}

\begin{figure*}
\centering
 \gridline{\fig{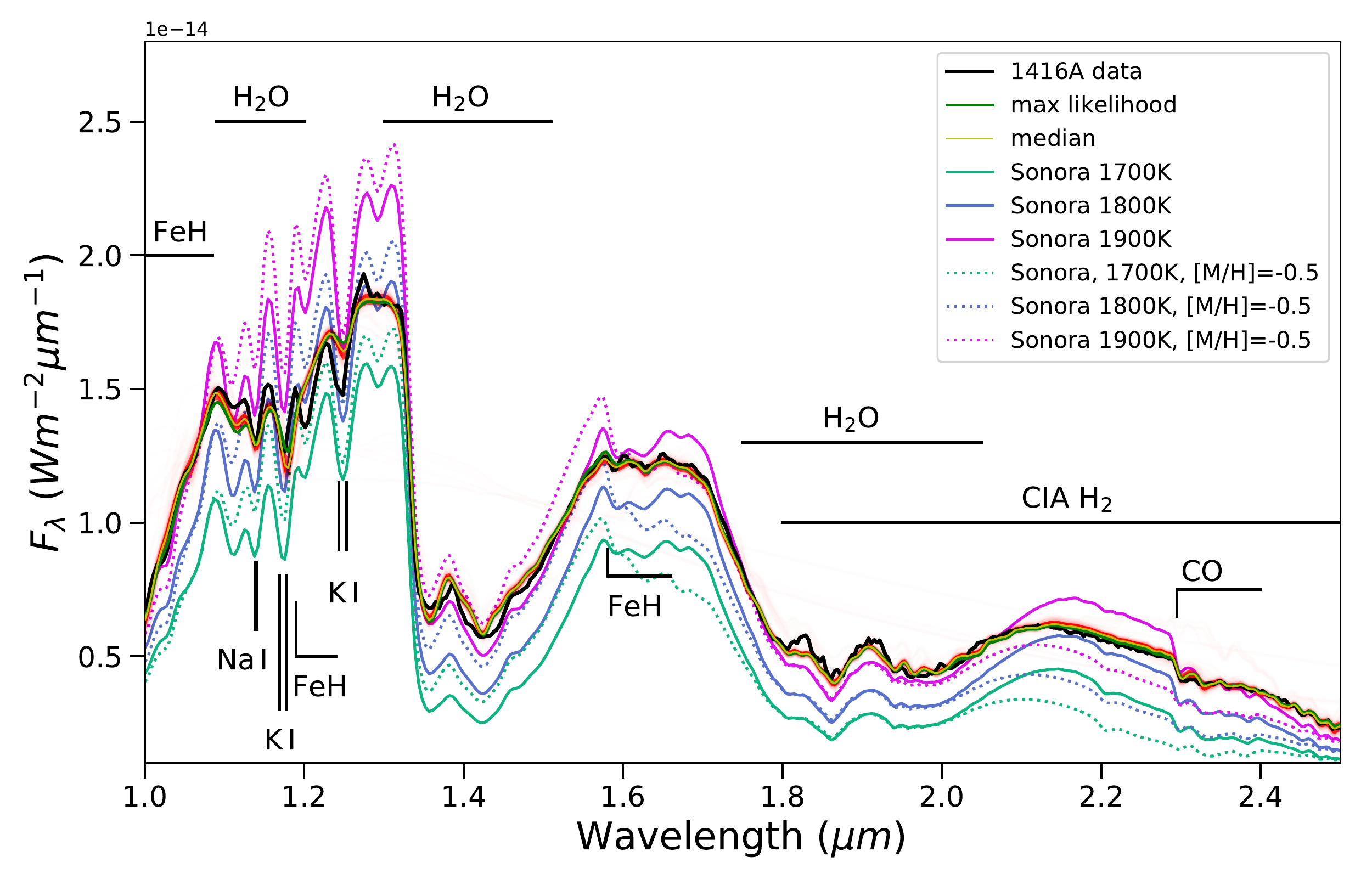}{0.5\textwidth}{\large(a)}
          \fig{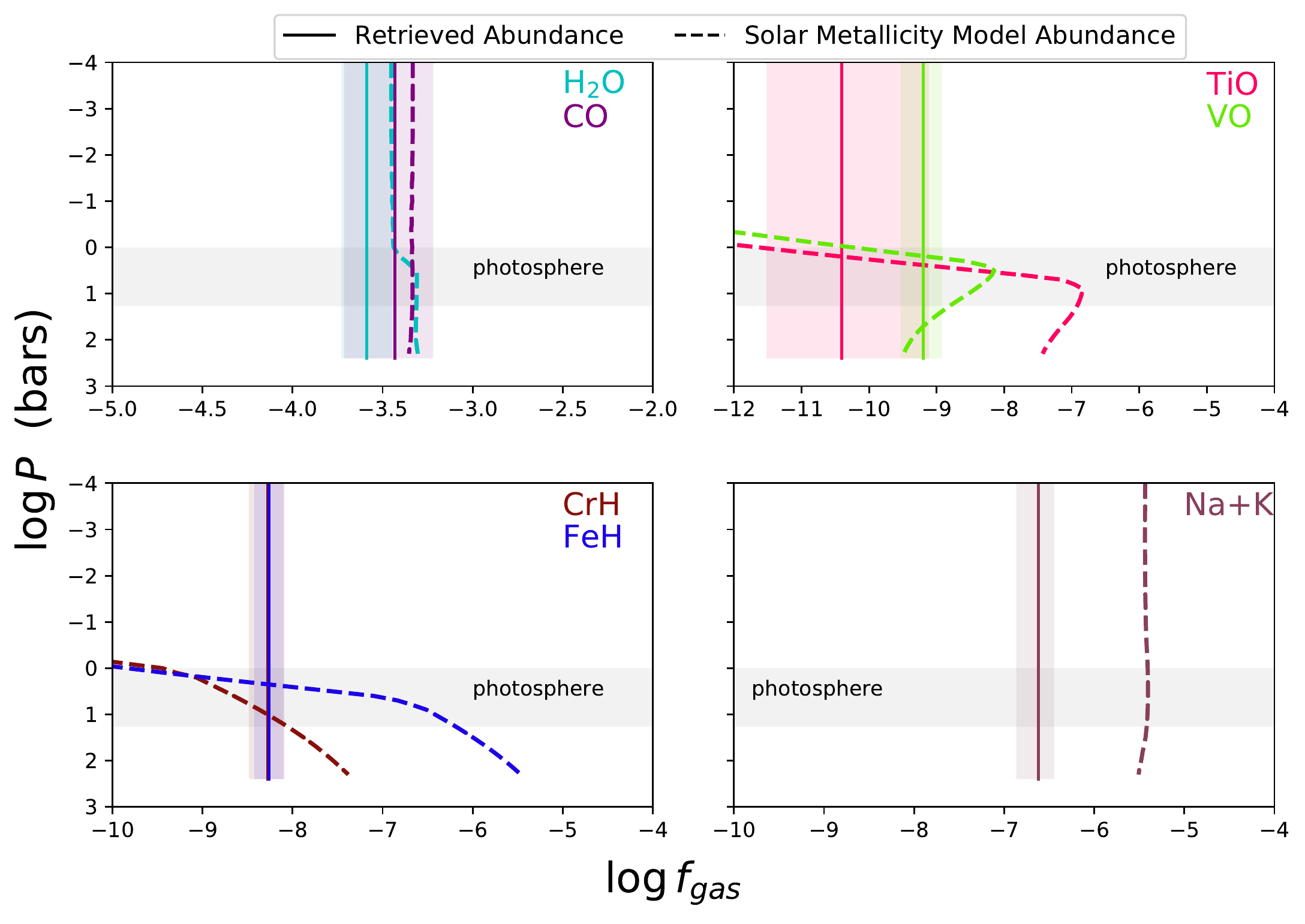}{0.5\textwidth}{\large(b)}} 
\caption{(a) Retrieved forward model spectra for the deck cloud model of J1416A. The maximum likelihood spectrum is shown in dark green, the median spectrum in yellow, and 500 random draws from the final 2000 samples of the EMCEE chain in red. The SpeX prism data is shown in black. For comparison the Sonora grid model solar metallicity spectra for log $g= 5.0$ and \Teff~$=1600$K, 1700K and 1800K (solid light green, teal, and blue), as well as [M/H]~$=-0.5$ for log $g= 5.0$ and \Teff~$=1800$K and 1900K (dotted blue and purple). These \Teff values bracket the range of the SED-derived and retrieval-derived \Teff. (b) Retrieved uniform-with-altitude mixing abundances for constrained gases compared to Solar metallictiy and C/O model abundances. The approximate location of the photosphere is shown in gray.}
\label{fig:1416A_decK_SPEC_VMR_burrows}
\vspace{0.5cm} 
\end{figure*}

\begin{figure*}[h!]
\gridline{\fig{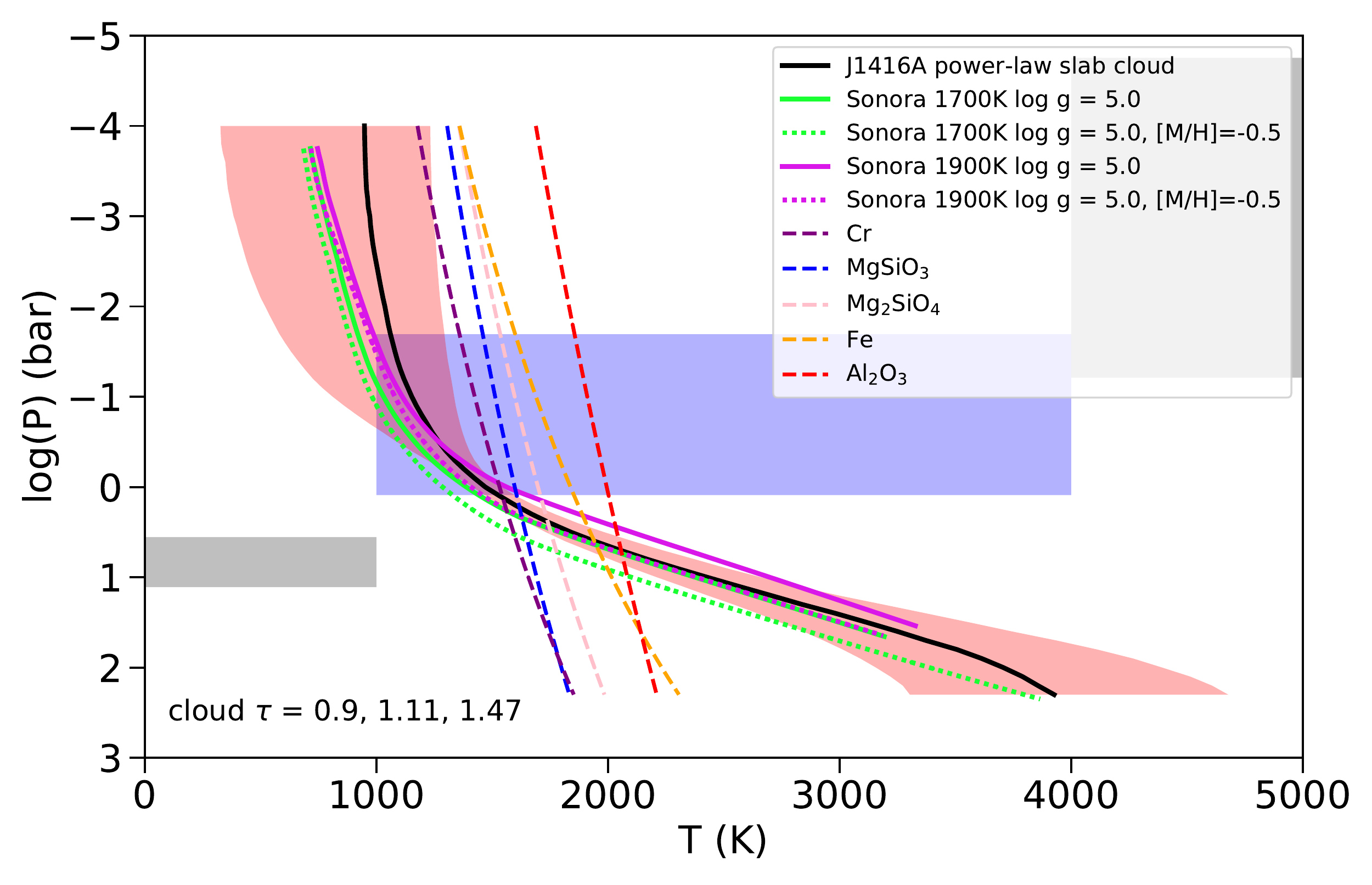}{0.5\textwidth}{\large(a)}
          \fig{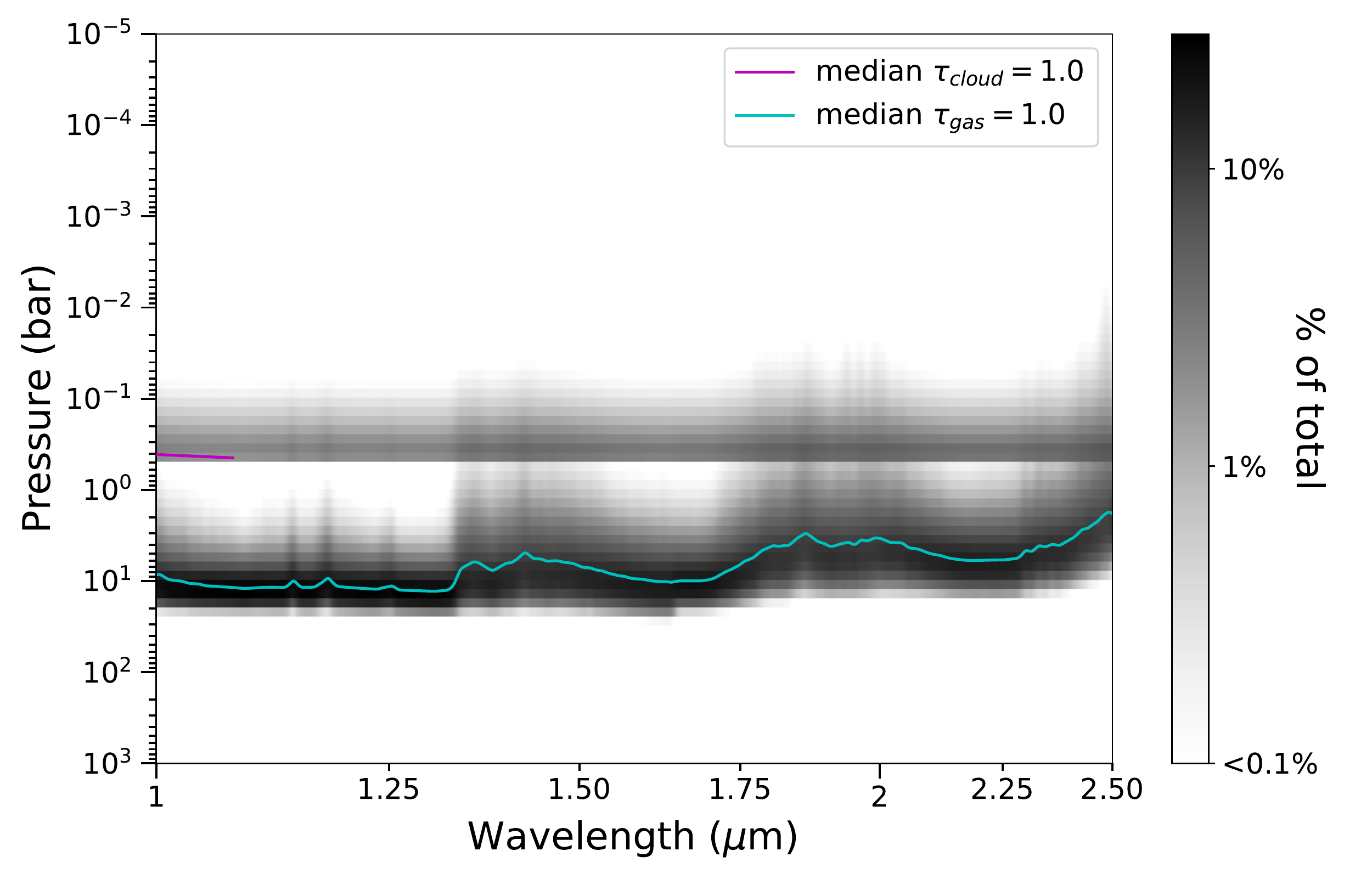}{0.5\textwidth}{\large(b)}} 
\caption{(a) Retrieved Pressure-Temperature Profile (black) compared to cloudless Sonora solar and low-metallicity model profiles similar to the SED-derived and retrieval-derived effective temperatures (neon green and purple). The median cloud slab height and location is shown purple with the 1 $\sigma$ shown in grey, indicating the ranges of height and base locations. Optical depth for the cloud is shown in the bottom left corner. The colored dashed lines are condensation curves for the listed species. (b) The contribution function associated with this cloud model, with the median cloud (magenta) and gas (aqua) at an optical depth of $\tau=1$ over plotted.}
\label{fig:1416A_slab_burrows_PT_profiles}
\vspace{0.5cm} 
\end{figure*}

\begin{figure*}
  \hspace{-0.25cm}
   \includegraphics[scale=.22]{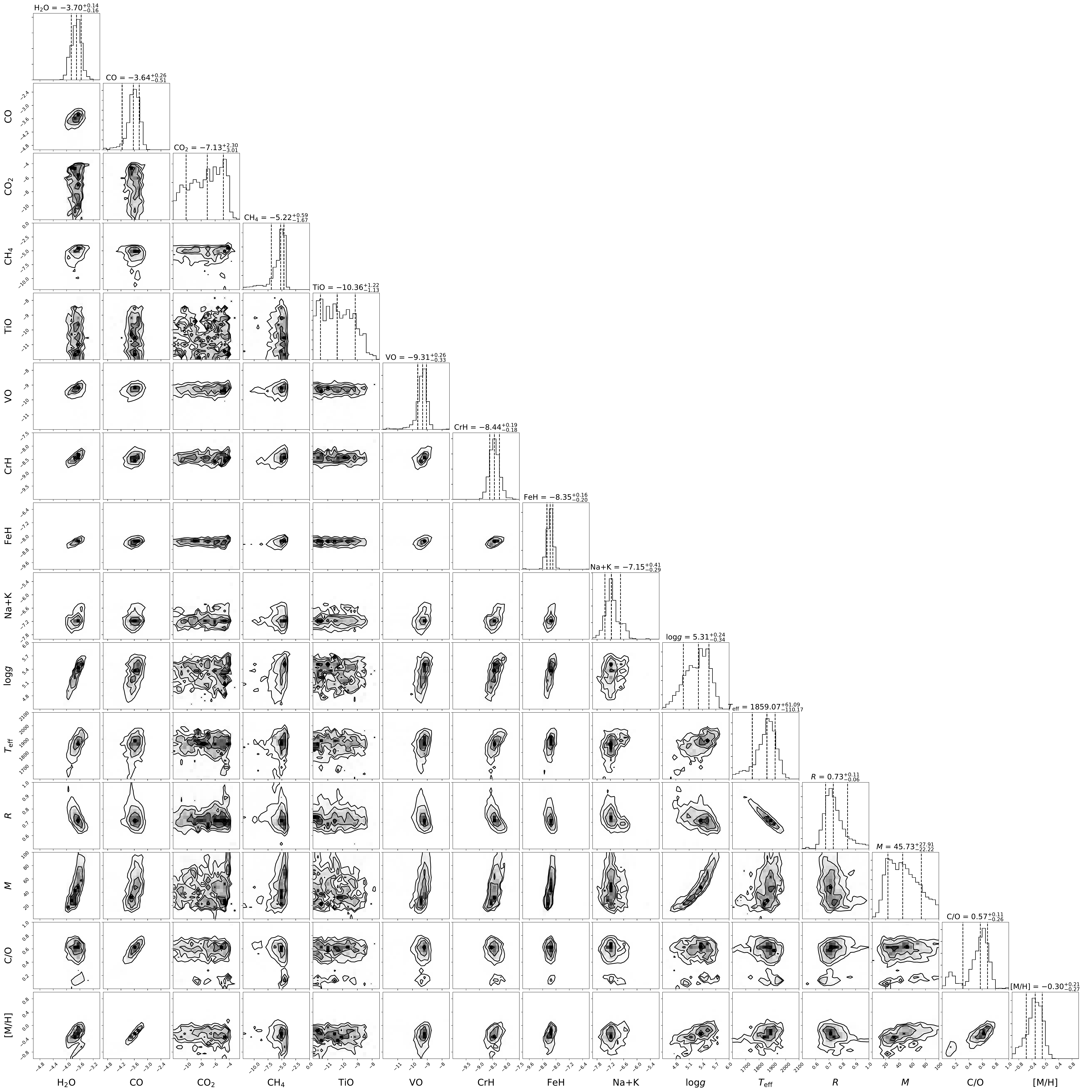}
\caption{J1416A power-law slab cloud posterior probability distributions for the retrieved parameters and extrapolated parameters. 1D histograms of the marginalized posteriors are shown along the diagonals with 2D histograms showing the correlations between the parameters. The dashed lines in the 1D histograms represent the 16\textsuperscript{th}, 50\textsuperscript{th}, and 84t\textsuperscript{th} percentiles, with the 68\% confidence interval as the width between the 16\textsuperscript{th} and 84\textsuperscript{th} percentiles. Parameter values listed above are shown as the median~$\pm1\sigma$. Gas abundances are displayed as log$_{10}$(X) values, where X is the gas. \Teff, radius, mass, C/O ratio, and {[M/H]} are not directly retrieved parameters, but are calculated using the retrieved $R^2/D^2$ and log($g$) values along with the predicted spectrum. Our derived C/O ratio is absolute, where Solar C/O is 0.55, while our [M/H] is relative to Solar. CO$_2$ and TiO abundances are not constrained and thus only provide upper limits.}
\label{fig:1416A_s1_89_burrows_postcorner}
\end{figure*}

\begin{figure*}
  \centering
  \hspace{-0.25cm}
   \includegraphics[scale=.36]{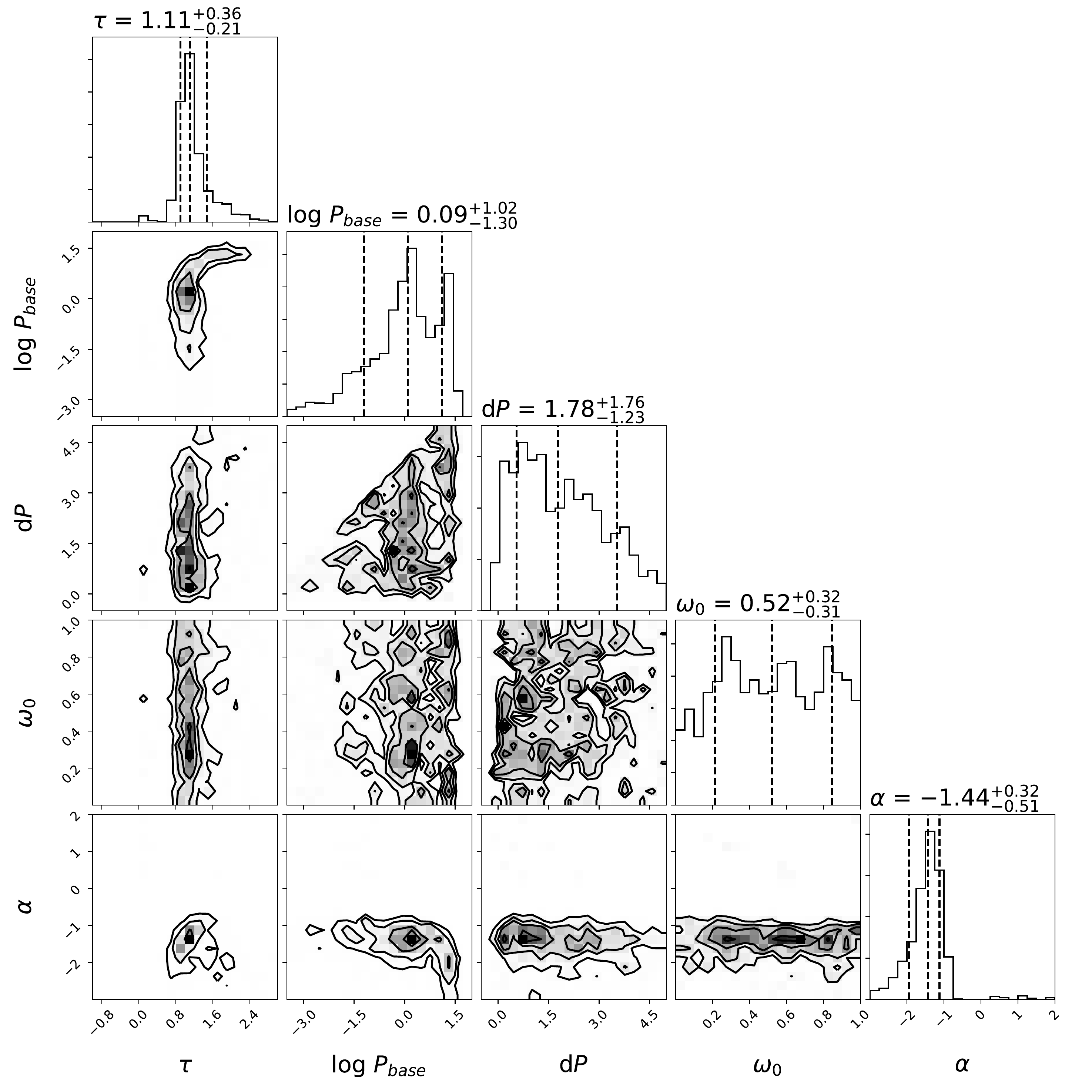}
\caption{J1416A power-law slab cloud posterior probability distributions for the cloud parameters. The cloud top pressure (log $P_{top}$) and the cloud height (d$P$) are shown in bars, and $\alpha$ is from the optical depth equation $\tau = \tau_0\lambda^\alpha$.}
\label{fig:1416A_s1_89_burrows_cloudcorner}
\end{figure*}

\begin{figure*}
\centering
 \gridline{\fig{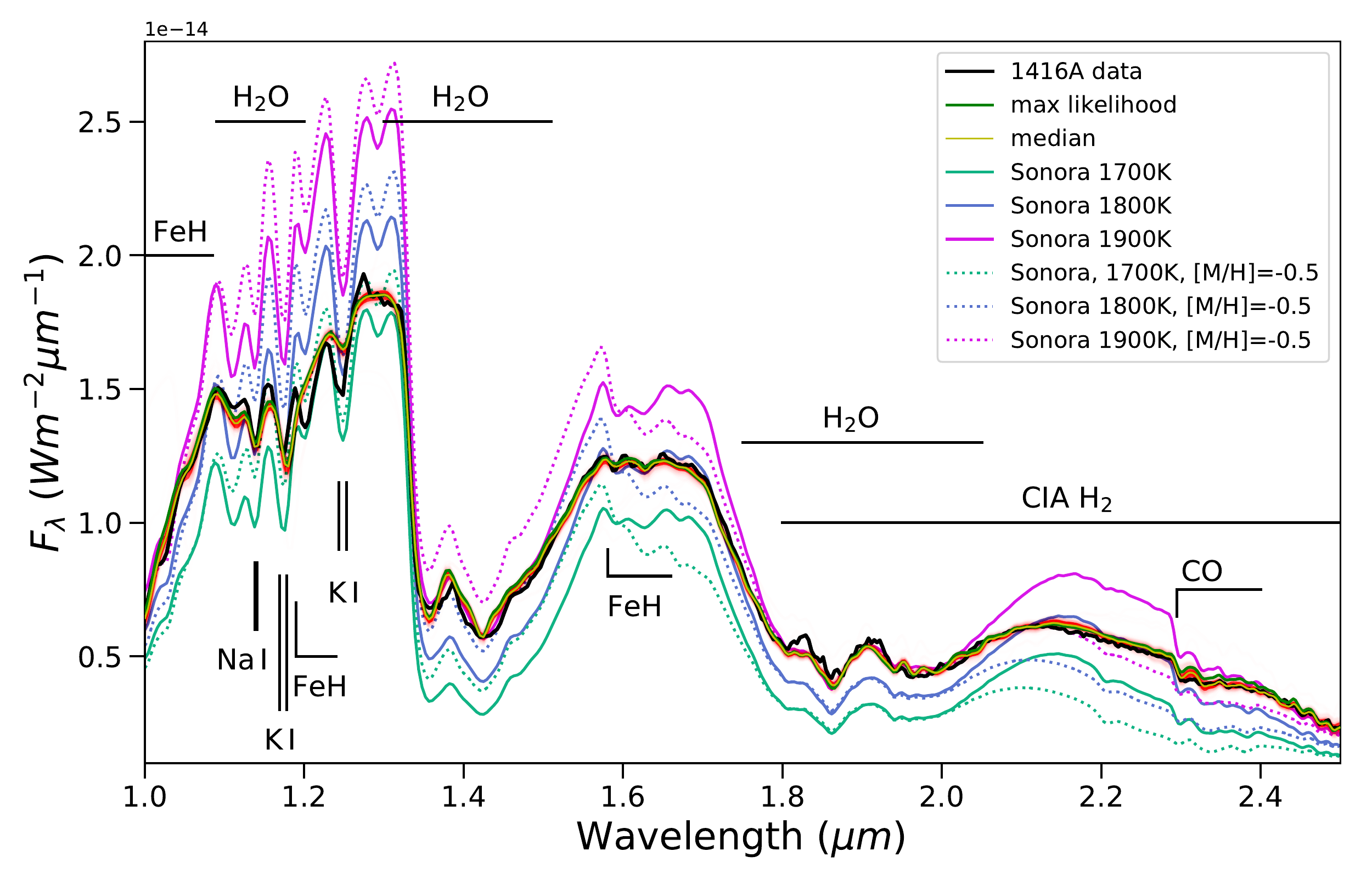}{0.5\textwidth}{\large(a)}
          \fig{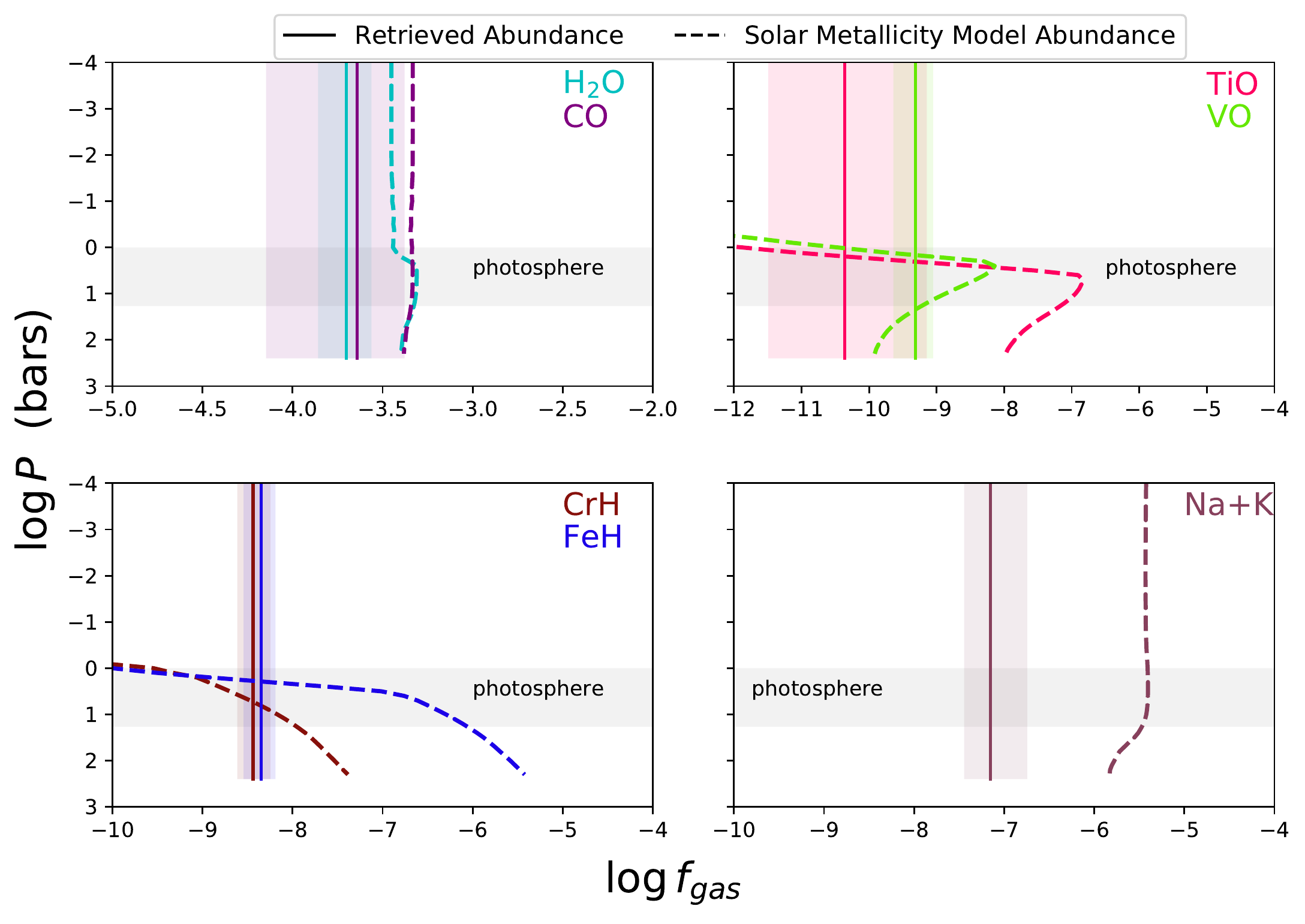}{0.5\textwidth}{\large(b)}} 
\caption{(a) Retrieved forward model spectra for the slab cloud model of J1416A. The maximum likelihood spectrum is shown in dark green, the median spectrum in yellow, and 500 random draws from the final 2000 samples of the EMCEE chain in red. The SpeX prism data is shown in black. For comparison the Sonora grid model solar metallicity spectra for log $g= 5.0$ and \Teff~$=1600$K, 1700K and 1800K (solid teal, blue, and purple), as well as [M/H]~$=-0.5$ for log $g= 5.0$ and \Teff~$=1800$K and 1900K (dotted teal, blue, and purple). These \Teff values bracket the range of the SED-derived and retrieval-derived \Teff. (b) Retrieved uniform-with-altitude mixing abundances for constrained gases compared to Solar metallictiy and C/O model abundances. The approximate location of the photosphere is shown in gray.}
\label{fig:1416A_slab_SPEC_VMR_burrows}
\vspace{0.5cm} 
\end{figure*}

Here we show the resultant figures for the power-law deck (Figures~\ref{fig:1416A_deck_burrows_PT_profiles}--\ref{fig:1416A_decK_SPEC_VMR_burrows}) and slab (Figures~\ref{fig:1416A_slab_burrows_PT_profiles}--\ref{fig:1416A_slab_SPEC_VMR_burrows}) clouds using the Burrows alkali cross-sections. These models are also indistinguishable from the winning models with their corresponding BIC values are listed in Table \ref{tab:1416AModels} in Section \ref{sec:Retrieval_Models_A}.

\subsubsection{Deck Cloud Alternative Alkalies}

\clearpage
\subsubsection{Slab Cloud Alternative}

\clearpage
\subsection{J1416B Burrows Models}\label{sec:1416B_Burrows_alk}
\begin{figure*}[h!]
 \gridline{\fig{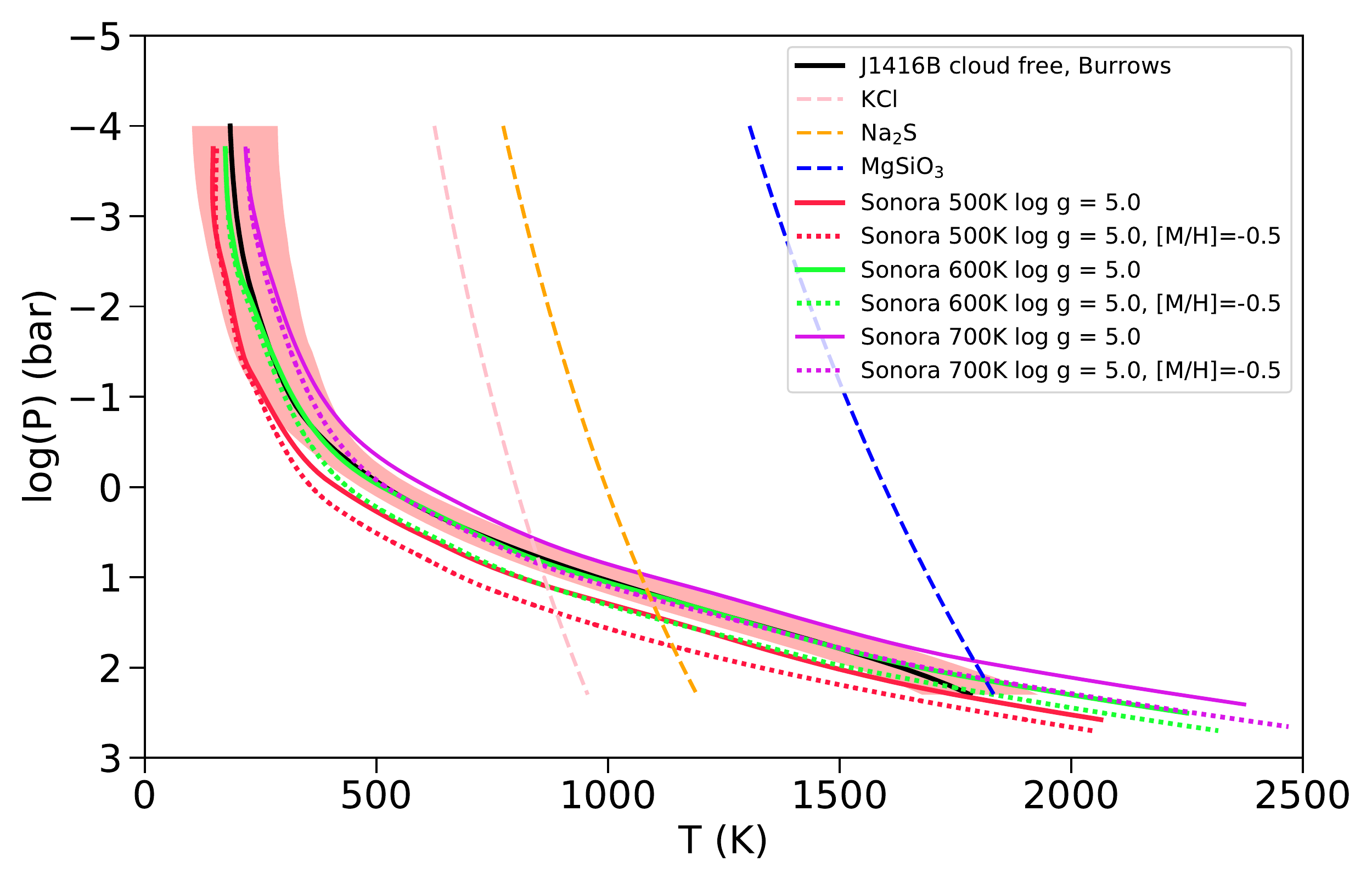}{0.5\textwidth}{\large(a)}
          \fig{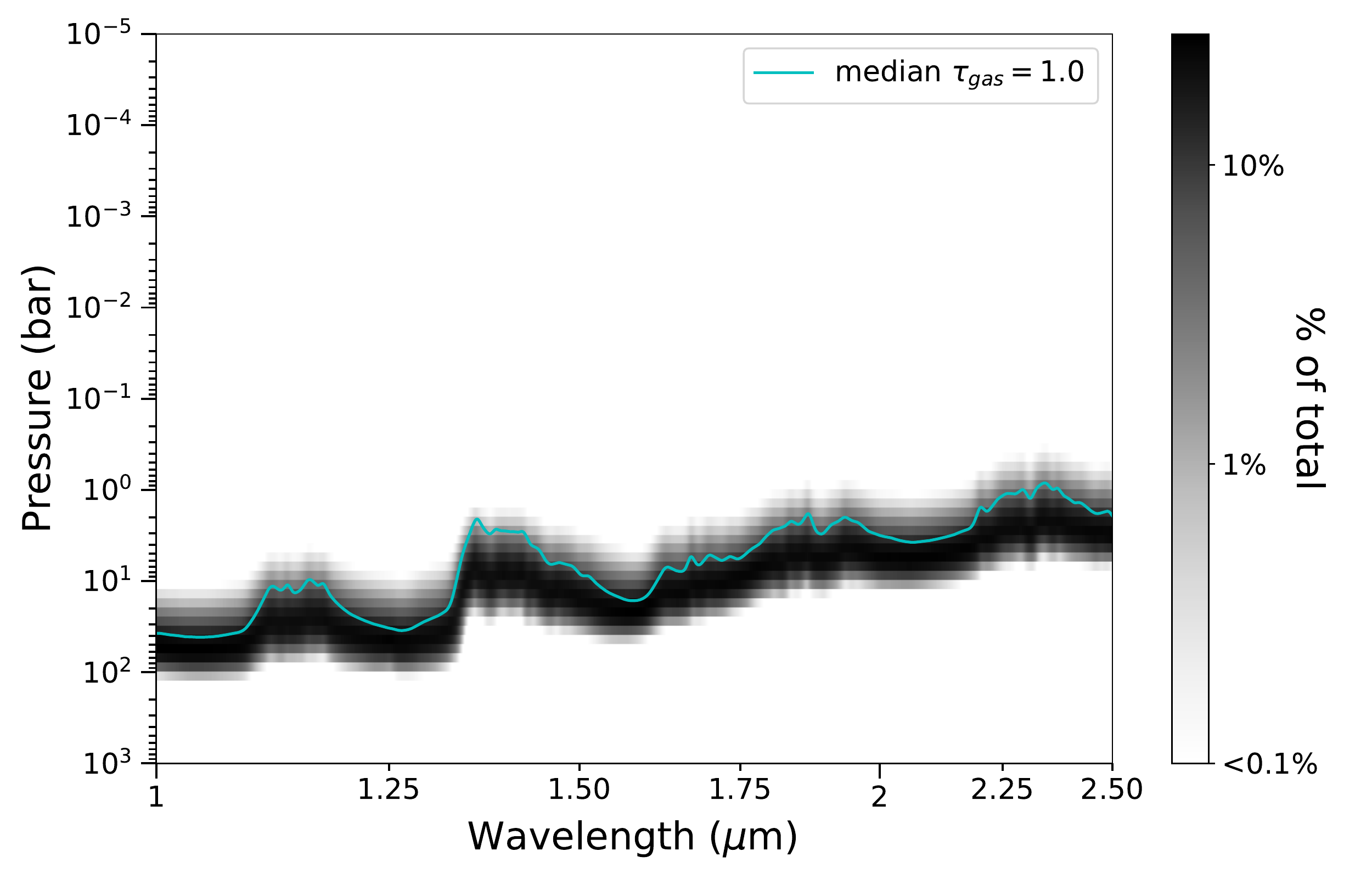}{0.5\textwidth}{\large(b)}} 
\caption{(a)Retrieved Pressure-Temperature Profile (black) compared to cloudless Sonora solar and low-metallicity model profiles (neon green, purple and bright pink). (b) Contribution plot with maximum likelihood gas at $\tau= 1$.}
\label{fig:1416b_PT_profile_burrows}
\vspace{0.5cm} 
\end{figure*}

\begin{figure*}
  \hspace{-0.25cm}
   \includegraphics[scale=.28]{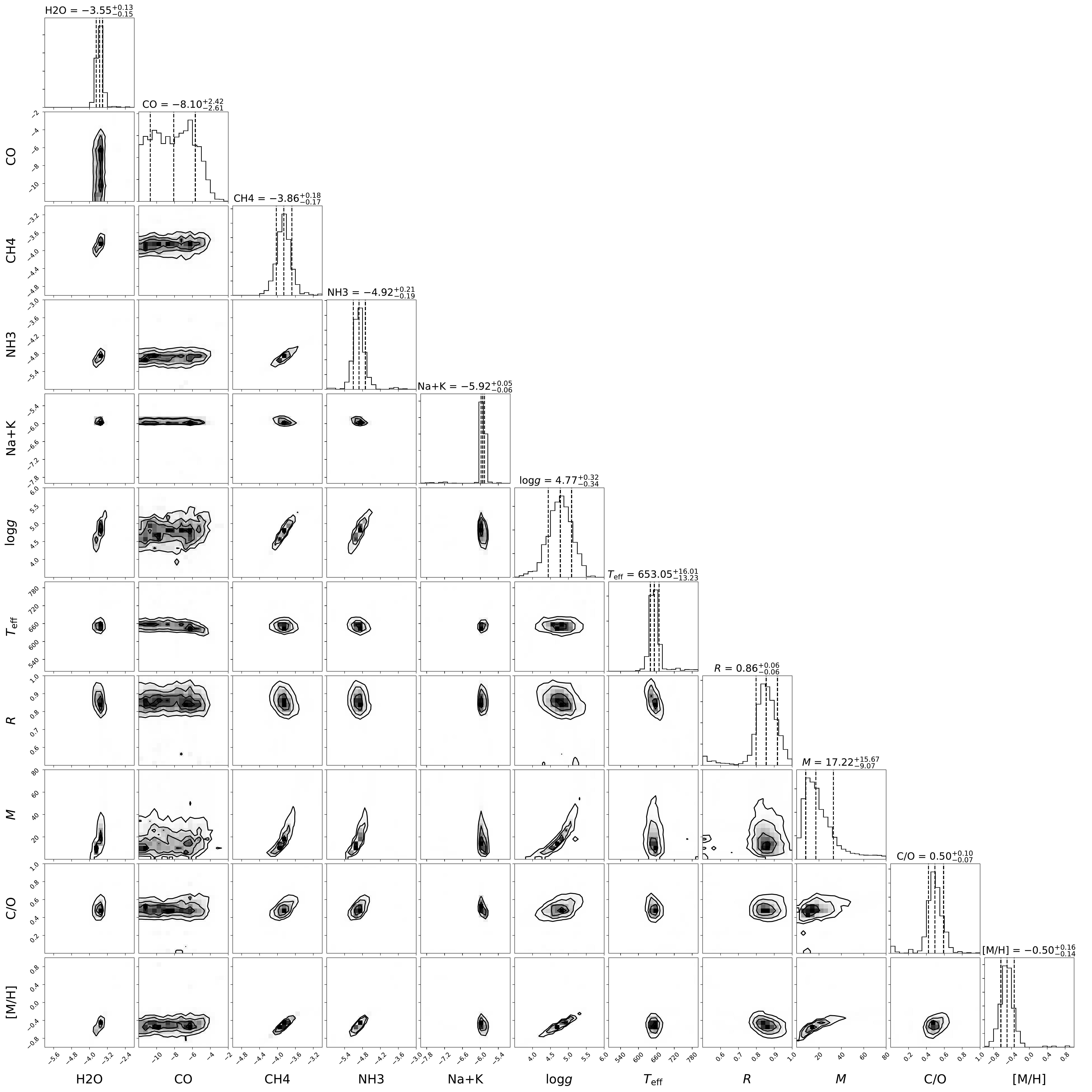} 
\caption{J1416B cloud-free posterior probability distributions for the retrieved parameters using the Burrows alkalis. \Lbol, \Teff, radius, mass, C/O ratio, {[Fe/H]}, and {[M/H]} are not directly retrieved parameters, but are calculated using the retrieved $R^2/D^2$ and log $g$ values along with the predicted spectrum. CO abundance is not constrained and thus only provides an upper limit.}
\label{fig:1416b_nc_burrows_postcorner}
\vspace{0.5cm} 
\end{figure*}

\begin{figure*}
\centering
 \gridline{\fig{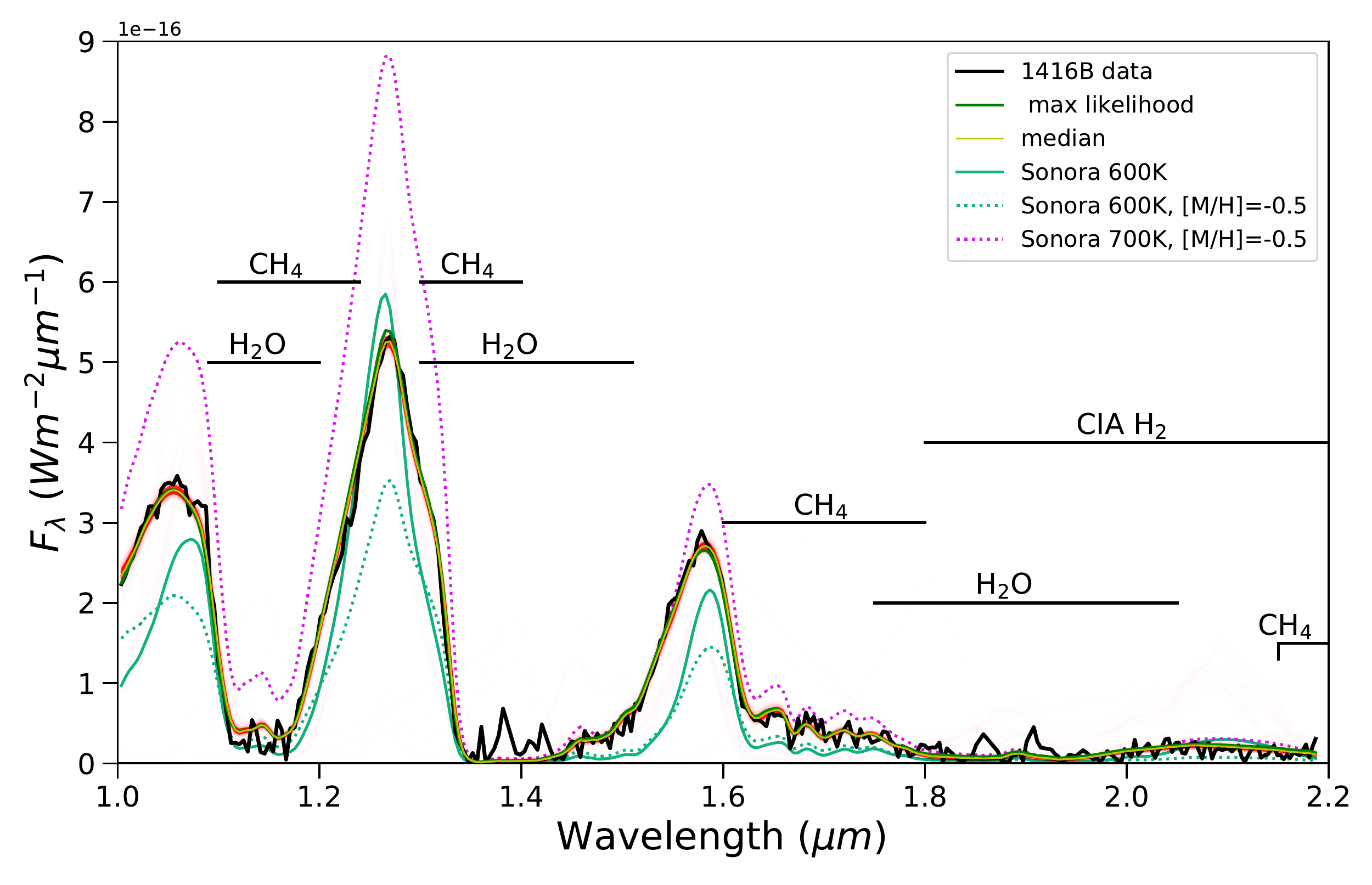}{0.5\textwidth}{\large(a)}
          \fig{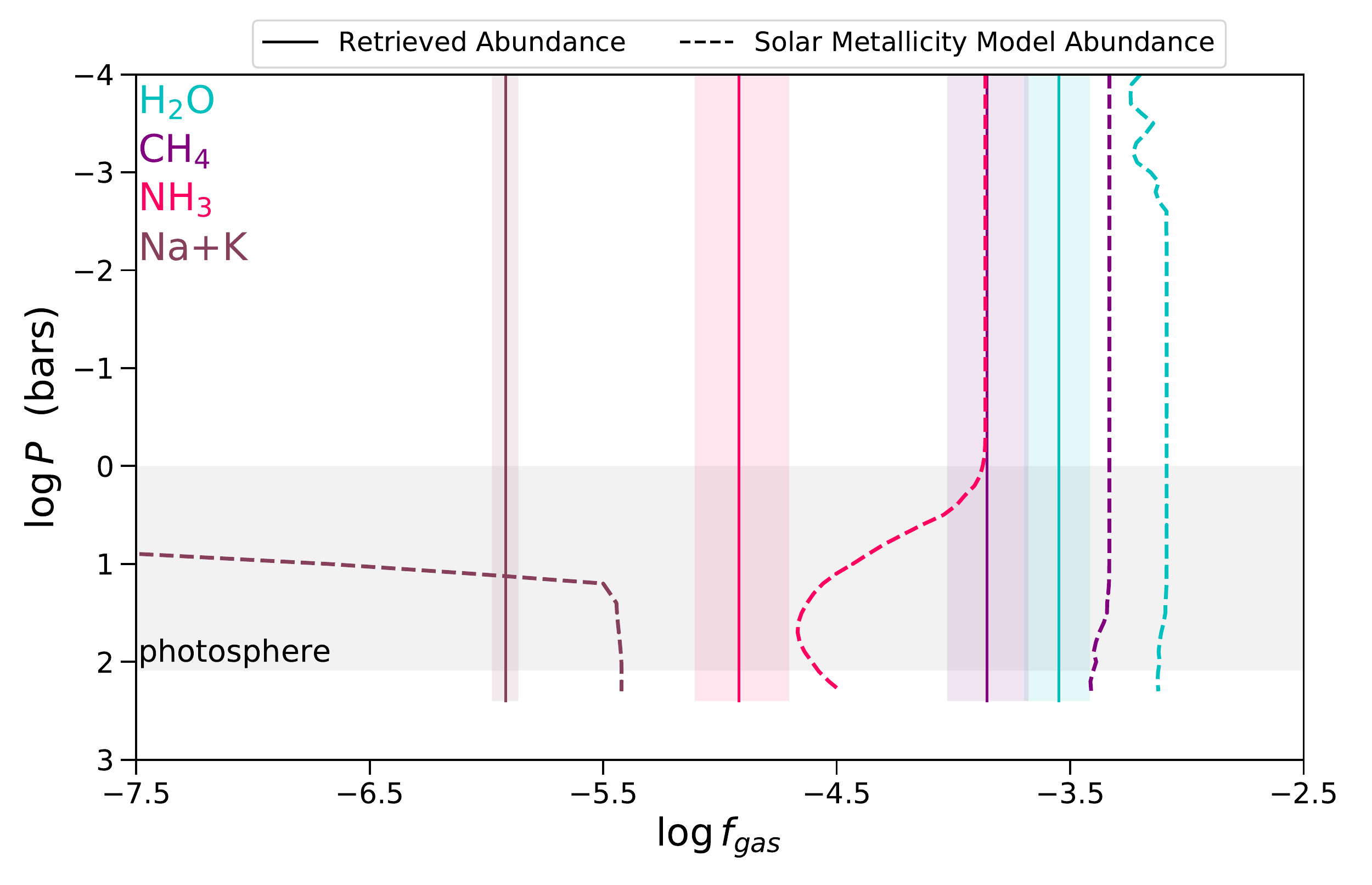}{0.5\textwidth}{\large(b)}} 
\caption{(a) Data (in black) compared to the retrieved maximum likelihood (in green) and median (in yellow) spectra. In red we show 500 random draws from the final 5,000 walkers of the converged MCMC chain. Sonora solar and low-metallicity model spectra are shown in teal and purple, respectively. (b) Retrieved uniform-with-altitude gas abundances for the cloudless Burrows alkali model compared to solar abundances.}
\label{fig:1416B_nc_SPEC_VMR_burrows}
\vspace{0.5cm} 
\end{figure*}

Figures~\ref{fig:1416b_PT_profile_burrows}--\ref{fig:1416B_nc_SPEC_VMR_burrows} show the cloud-free Burrows alkali cross-section model for J1416B, which present a better fit to the data, however produces inconsistent alkali abundances between J1416A and J1416B.

\clearpage
\bibliographystyle{yahapj}
\bibliography{references}

\end{document}